\documentclass[twocolumn,superscriptaddress,showpacs,prd,
aps,amsmath,amssymb,nofootinbib,eqsecnum]{revtex4-1}
\usepackage{hyperref}
\usepackage{graphicx}%https://www.overleaf.com/project/5d999fc231c7f60001a7f52d
\usepackage{subcaption}
\usepackage{bm} % bold math
\usepackage{mathrsfs} % -> \mathscr
\usepackage{url}
\usepackage{color}
\usepackage{comment}
\newcommand{\be}{\begin{equation}}
\newcommand{\ee}{\end{equation}}
\newcommand{\bea}{\begin{eqnarray}}
\newcommand{\eea}{\end{eqnarray}}

\newcommand{\se}{\mathfrak{S}}

\newcommand{\eps}{\epsilon}

\usepackage{amsmath,amsfonts,amssymb,amsthm}
\usepackage{mathtools}
\usepackage{commath}

\begin{document}

\title{A  Parametrized Equation of State for Neutron Star Matter with Continuous Sound Speed}
\author{Michael F. O'Boyle} 
\email{moboyle2@illinois.edu}
\affiliation{
Department of Physics, University of Illinois at Urbana-Champaign, Urbana, IL 61801, USA}
\author{Charalampos Markakis} 
\email{c.markakis@damtp.cam.ac.uk}
\affiliation{
DAMTP, University of Cambridge, Wilberforce Rd, Cambridge CB3 0WA, UK}
\affiliation{School of Mathematical Sciences, Queen Mary University of London, Mile End Rd, 
London E1 4NS, UK}
\affiliation{NCSA, University of Illinois at Urbana-Champaign, Urbana, IL 61801, USA}

%\author{Roland Haas} 
%\email{charalampos.markakis@uni-jena.de}
%\affiliation{
%NCSA, University of Illinois at %Urbana-Champaign, Urbana, IL 61801, USA
%}

\author{Nikolaos Stergioulas} 
\email{niksterg@auth.gr}
\affiliation{Department  of  Physics,  Aristotle  University  of  Thessaloniki,
University  Campus,  54124,  Thessaloniki,  Greece}

\author{Jocelyn S. Read} 
\email{jread@fullerton.edu}
\affiliation{GWPAC, California State University Fullerton, Fullerton, CA 92831, USA}

\date{\today}  

\begin{abstract} 
We present a generalized piecewise polytropic parameterization for the neutron-star equation of state  using an ansatz that imposes  continuity in not only pressure and energy density, but also in the speed of sound. %predicted by candidate equations of state. 
The universe of candidate equations of state is shown to admit preferred dividing densities, determined by minimizing an error norm consisting of integral astrophysical observables. Generalized piecewise polytropes accurately reproduce astrophysical observables, such as mass, radius, tidal deformability and mode frequencies, as well as thermodynamic quantities, such as the adiabatic index. This makes the new EOS useful for  parameter estimation from gravitational waveforms.
Since they are differentiable, generalized piecewise polytropes can improve pointwise convergence in numerical relativity simulations of neutron stars. Existing implementations of piecewise polytropes can easily accommodate this generalization with the same number of free parameters. Optionally, generalized piecewise polytropes can also accommodate adjustable jumps in sound speed, which allows them to capture phase transitions in neutron star matter.

\end{abstract} 

%\pacs{04.20.-q, 04.40.Dg, 04.40.Nr, 52.30.Cv, 95.30.Qd}

\maketitle

\section{Introduction}
A long outstanding problem in nuclear physics is determining the correct thermodynamic equation of state (EOS) for cold matter above nuclear saturation density \cite{RevModPhys.89.015007,Baym:2017whm}. The only locations in the universe where such matter is believed to exist are the cores of neutron stars (NS), remnants of the gravitational collapse of moderately massive main sequence stars. A current goal of relativistic astrophysics is to use NS measurements to constrain the nuclear EOS. Past studies have relied on electromagnetic techniques, such as Shapiro time delay mass measurements \cite{Antoniadis:2013pzd,Arzoumanian:2017puf} and thermal \cite{Fortin:2014mya,Ozel:2015fia} and X-ray \cite{Bogdanov:2012md,Verbiest:2008gy} radius determinations. In principle, rotation frequencies can provide additional constraints. However, all observed NSs are well described by the slowly-rotating approximation \cite{Haensel_2009}, so little information can be gathered.

Now, with the dawn of gravitational wave (GW) astronomy, it is possible to probe NSs by studying the waveforms produced by binary inspirals \cite{PhysRevLett.119.161101,PhysRevLett.121.161101,Abbott_2020,PhysRevLett.120.172703,wysocki2020inferring}. GWs from such events are strongly influenced by NS masses and tidal deformabilities \cite{Buonanno:2009zt}, a parameter quantifying how the NS deforms when an external gravitational field is applied \cite{Flanagan:2007ix,Yagi:2013awa,VanOeveren:2017xkv}. The deformability depends on the matter in the star, so GW observations provide an additional means to study the NS EOS \cite{Wade:2014vqa,Vivanco:2019qnt}. 

Realistic EOSs are constructed from nontrivial micophysics, while an observation can reasonably constrain just a few parameters. Read \textit{et al.} \cite{Read:2008iy,Read:2009yp, markakisNeutronStarEquation2009, markakisInferringNeutronStar2012, readMatterEffectsBinary2013} developed a formalism modeled after one used by Vuille and Ipser \cite{Vuille:1999} where the high density region is partitioned into three intervals and a polytropic (power law) form for the pressure vs mass density curve is applied in each interval. The result is a four parameter ansatz termed \textit{piecewise polytropes} (PPs). This approximation reproduces the EOS it tries to capture fairly well, and it makes reasonably accurate predictions of certain integral observables (mass, radius, and moment of inertia). However, the speed of sound $c_{\rm s}$ in this approximation is discontinuous. The tidal deformability is sensitive to $c_{\rm s}$ \cite{Hinderer:2007mb}, and it affects oscillations \cite{Kastaun:2008jr,westernacher-schneiderHamiltonJacobiHydrodynamics2020} and hydrodynamics \cite{demetrios_christodoulou_compressible_2014,markakisHamiltonianHydrodynamicsIrrotational2014} in numerical simulation.

Lindblom introduced an alternative to PPs termed a \textit{spectral} expansion, where the logarithm of adiabatic index $\gamma$ is fitted with polynomials in the logarithm of pressure. The advantage of this formalism is its smoothness over the whole core region \cite{Lindblom:2010bb}. The LSC uses this formalism when performing parameter estimation on the neutron star EOS from gravitational wave observations \cite{Abbott:2018exr}. However, to recover the primitive variables required for hydrodynamic simulations (pressure and energy density), one must integrate the expansions of the adiabatic index. Closed-form expressions do not exist for the relevant integrals, so a numeric quadrature must be performed every time a thermodynamic quantity is evaluated. 

Foucart \textit{et al.} have recently performed a full numerical relativity simulation of a binary neutron star inspiral using Lindblom's spectral formalism for the cold EOS. They also introduced a procedure for matching the parameterized core to a known crust EOS that ensures both continuity and differentiability. However, they report that a few-parameter spectral expansion cannot accurately reproduce the observable curves predicted by the original EOS (e.g. the mass-radius relation). Instead, they demand that the expansion accurately reproduce the maximum mass and radius of a 1.35 $M_{\odot}$ star predicted by the original EOS \cite{Foucart:2019yzo}.

This work introduces an extension of the piecewise polytrope formalism that ensures a continuous sound speed and accurately reproduces the observables predicted by the original EOS. The result is an algebraically simple expression that can be easily implemented in simulations and avoids the difficulties of a non-differentiable formalism. It thus has the simplicity of standard PPs and the desirable properties of the spectral expansion.

In Section \ref{sec:Barotropic} we review the thermodynamics and hydrodynamics of barotropic fluids; we make several observations that will provide the physical motivation of our formalism. Section \ref{sec:EOS} summarizes the piecewise polytrope formalism and discusses the main difficulties encountered in its application. We derive an generalization of the polytropic EOS from thermodynamic considerations in Section \ref{sec:OurEOS} then show how it can be used to construct a differentiable piecewise formalism. Section \ref{sec:Results} illustrates the main advantages of this formalism when it is applied to cold nuclear EOS candidates, namely accurate observable reproduction and a smooth, better fitting curve.

Throughout this work, we follow the convention of absorbing the speed of light $c$ into the definition of pressure and energy density \cite{Read:2008iy}. As a result, rest-mass density $\rho$, energy density $\eps$, and pressure $p$ have the same cgs unit ($\rm {g/cm^3}$) and specific enthalpy $h$ is dimensionless.

\section{Barotropic fluids}\label{sec:Barotropic}
\subsection{Thermodynamics}
We will assume that neutron-star matter consists of a  perfect fluid. Moreover, we assume that the fluid is \emph{simple}: i.e., that all the thermodynamic quantities depend only  on the 
proper baryon number density $n$ and on the entropy density $s$. The relation
\be \label{e:EOS}
  \eps = \eps(n,s) . 
\ee
is the fluid's EOS. The baryon chemical potential $\mu$ and the temperature $T$  are  defined by
\be \label{e:def_T_mu}
T: = {\left. {\frac{{\partial \epsilon }}{{\partial s}}} \right|_n}\quad \;\,{\rm{and}}\quad \;\,\mu : = {\left. {\frac{{\partial \epsilon }}{{\partial n}}} \right|_s}
\ee
The first law of thermodynamics can be written as
\be \label{e:deps1stlaw}
    d\eps =\mu \,dn+Tds.
\ee
As a consequence, $p$ is a function of $(n,s)$ entirely determined by (\ref{e:EOS}): 
\be \label{e:p_EOS}
  p = -\eps + T  s + \mu \,n . 
\ee
Let us introduce the specific enthalpy, 
\be \label{e:def_h}
  h := \frac{\eps+p}{\rho} = \frac{\mu}{m \, } + T \se ,
\ee
where $m$ is
the baryon rest mass, $\rho$ is the rest-mass density
\be 
  \rho := m \, n,
\ee
and  $\se$ is the specific entropy:
\be \label{e:def_S}
  \se := \frac{s}{\rho} .
\ee
The second equality in (\ref{e:def_h}) is an immediate consequence of (\ref{e:p_EOS}). From Eqs.~\eqref{e:deps1stlaw}--\eqref{e:def_S},
we obtain the  thermodynamic relations
\begin{subequations} \label{eq:depsdp}
\bea
d\eps &=&h\,d\rho+\rho T d\se \\
dp&=&\rho(dh-Td\se).
\eea
\end{subequations}
Since $h$ is as a function of $(n,s)$ or, equivalently, $(\rho,\se),$ differentiating it yields:
\be
dh=\frac{h}{\rho} \, c_{\rm s}^2 \, d\rho+  \frac{\partial h}{\partial \se} d\se
\ee
where the speed of sound $c_{\rm s}$ is defined by
\be  \label{eq:soundspeed}
c_{\rm{s}}^2 = {\left. {\frac{{\partial p}}{{\partial \epsilon }}} \right|_\se} = \frac{\rho }{h}{\left. {\frac{{\partial h}}{{\partial \rho }}} \right|_\se}.
\ee

If, as assumed in \S \ref{sec:EOS} onwards,  the temperature is much lower than the Fermi temperature,  the neutron-star matter is fully degenerate and the EOS is \textit{barotropic}. That is, the thermodynamic variables  $p$ and $\eps$ are functions of $n$ only. 
Then, the thermodynamic relations \eqref{eq:depsdp} simplify to
\begin{subequations} \label{eq:depsdpthermodynamic}
\bea 
d\eps &=& h \, d\rho \label{eq:depsdpthermodynamic1}\\
 dp &=& \rho \, dh  \label{eq:depsdpthermodynamic2}
\eea
\end{subequations}
These relations may be used to express $\rho$, $p$, and $\eps$ as functions of the specific enthalpy $h$.

A convenient parameter that characterizes an EOS is the adiabatic index. For a barotropic fluid, we define this quantity to be
\begin{equation}
    \gamma := \frac{d \log p}{d \log \rho} = \frac{\rho}{p}\frac{dp}{d\rho}
\end{equation}

\subsection{Hydrodynamics}  \label{sec:hydrodynamics}
A perfect fluid is characterized by the energy-momentum tensor
\be \label{e:energymomentumfluid}
T_{\beta}^{\alpha} = (\eps + p) \, u^\alpha u_\beta + p \,\delta^\alpha_{\beta} ,
\ee
where $u^\alpha$ is the four-velocity, $\eps$ is the proper energy density, and $p$ the fluid pressure. 
Using Eqs.~\eqref{e:deps1stlaw}--\eqref{eq:depsdpthermodynamic},
the conservation law of the  fluid energy-momentum tensor \eqref{e:energymomentumfluid} can be written as:
\begin{align} \label{e:cons_enermom}
  \nabla_\alpha T^\alpha_{\beta}  =& \rho u_\beta u^\alpha\nabla_\alpha h+h u_\beta\nabla_\alpha(\rho u^\alpha) \\
+&\rho h u^\alpha \nabla_\alpha u_\beta+\rho\nabla_\beta h - \rho T \nabla_\beta \se= 0 \nonumber  
\end{align}
where $\nabla_\alpha$ is the  covariant derivative compatible with the spacetime metric $g_{\mu\nu}$. Given the rest-mass conservation law 
\be   \label{eq:continuityeqn}
\nabla_\alpha (\rho u^\alpha)=0,
\ee
Eq. \eqref{e:cons_enermom} yields the relativistic Euler
equation:
\bea  \label{eq:Euler}
   u^\alpha \nabla_\alpha(h u_\beta) +\nabla_\beta h = T \nabla_\beta  \se. 
\eea
If the fluid is also taken to be barotropic, then the right hand side of this equation vanishes:
\bea  \label{eq:EulerBarotropic}
   u^\alpha \nabla_\alpha(h u_\beta) +\nabla_\beta h =0. 
\eea
Thus, only the specific enthalpy $h$ is needed for the hydrodynamics sector. In light of this, it is advantageous to rewrite the energy momentum tensor as
\be \label{e:energymombar}
T^\alpha_{\beta} = \rho(h) \, h \, u^\alpha u_\beta + p(h) \,\delta^\alpha _\beta .
\ee
In this form, the gravitational sector only requires thermodynamic functions of $h$. This is true for simulations, as well as for the construction of initial data \cite{Shibata:1998um,Gourgoulhon:2000nn,Taniguchi:2001qv}. Thus, for barotropic fluids, the form of the EOS expressing $p$, $\rho$ and $\eps$ as functions of $h$ can be regarded as fundamental.

Although, at first glance, the sound speed $c_{\rm s}$ does not appear in the hydrodynamic  equations, the characteristics of the system depend  explicitly on $c_{\rm s}$  \cite{demetrios_christodoulou_compressible_2014}. The dependence of the hydrodynamic equations on $c_{\rm s}$ can be made explicit by using Eqs.~\eqref{eq:soundspeed} to rewrite the continuity equation  \eqref{eq:continuityeqn} as an evolution equation for the specific enthalpy $h$ (rather than rest-mass density $\rho$): 
\be \label{eq:acousticcontinuity}
a^{\alpha \beta }{\nabla _\alpha }(h\,u_\beta ) = 0
\ee
where
\be
{a^{\alpha \beta }} = {g^{\alpha \beta }} + (1 - c_{\rm{s}}^{ - 2}){u^\alpha }{u^\beta }
\ee
is the inverse of the acoustic metric
$a_{\alpha\beta}$ (obtained from ${a^{\alpha \beta }}{a_{\beta \gamma }} = \delta^\alpha _{\gamma})$ \cite{Visser:1997ux}.
The null cones of the acoustic metric are the sound cones \cite{demetrios_christodoulou_compressible_2014}.  Because the evolution of a fluid depends explicitly on the sound speed, a numerical evolution that uses a parametrized EOS may differ significantly from an evolution with the original tabulated EOS if the sound speed $c_{\rm s}$ is not modelled accurately. In addition, neutron-star oscillation modes (such as p-modes or radial modes) and their frequencies are  sensitive to the value of sound speed 
\cite{Kastaun:2008jr}.

In addition, discontinuities in $c_{\rm s}$, whether introduced physically (via a phase transition) or artificially (via a piecewise polytropic EOS approximation), typically cause  artifacts in numerical simulations which limit point-wise convergence \cite{Foucart:2019yzo}. Moreover, as shown in Refs.
\cite{levequeFiniteVolumeMethods2002,vossExactRiemannSolution2005,PhysRevD.83.064002}, when the fluxes of hydrodynamic conservation laws are non-smooth, split waves and composite structures may be present in the solutions. In these cases a numerical solution may not converge to the physically correct solution.

%Numerical computation of  mode frequencies using a shooting method typically fails [Kostas Kokkotas, personal communication] when using piecewise polytropes, due to the discontinuities
%in $c_{\rm s}$. 

It is thus desirable to have a parametrized EOS approximation that faithfully reproduces not only $\rho$, $p$ and $\epsilon$, but also $c_{\rm s}$, as continuous functions of $h$ \cite{annalaEvidenceQuarkmatterCores2020}.  

\section{Piecewise Polytropes} \label{sec:EOS}

%In Sec.~\ref{sec:EOS} 

\subsection{The Formalism}
We review the piecewise polytrope formalism introduced by Read \textit{et al.} in \cite{Read:2008iy}. A polytrope is a  power-law EOS of the form
\be  \label{eq:polytrope}
  p = K \rho^{\Gamma}
\ee
where $K$ is the polytropic constant and  $\Gamma=1+1/n$ is the adiabatic index. Substituting Eq.~\eqref{eq:polytrope} into Eq.~\eqref{eq:depsdpthermodynamic2}, integrating,  solving for $h(\rho)$ and inverting yields
\be  \label{eq:rhoofhpoly}
\rho(h ) = \left[\frac{h-1 }{K (1+n)}\right]^n
\ee
Substituting this equation of state back into Eq.~\eqref{eq:depsdpthermodynamic2}
or Eq.~\eqref{eq:polytrope}
yields 
\be \label{eq:pofhpoly}
 p(h) = K \left[\frac{h-1 }{K (1+n)}\right]^{1+n}
\ee
Eq.~\eqref{e:def_h} may be used to obtain the energy density:
\be \label{eq:epsilonofhpoly}
\eps (h)  =\rho(h ) \left[1+\frac{n(h-1) }{1+n}\right] 
\ee
In a piecewise polytropic approximation, one applies the above EOS $\rho(h)$, shifted by a constant $a_i$ along the $h$ axis, in a set of intervals $h_{i-1} \le h \le h_{i}$:
\begin{subequations}
    \begin{align}
        \rho(h ) &= \left[\frac{h-1-a_i }{K_i (1+n_i)}\right]^{n_i}, \\
        p(h)  &= K_i \left[\frac{h-1-a_i }{K_i (1+n_i)}\right]^{1+n_i}, \\
        \eps (h)  &= \rho(h ) \left[1+\frac{n_i(h-1)+a_i}{1+n_i}\right]
    \end{align}
\end{subequations}
with the constants $K_i$ and $a_i$  determined by continuity of the above functions at each junction.

\subsection{Difficulties}\label{sec:PPproblems}
The PP formalism provides a convenient parameterization for numerical relativity simulations because it only involves simple algebraic expressions. However, the overall parameterization does not enforce differentiability at the dividing densities. This  causes reflections in hydrodynamic simulations that are not predicted by the original EOS.

\begin{figure*}[!htpb]
        \centering
        \begin{subfigure}{0.3\textwidth}
            \centering
            \includegraphics[width=\textwidth]{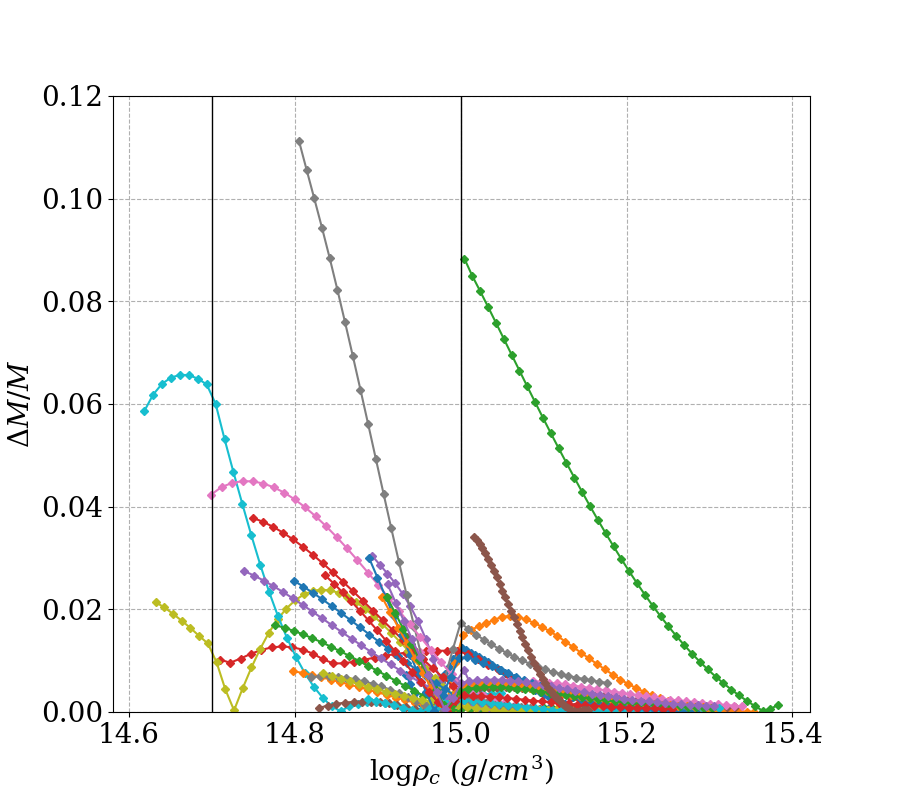}
            \caption{PP Mass Errors}
            \label{fig:PPMassError}
        \end{subfigure}
        \hfill
        \begin{subfigure}{0.3\textwidth}
            \centering
            \includegraphics[width=\textwidth]{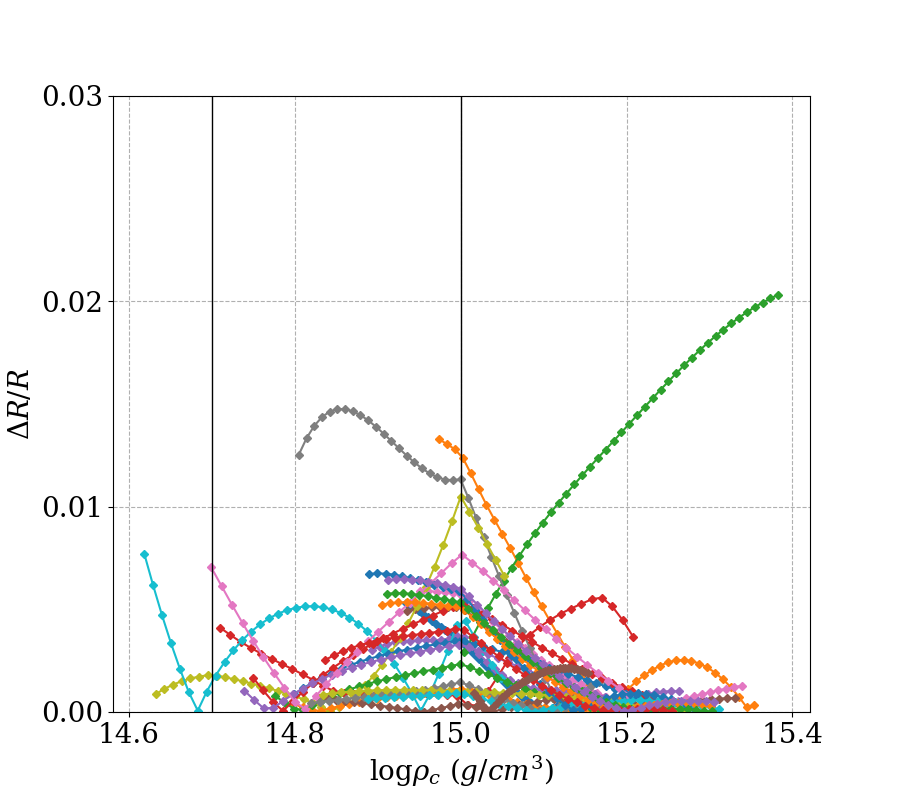}
            \caption{PP Radius Errors}
            \label{fig:PPRadiusError}
        \end{subfigure}
        \hfill
        \begin{subfigure}{0.3\textwidth}
            \centering
            \includegraphics[width=\textwidth]{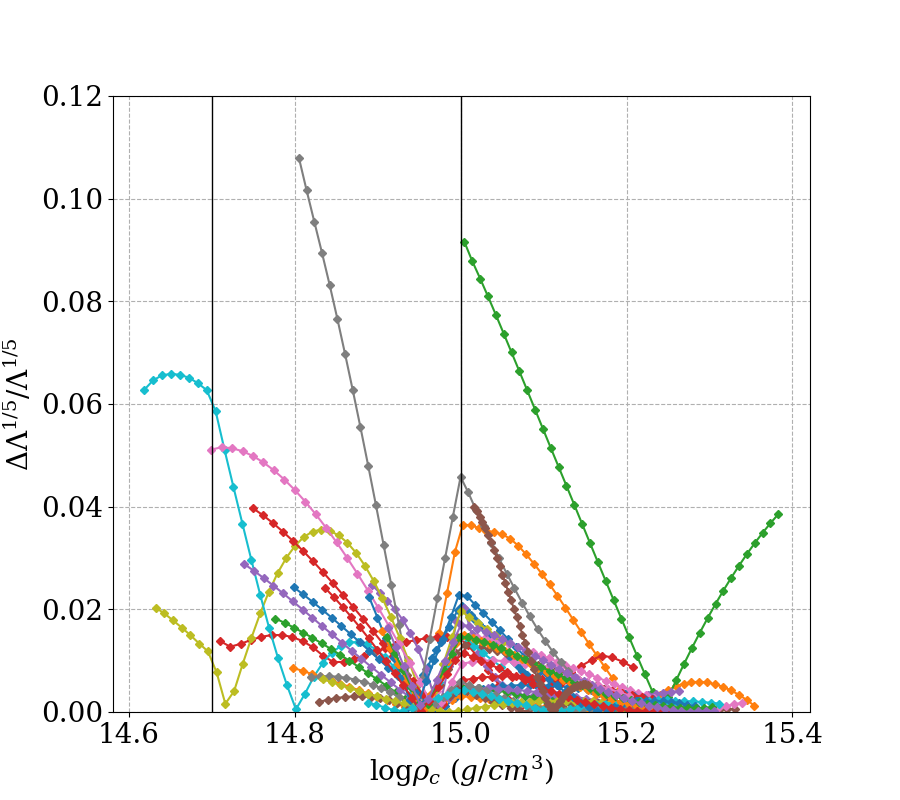}
            \caption{PP Tidal Deformability Errors}
            \label{fig:PPDeformError}
        \end{subfigure}
        
        \vskip \baselineskip
        \begin{subfigure}{0.3\textwidth}
            \centering
            \includegraphics[width=\textwidth]{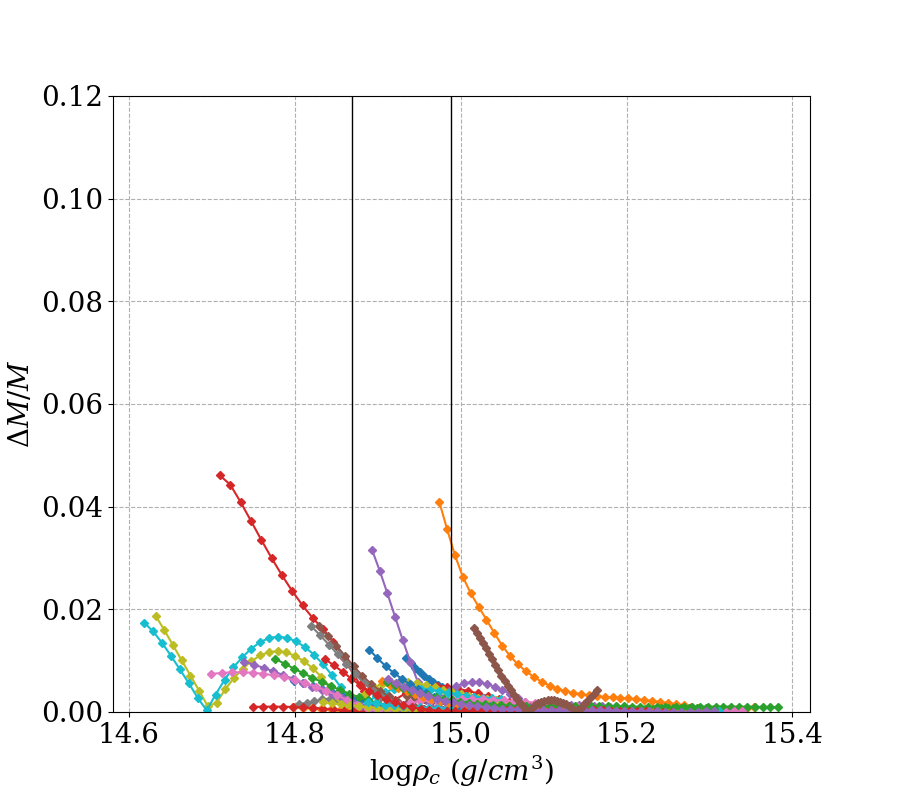}
            \caption{GPP Mass Errors}
            \label{fig:GPPMassError}
        \end{subfigure}
        \hfill
        \begin{subfigure}{0.3\textwidth}
            \centering
            \includegraphics[width=\textwidth]{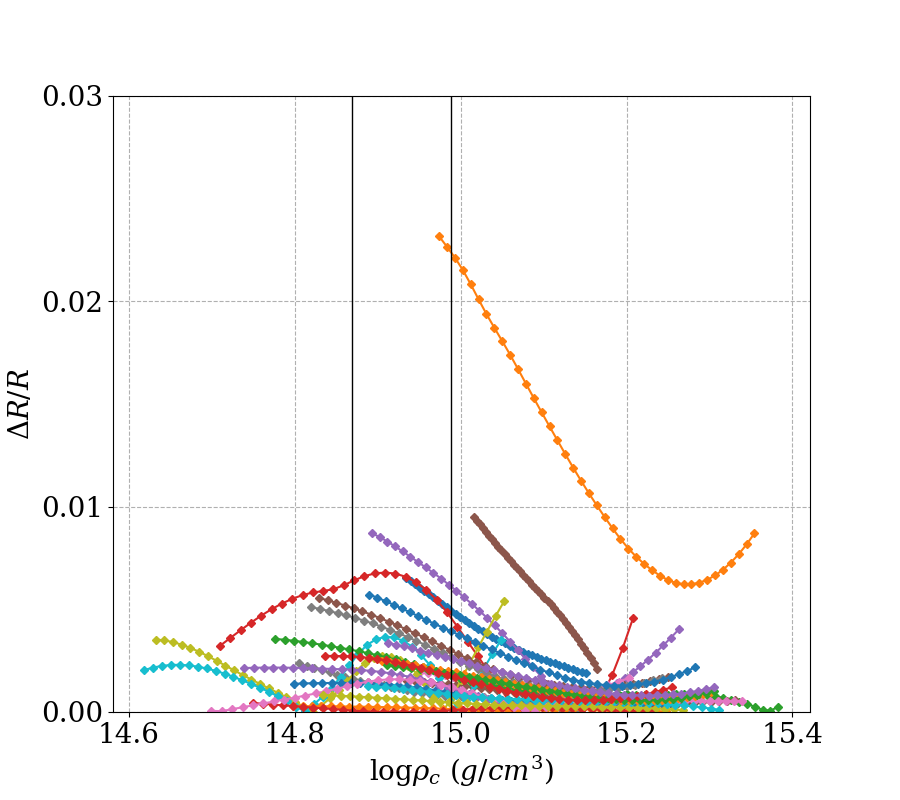}
            \caption{GPP Radius Errors}
            \label{fig:GPPRadiusError}
        \end{subfigure}
        \hfill
        \begin{subfigure}{0.3\textwidth}
            \centering
            \includegraphics[width=\textwidth]{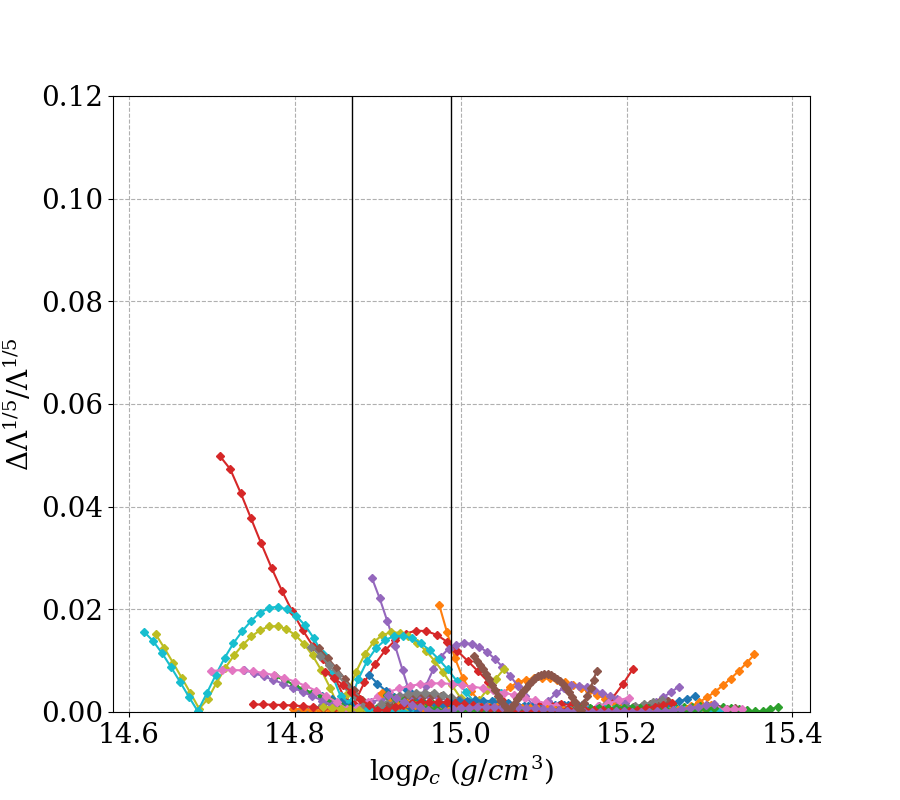}
            \caption{GPP Tidal Deformability Errors}
            \label{fig:GPPDeformError}
        \end{subfigure}
        
        \caption{\label{fig:ObsErrors} For each candidate EOS, a sequence of TOV\ stellar models is generated with the same central densities $\rho_c$, plotted on the horizontal axis. The relative error between  observables predicted by each fit and the tabulated EOSs is shown on the vertical axis. GPP fits significantly improve  accuracy in mass and tidal deformability, but not radius. }
\end{figure*}

In addition, Read \textit{et al.} demonstrated that PPs can yield low pointwise errors compared to tabulated EOSs when computing integral observables (i.e. mass, radius, moment of inertia, and tidal deformability).
However, when the whole range of stellar models predicted by a candidate EOS is considered, larger errors may result. This is demonstrated in Fig. \eqref{fig:ObsErrors}, where the mass, radius and tidal deformability of stars with the same central density,  computed from both a tabulated EOS and its PP fit, are compared. Dimensionless tidal deformability $\Lambda$ is related to the tidal Love number $k_2$ and compactness $C=M/R$ by $\Lambda=k_2/C^{5}$ \cite{VanOeveren:2017xkv}. Because of this sensitive dependence on $C$, $\Lambda$ can vary by many orders of magnitude, making numerical calculations cumbersome. We instead consider errors in $\Lambda^{1/5}$, which are of the same order as errors in mass and radius. A PP parameterization can result in errors as large as 11\% in mass and $\Lambda^{1/5}$. 

A third problem arises when one tries to recover the neutron star EOS from a GW measurement. Work by Fasano \textit{et al.} \cite{Fasano:2019zwm} and Gamba \textit{et al.} \cite{Gamba:2019kwu} has recently demonstrated that recovery of PP parameters by Bayesian inference leads to very large confidence intervals. The origin of this difficulty is the piecewise nature of the parameterization. The confidence intervals on the EOS parameters are assigned to reproduce the confidence intervals on the observable quantities obtained from the waveform. However, it is possible that the central densities required to produce the likely masses lie below the uppermost dividing density. In this case, no information about the last exponent can be gleaned. 

\section{Generalized piecewise polytropes with continuous sound speed}\label{sec:OurEOS}

\subsection{The Formulation}

In light of the advantages and disadvantages described above, it is desirable to seek an improved  parametrized EOS with the following properties:
\\ (i) The EOS parameters can be used to reproduce the integral observables of the original realtistic EOS accurately.
\\ (ii) The sound speed must be continuous across the dividing densities. 
\\(iii) The parametrized EOS should have a relatively simple analytical form that can be efficiently evaluated in numerical evolutions.
\\ (iv) The advantages of piecewise polytropes should be retained. In particular, the number of freely specifiable parameters should be the same as for piecewise polytropes.
\\

For barotropic fluids,  these requirements can be met by considering the fundamental EOS to be the functional $\rho(h)$ rather than $p(\rho)$.\footnote{For instance, to compute the EOS of an ideal degenerate Fermi gas, one first integrates the Fermi distribution to obtain the rest mass density $\rho$ as a function of the dimensionless Fermi energy which, in the fully degenerate case, coincides with specific enthalpy $h$. All other thermodynamic quantities such as pressure and energy density then follow from the first law of thermodynamics and the Gibbs-Duhem relation.} The pressure and energy density can then be obtained by integrating Eq.~\eqref{eq:depsdpthermodynamic}.
\footnote{Recall also that $h$ is the fundamental thermodynamic quantity that appears in the barotropic Euler equations, as detailed in Sec. \ref{sec:hydrodynamics}, and that  solving for neutron star structure only requires $\rho(h)$ \cite{Shibata:1998um,Gourgoulhon:2000nn,Taniguchi:2001qv}. Pressure and energy density  appear in the source of the Einstein equations.}

We begin with the same polytropic ansatz for $\rho(h)$ considered by Read \textit{et al.}:
\begin{equation}\label{eq:RhoOfHGeneral}
    \rho(h) = \left[\frac{h-1-a}{K(1+n)}\right]^{n}.
\end{equation}
 We then apply the thermodynamic identity Eq.~\eqref{eq:depsdpthermodynamic2} and integrate to obtain the pressure,
\begin{equation}\label{eq:ourpofh}
    p(h)  = K\left[\frac{h-1-a}{K(1+n)}\right]^{1+n} + \Lambda,
\end{equation}
where $\Lambda$ is a constant of integration. For  a classical polytrope, $a$ would be set to zero and the boundary condition $p\rightarrow0$
as $h\rightarrow1$ would require $\Lambda=0$. However, we intend to use this form to parameterize the high density region of the EOS away from the star's surface, and cold, dense nuclear matter is not a dilute classical gas. So there is no  reason \textit{a priori} to demand $\Lambda=0$. In fact, this  additional parameter gives us the freedom to demand continuity in sound speed. Substituting Eq.~\eqref{eq:ourpofh} into \eqref{e:def_h}
%Eq.~\eqref{eq:depsdpthermodynamic1}
 yields  the energy density:
\begin{equation}\label{ourepsilonofh}
    \eps (h)  =\rho(h) \left[1+\frac{n(h-1)+a}{1+n}\right]-\Lambda.
\end{equation}

We term the set of relations \eqref{eq:RhoOfHGeneral}-\eqref{ourepsilonofh} a \textit{generalized polytropic EOS}. Expressing  thermodynamic quantities as  functions of  specific enthalpy is needed for constructing TOV sequences or initial data in simulations. However, expressing $p$ and $\eps$ as functions of $\rho$ yields simpler expressions:
\begin{align}
    p(\rho) =& K \rho^{\Gamma} + \Lambda \label{eq:ourpofrho}\\
    \eps(\rho) =& \frac{K}{\Gamma-1}\rho^{\Gamma}+(1+a)\rho-\Lambda \label{eq:ourepsofrho}
\end{align}
We have defined $\Gamma= 1 + 1/n$, as in the standard polytrope formalism. This form also facilitates interpretation of the parameters. If the form \eqref{eq:ourpofrho} is used throughout the interior, then $\Lambda$ would be related to the density at the surface of the star where $p=0$. From \eqref{eq:ourepsofrho}, $a$ may be interpreted as a bulk binding energy. $\Gamma$ is no longer the adiabatic index, though: it is merely the exponent of the rest mass density.\footnote{We use the lowercase $\gamma$ to denote adiabatic index in this work.}

We develop a generalized piecewise polytropic parameterization following Read \textit{et al.} \cite{Read:2008iy}. The range of mass densities above a point $\rho_0$ is partitioned by $N$ dividing densities denoted by $\rho_1,\dots, \rho_N$. The EOS in each interval $[\rho_{i-1},\rho_i]$ is characterized by a set of parameters $K_i, \Gamma_i, a_i, \Lambda_i$. We now impose continuity and differentiability at each dividing density
in each of the thermodynamic quantities and use this to constrain the parameters.

Consider the pressure $p(\rho)$ first. Imposing \textit{differentiability} at a dividing density $\rho_i$ leads to the relation
\begin{equation}\label{eq:ourkequation}
    K_{i+1} = K_i \frac{\Gamma_i}{\Gamma_{i+1}}\rho_i^{\Gamma_i-\Gamma_{i+1}}.
\end{equation}
(Contrast this with piecewise polytropes, where the constants $K_i$ were used to impose \textit{continuity} at $\rho_i$.) The additional constant $\Lambda$ is used to also impose \textit{continuity} at this point. With Eq.~\eqref{eq:ourkequation}, this yields the relation
\begin{equation}\label{eq:ourlambdaequation}
    \Lambda_{i+1}=\Lambda_i+\bigg(1-\frac{\Gamma_i}{\Gamma_{i+1}}\bigg)K_i\rho_i^{\Gamma_i}
\end{equation}
The pressure $p(\rho)$ is now continuous and differentiable at the dividing densities.

We now turn to  the energy density $\epsilon(\rho)$. We demand differentiability at the dividing density \(\rho_i \) which, with Eq.~\eqref{eq:ourkequation}, leads to the relation
\begin{equation}\label{eq:ouraequation}
    a_{i+1} = a_i+\Gamma_i \frac{\Gamma_{i+1}-\Gamma_i}{(\Gamma_{i+1}-1)(\Gamma_i-1)}K_i \rho_i^{\Gamma_i -1}.
\end{equation}
It can be shown that Eqs.~\eqref{eq:ourkequation}, \eqref{eq:ourlambdaequation} and \eqref{eq:ouraequation}
imply continuity of $\epsilon(\rho)$  as well. Since $p(\rho)$ and $\epsilon(\rho)$ are both differentiable, the sound speed $c_s$ is continuous by virtue of Eq.~\eqref{eq:soundspeed}.

We have shown how to enforce continuity and differentiability across dividing densities. The only remaining question is how to match the parameterized core to a known crust EOS. To ensure differentiability, we take the derivative of Eq. \eqref{eq:ourpofrho} in the first segment and demand continuity with $dp/d\rho$ in the crust. Like Ref. \cite{Read:2008iy}, we treat $K_1$
\footnote{The pressure  $p_1$ at the first dividing density $\rho_1$ has astrophysical significance, so it is used as the first parameter for piecewise polytropes \cite{Read:2008iy}. Varying $p_1$ is equivalent to varying $K_1$, and we use the later as first parameter here, as this simplifies the fitting procedure  for generalized piecewise polytropes.}
% (If we used $p_1$, we would need to know $\Gamma_1, \Lambda_1,\rho_1$ The reason is that gor generalized piecewise polytropes  it is  to determine $p_1$ from $K_1$ than the other way around.} 
and $\Gamma_i$ as free parameters that shift the logarithmic $dp/d\rho$ curve up and down or change its slope\footnote{The choice of fit parameters is non-unique. As discussed above, it is possible to substitute $p_1$ for $K_1$. In addition, users of PPs in the context of parameter estimation often use the pressures at each dividing density $\log p_i$ instead of the $\Gamma_i$ \cite{Ozel:2018,Lattimer:2015nhk,Lattimer:2017,De:2018uhw,Carney:2018sdv}. We use $K_1$ and the $\Gamma_i$ primarily for mathematical convenience.}. We then look for a density where the two curves intersect and designate it as $\rho_0$. We now use $\Lambda_1$ to ensure continuity. This ensures that $p(\rho)$ is continuous and differentiable at $\rho_0$.  That is to say, if we let $p_c(\rho)$ denote the crust EOS, we solve for a $\rho_0$ such that
\begin{equation} \label{eq:ourcrust1}
    \frac{dp_c}{d\rho}(\rho_0) = K_1\Gamma_1 \rho_0^{\Gamma_1 -1};
\end{equation}
then we compute
\begin{equation} \label{eq:ourcrust2}
    \Lambda_1=p_c(\rho_0)-K_1 \rho_0^{\Gamma_1}.
\end{equation}
A similar procedure is followed to determine $a_1$ by demanding continuity in $\epsilon(\rho)$:
\begin{equation} \label{eq:ourcrust3}
    a_1 = \frac{\eps_{c}(\rho_0)}{\rho_0} - \frac{K_1}{\Gamma_1-1}\rho_0^{\Gamma_1-1}+\frac{\Lambda_1}{\rho_0}-1
\end{equation}

By taking $K_1$ and $\Gamma_1$ as free parameters, we have seen that $\Lambda_1$ and $a_1$ are fixed by demanding continuity and differentiability with the crust EOS. Ostensibly, the dividing densities are also free parameters, but we will show below that there is a single set of  astrophysically motivated dividing densities for all candidate EOSs. Thus, the constraint relations \eqref{eq:ourkequation}-\eqref{eq:ouraequation} are all that is needed to compute the parameters $K,$ $\Lambda$, and $a$ in the remaining segments. There is no requirement on the remaining $\Gamma_i$, which  may be used as fit parameters. Thus, as with PPs, we use three segments so the only parameters to be  fitted are $K_1$, and $\{\Gamma_i\}$. All other parameters are determined by continuity and differentiability.

The meaning of the parameters is simplest to discern when the thermodynamic quantities are given as functions of rest-mass density. In addition, many EOS tables contain $p$ and $\epsilon$ as functions of $n$ or $\rho$. But, for completeness, we provide the quantities as functions of specific enthalpy:
\begin{subequations}
    \begin{align}
        \rho(h ) &= \left[\frac{h-1-a_i }{K_i (1+n_i)}\right]^{n_i}, \\
        p(h)  &= K_i \left[\frac{h-1-a_i }{K_i (1+n_i)}\right]^{1+n_i} + \Lambda_i, \\
        \eps (h)  &=\rho(h ) \ \left[1+\frac{n_i(h-1)+a_i}{1+n_i}\right]-\Lambda_i
    \end{align}
\end{subequations}
Note that these relations only differ from the polytropic forms by the constant offset $\Lambda$. So, modifying existing PP codes to accommodate this formalism is trivial. However, it should also be noted that the constraint equations \eqref{eq:ourkequation} and \eqref{eq:ouraequation}
differ from their PP counterparts.

For the reasons given in \cite{Read:2008iy}, it is most convenient to have four total free parameters in an EOS\ parameterization.
So, we divide the core region into three sections with two dividing densities, making $N=3$. The free parameters are $K_{1}$ and the exponents $\{\Gamma_1,\Gamma_2, \Gamma_3\}$. 

\subsection{Fitting Candidate EOSs}
As with piecewise polytropes, our parameterization is fitted to a candidate equation of state using the method of least squares: the parameters $K_1$ and $\Gamma_i$ were chosen to minimize the error function
\begin{align} \label{eq:OurLeastSquare}
    \begin{split}
        E(&K_1, \{\Gamma_i\})=\\
        &\sqrt{ \frac{1}{\rho_{\rm{u}} - \rho_{\rm{l}}} \sum_{i=1}^N \sum_{\rho \in [\rho_{\rm{l}},\rho_{\rm{u}}]} \Big(p_{\rm{true}}(\rho)-K_i\rho^{\Gamma_i}-\Lambda_i \Big)^2 \Delta \rho }
    \end{split}
\end{align}
The constants $K_i$ and $\Lambda_i$ are not independent of the other segments, so this is a constrained minimization problem. To create fits that accurately reproduced the integral observables of an EOS, it was decided to sum only over the central densities of stellar models predicted by the original EOS with astrophysically plausible masses. 

It is well-known that the NS maximum mass is a consequence of relativistic gravity and that different EOSs make different predictions for its value. Consistency of a candidate's prediction with the observed value is an important criterion for assessing the candidate's feasibility \cite{Ozel:2015fia,Abbott:2018exr}. We denote the central density that yields the maximum mass in the above expression by $\rho_{\rm{u}}$. In contrast, the NS minimum mass is sensitive to the details of the formation channel which is still not fully understood \cite{Burrows:2012ew,Janka:2016fox,Suwa:2018uni}. For the purpose at hand, it is important to note that the central density of low mass stars may be less than $\rho_0$. So, the lowest central density $\rho_{\rm{l}}$ was selected to give a $1.25 M_\odot$ star, ensuring the above summation was only over densities covered by the parameterization.

$\Lambda_i=0$ in the PP formalism, so the EOS could be made linear by considering  $\log p(\log \rho)$ instead of $p(\rho)$. This made the least squares problem linear and enabled a direct calculation of the parameters. The $\Lambda$ parameter in our formalism does not allow this, so the problem is fully nonlinear. The widely differing magnitudes of the $K_i$, $\Gamma_i$, and $\Lambda_i$  combined with the nontrivial constraint equations \eqref{eq:ourkequation} and \eqref{eq:ouraequation} made the problem too difficult for software optimization routines. 

We found that a standard gradient descent algorithm to minimize $E$ was the most effective\footnote{We experimented with other algorithms, like Nesterov's accelerated descent method \cite{aleks2016nesterovs} and a pseudo-Newton method outlined in \cite{goodfellow_deep_2016}. However, the differentiability requirement restricted the variability of the fit parameters. So, these sophisticated methods caused larger parameter variations than could be accommodated. A standard gradient descent allowed the step size to be tuned so differentiability could be maintained throughout the procedure.}.
Let the vector $\mathbf{x}=(K_1,\Gamma_1,\dots,\Gamma_N)$. From an initial guess $\mathbf{x}_0$, the minimum is obtained by iterating
\begin{equation}
    \mathbf{x}_{k+1}=\mathbf{x}_k-\eta\nabla E(\mathbf{x}_k)
\end{equation}
until a tolerance is reached \cite{goodfellow_deep_2016}. A single value  of the parameter $\eta$ was found to be effective for fitting all candidate EOSs. The necessary first derivatives were approximated with second order finite difference expressions.

\subsection{Determination of the Dividing Densities}\label{sec:RhoDiv}

As mentioned above, the dividing densities $\rho_i$ could also be taken as free parameters, since there are no obvious constraints. Moreover, the microphysics of cold nuclear matter is still uncertain, so there is no clear physical motivation for these quantities. However, as with piecewise polytropes, there fortuitously exist choices of matching densities that minimize the average error across all candidate equations of state. This allows us to  fix the two dividing densities at their preferred values, reducing the number of free parameters from six to four. 

The two dividing densities were chosen such that an average error across all EOS\ fits is minimized. Instead of using a standard $L^2$ error function, though, we chose an error function to address the difficulties described with PPs in \S \ref{sec:PPproblems}. A parameterized EOS is a meaningful substitute to the original EOS only if it  makes the same predictions about stellar structure.

To ensure the fits would accurately reproduce observables, we used an $L^2$ error norm in observable quantities:
\begin{align} \label{eq:ObservableError}
\begin{split}
    E_{}(&\rho_1,\rho_2;K_1,\{\Gamma_j\})=\\
    &\sqrt{\sum_{\rho_i}\Bigg[\bigg(\frac{\Delta M_i}{M_i}\bigg)^2 + \bigg(\frac{\Delta R_i}{R_i}\bigg)^2 + \bigg(\frac{\Delta \Lambda^{1/5}_i}{\Lambda^{1/5}_i}\bigg)^2 \Bigg]}
\end{split} 
\end{align}
The above sum was performed over ten central densities in the range of stellar models for each EOS. Each $\Delta$ denotes the difference between the quantities predicted by the fit and the original tabulated EOS, and they are each normalized by the quantity predicted by the original EOS. We chose to fit mass, radius, and tidal deformabilty with application to GW astrophysics in mind. The masses of the two objects have the highest order effect on the gravitational waveform of a binary inspiral. Terms involving the tidal deformability are the leading order deviation from point mass (black hole) waveforms \cite{Buonanno:2009zt} and combine with the mass to provide information about the objects' EOSs \cite{Wade:2014vqa}. The  choice of observables and their relative weighting in the error norm  \eqref{eq:ObservableError} is somewhat subjective. To check the sensitivity of our results to the choice of norm, we used both $L^{4}$ and $L^{6}$ norms (as a proxies to a supremum norm) and did not observe noticeable change in the optimal dividing densities. Additional  astrophysical observables, such as quadrupole moment $Q$ or moment of intertia $I$ would also be straightforward to include in the  error norm. Because of $I$-Love-$Q$ relations between these observables and tidal deformability \cite{Yagi:2013awa}, we do not expect  the error minimization procedure to be affected significantly by such a choice, but we leave this for investigation in future work. 

To select dividing densities which minimize the observable error in \eqref{eq:ObservableError} across all candidate EOSs considered, we used the following algorithm:
\\
\\ (1) Hold one density fixed 
\\ (2) Vary the second density. Perform last square fits to each candidate EOS such that the error function \eqref{eq:OurLeastSquare} is minimized at each value of the varied density.
\\ (3) Compute the observable error function \eqref{eq:ObservableError} for the fits obtained in the previous step, and compute its average across all EOSs at each dividing density.
\\ (4) Select the density which achieves minimum global error.
\\ (5) Hold this density fixed and revise the first density by following Steps (2) through (4).
\\ (6) Repeat until desired tolerance is met.
\\

\begin{figure}
    \centering
    \includegraphics[width=\linewidth]{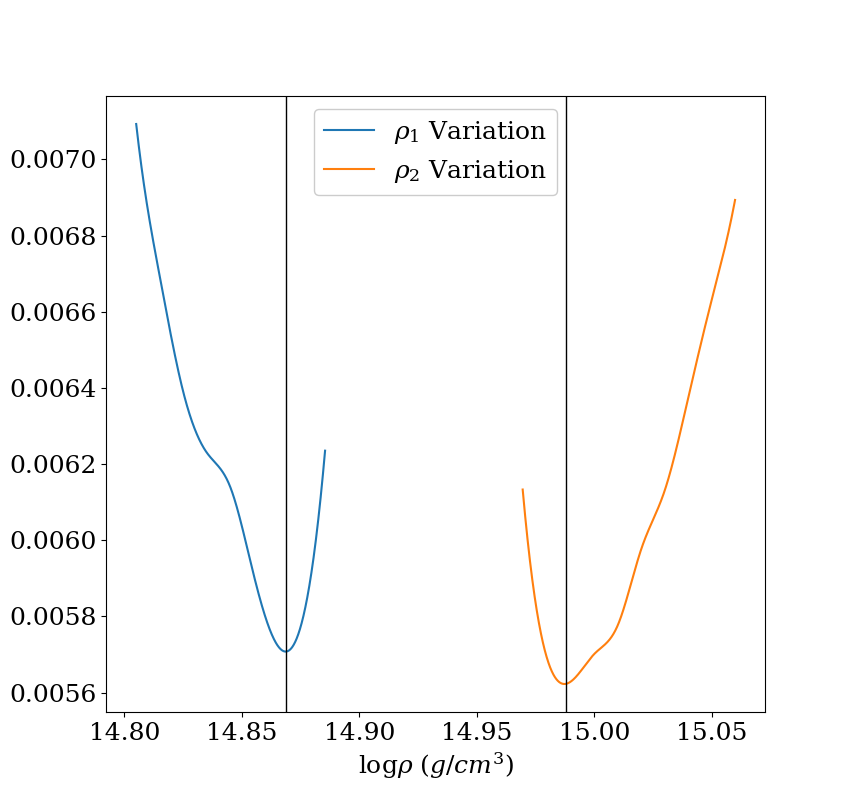}
    \caption{The observable error function defined in Eq. \eqref{eq:ObservableError} averaged over all candidate EOSs as the dividing densities are varied. Note that clear minima emerge, indicating preferred dividing densities.}
    \label{fig:rho_div}
\end{figure}

As illustrated in Fig. \eqref{fig:rho_div}, the preferred dividing densities which emerge are $10^{14.87} ~ \mathrm{g/cm^3}$ and $10^{14.99} ~ \mathrm{g/cm^3}$. We expect these dividing densities to be sensitive to the choice of the observable error function used in the minimization procedure. As stated above, the form of Eq. \eqref{eq:ObservableError} was selected so the resulting fits make accurate predictions that are relevant to GW applications, which we demonstrate in the next section. Changing the form of the error function would likely change the resulting fits. Exploring different error functions, which could be obtained by weighting the observables differently, introducing other observables, or using error norms besides $L^2$, and their impact of the quality of the resulting fits would make an interesting future study.

\section{Results}\label{sec:Results}
We applied this procedure to the cold nuclear matter EOS candidates listed in Table A1 of Ref. \cite{Abbott_2020}. 
The tables for SLy(4) and QHC19 were obtained from CompOSE \cite{typelEoSCatalog}, while the remaining EOS tables were obtained from the Feryal \"Ozel catalog \cite{ozelEoSCatalog}.
The obtained parameters are listed in Appendix A. While the low density EOS is generally agreed upon \cite{Baym:2017whm}, there are small deviations between candidates. So, to avoid uncertainties associated with imposing one crust EOS on all candidates, we chose to quantify error by matching the piecewise core to the low density region\ provided with each EOS table.

To use these parameters in a simulation, one needs only to match them to an arbitrary crust. Most numerical relativity simulations cannot resolve the low density regions of the stars, so a popular choice is a single polytropic piece with $\Gamma = 1.35692$ \cite{Read:2009yp,Kyutoku:2010zd,Kyutoku:2011vz,Hotokezaka:2012ze,Foucart:2019yzo}. However, if detailed knowledge of the low density EOS\ is required (e.g. for studying ejecta or neutrino emission in accretion disks surrounding a post-merger remnant \cite{Siegel:2017nub,Fernandez:2018kax,Shibata:2019wef}), we provide a set of parameters that fit the low density part of the SLy(4) EOS in Appendix B. Instructions for fitting the provided parameters to a desired crust are provided in Appendix A. The effect of the crust on the EOS inference from gravitational wave observations has been studied by Gamba et al. \cite{Gamba_2019}.

\subsection{Fidelity of Integral Observables}
A significant motivation for pursuing this work was to obtain a formalism that accurately reproduces the integral observables predicted by the original EOS. The dividing density procedure outlined in \S \ref{sec:RhoDiv} was followed to enforce this condition. The effectiveness of this procedure is illustrated in Fig. \eqref{fig:ObsErrors}, where the relative errors in mass, radius, and rescaled tidal deformability are plotted for each EOS as a function of the stellar models' central densities.  We compare our GPP formalism to standard PPs with the dividing densities reported in \cite{Read:2008iy}. GPPs are able to capture the mass and rescaled tidal deformability with great accuracy: most fits have errors bounded at 2\%. However, the errors in radius were comparable for both GPPs and PPs.

An accurate empirical relationship has been found between radius and tidal deformability \cite{Annala:2017llu}, so the different functional forms of their error curves in Fig. \ref{fig:ObsErrors} seems puzzling. The reason is that the relationship was calculated for stars of the same mass, while we consider mass, radius, and deformability as functions of the central density. This means that the deformability error reflects the errors in both mass and radius, since it is a derived quantity. Moreover, since the mass error is larger in magnitude than the radius error, we expect it to dominate the deformability error, as reflected in the similar magnitudes of their error curves in Fig. \ref{fig:ObsErrors}.
 
\subsection{Thermodynamic Quantities}
\begin{figure*}[!htpb]
        \centering
        \begin{subfigure}{0.4\textwidth}
            \centering
            \includegraphics[width=\textwidth]{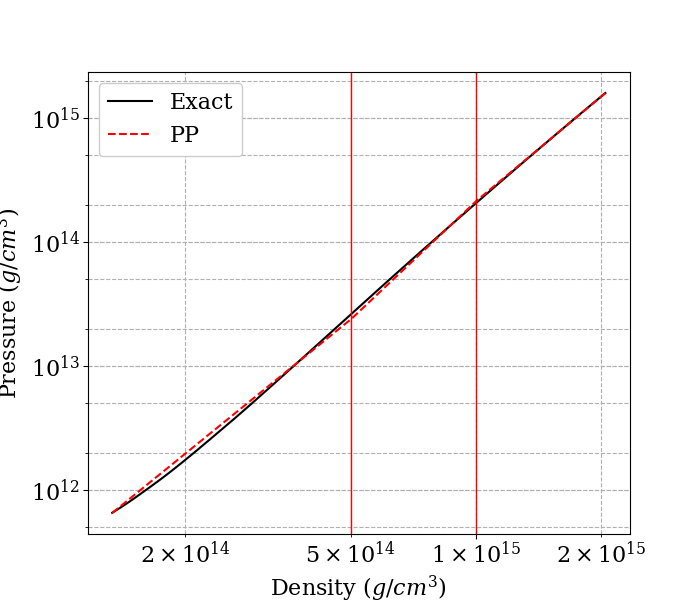}
            \caption{Pressure vs Density, PP fit to SLy(4)}
            \label{fig:PPpresSLY}
        \end{subfigure}
        \hfill
        \begin{subfigure}{0.4\textwidth}
            \centering
            \includegraphics[width=\textwidth]{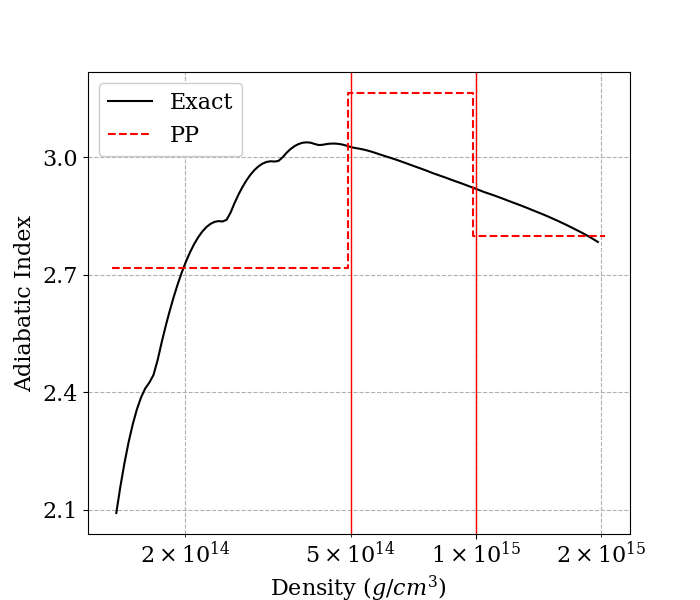}
            \caption{Adiabatic Index vs Density, PP fit to SLy(4)}
            \label{fig:PPgammaSLY}
        \end{subfigure}
        
        \vskip \baselineskip
        \begin{subfigure}{0.4\textwidth}
            \centering
            \includegraphics[width=\textwidth]{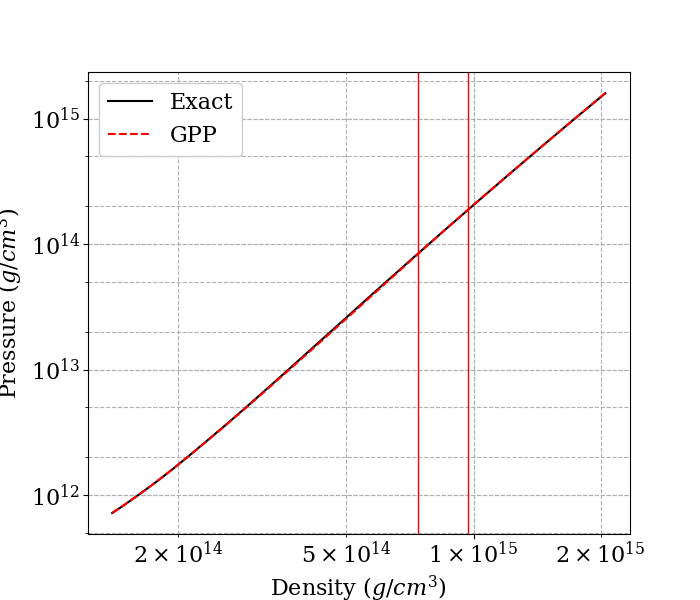}
            \caption{Pressure vs Density, GPP fit to SLy(4)}
            \label{fig:GPPpresSLY}
        \end{subfigure}
        \hfill
        \begin{subfigure}{0.4\textwidth}
            \centering
            \includegraphics[width=\textwidth]{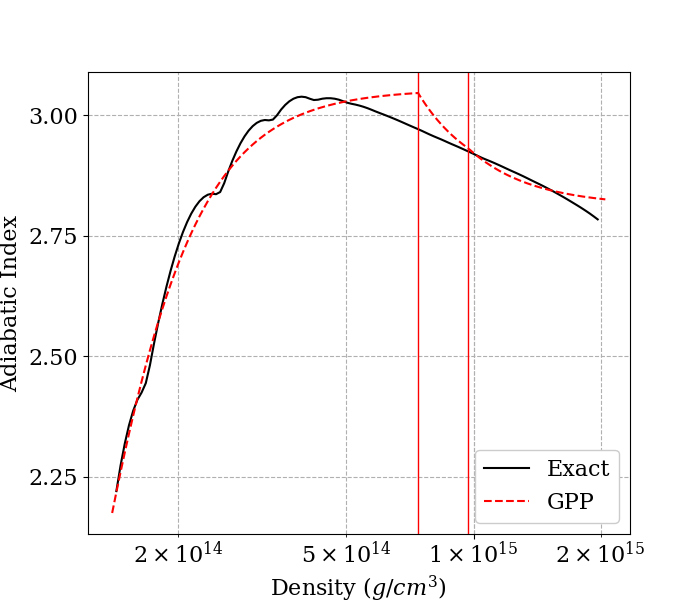}
            \caption{Adiabatic Index vs Density, GPP fit to SLy(4)}
            \label{fig:GPPgammaSLY}
        \end{subfigure}
        
        \caption{\label{fig:GPPSLY}  Thermodynamic quantities predicted by PP and GPP fits compared to the original EOS for SLy(4). The blue curve is the GPP predicted relation and the orange curve is the relation predicted from the SLy(4) table. The red vertical lines represent the dividing densities.}
\end{figure*}

\begin{figure*}[!htpb]
        \centering
        \begin{subfigure}{0.4\textwidth}
            \centering
            \includegraphics[width=\textwidth]{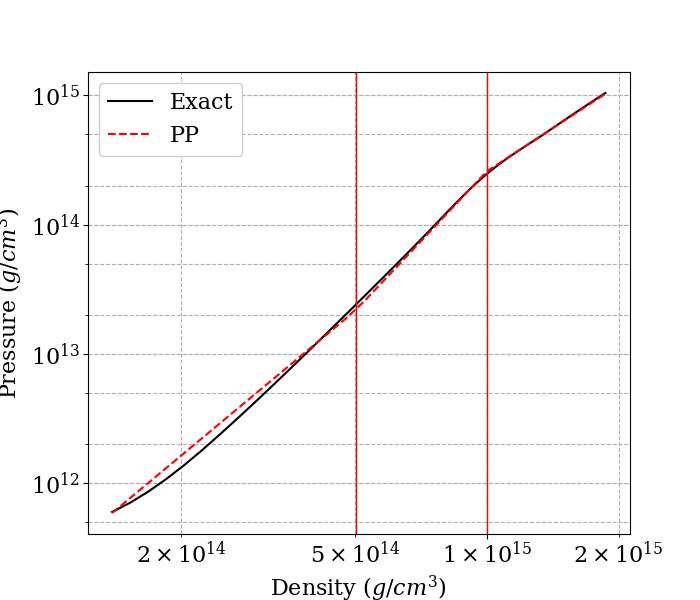}
            \caption{Pressure vs Density, PP fit to QHC19}
            \label{fig:PPpresQHC}
        \end{subfigure}
        \hfill
        \begin{subfigure}{0.4\textwidth}
            \centering
            \includegraphics[width=\textwidth]{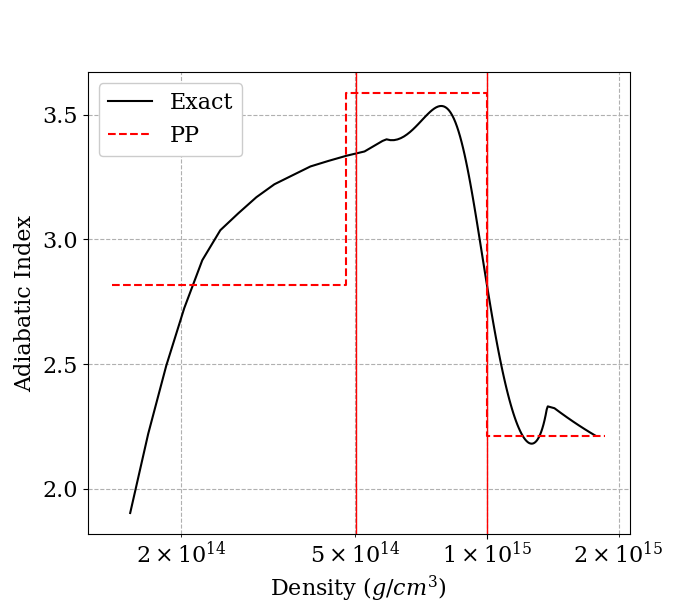}
            \caption{Adiabatic Index vs Density, PP fit to QHC19}
            \label{fig:PPgammaQHC}
        \end{subfigure}

        \vskip \baselineskip
        \begin{subfigure}{0.4\textwidth}
            \centering
            \includegraphics[width=\textwidth]{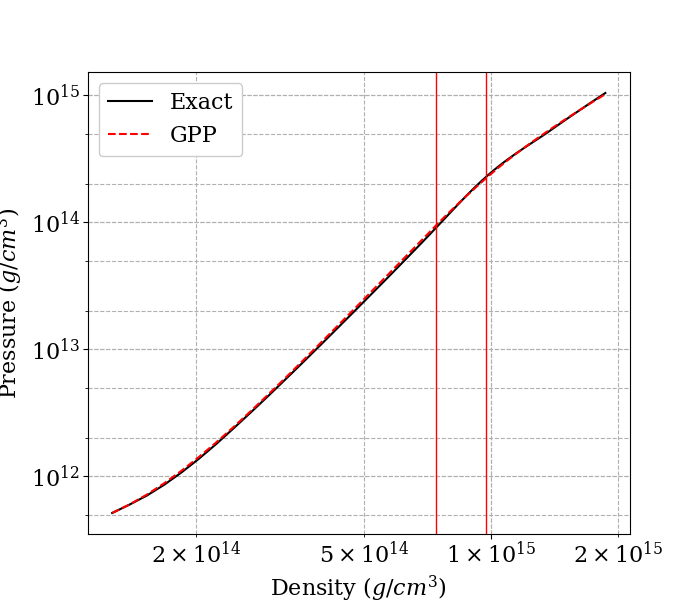}
            \caption{Pressure vs Density, GPP fit to QHC19}
            \label{fig:GPPpresQHC}
        \end{subfigure}
        \hfill
        \begin{subfigure}{0.4\textwidth}
            \centering
            \includegraphics[width=\textwidth]{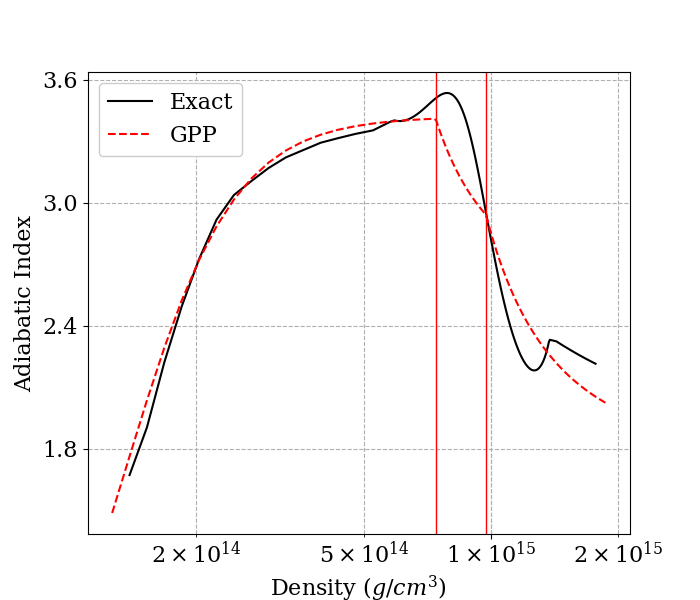}
            \caption{Adiabatic Index vs Density, GPP fit to QHC19}
            \label{fig:GPPgammaQHC}
        \end{subfigure}

        \caption{\label{fig:GPPQHC} Thermodynamic quantities predicted by PP and GPP fits compared to the original EOS for QHC19. The blue curve is the GPP predicted relation and the orange curve is the relation predicted from the QHC19 table. The red vertical lines represent the dividing densities.}
\end{figure*}

\begin{figure*}[!htpb]
        %\centering
        \begin{subfigure}{0.4\textwidth}
            \centering
            \includegraphics[width=\linewidth]{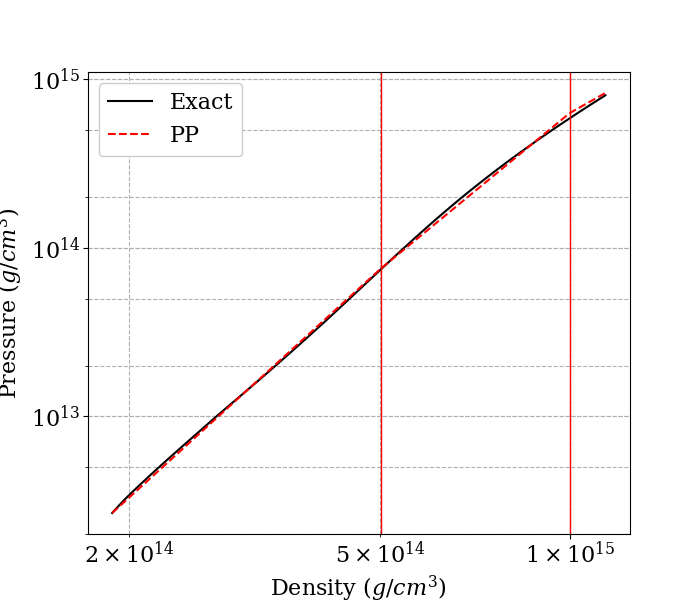}
            \caption{Pressure vs Density, PP fit to MS1b}
            \label{fig:PPpresMS1}
        \end{subfigure}
        \hfill
        \begin{subfigure}{0.4\textwidth}
            \centering
            \includegraphics[width=\textwidth]{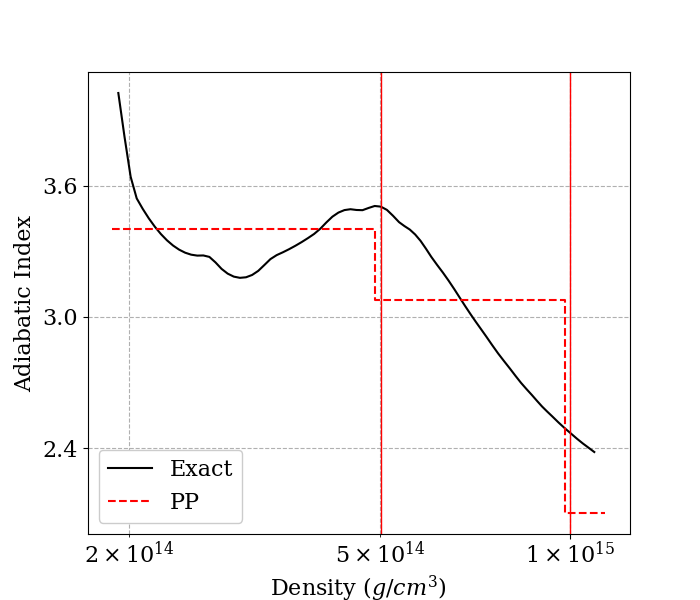}
            \caption{Adiabatic Index vs Density, PP fit to MS1b}
            \label{fig:PPgammaMS1}
        \end{subfigure}
        
        %\vskip \baselineskip
        %\hfill
        \begin{subfigure}{0.4\textwidth}
            \centering
            \includegraphics[width=\textwidth]{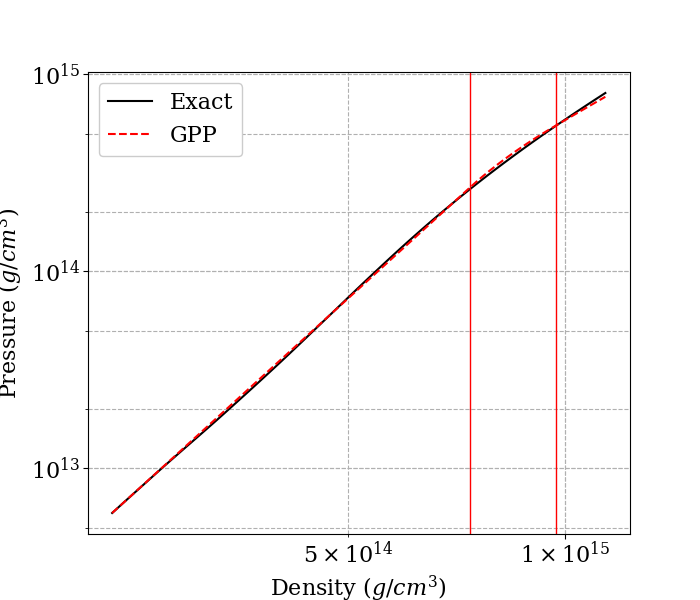}
            \caption{Pressure vs Density, GPP fit to MS1b}
            \label{fig:GPPpresMS1}
        \end{subfigure}
        \hfill
        \begin{subfigure}{0.4\textwidth}
            \centering
            \includegraphics[width=\textwidth]{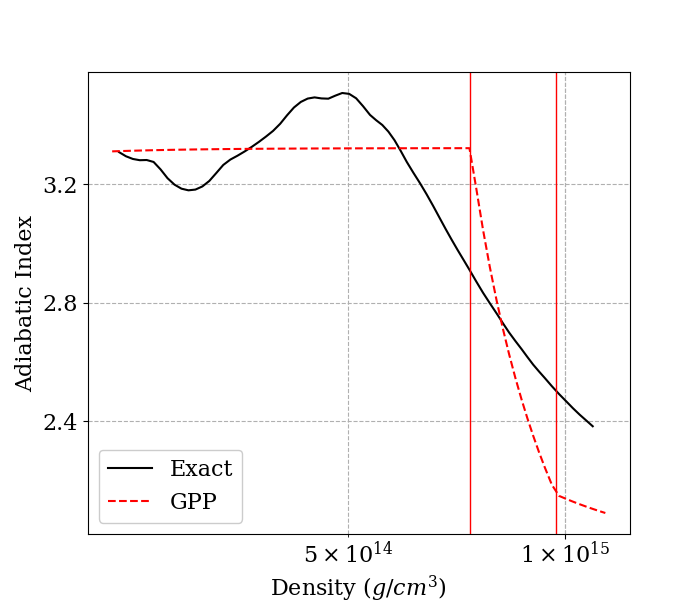}
            \caption{Adiabatic Index vs Density, GPP fit to MS1b}
            \label{fig:GPPgammaMS1}
        \end{subfigure}

        \caption{\label{fig:GPPMS1} Thermodynamic quantities predicted by a GPP fit compared to the original EOS for MS1b. The blue curve is the GPP predicted relation and the orange curve is the relation predicted from the MS1b table. The red vertical lines represent the dividing densities.}
\end{figure*}

The other major motivation of this work was to obtain a parameterized formalism that was smooth. We enforced constraints that imposed a continuous sound speed, and simultaneously enforced a smooth adiabatic index. We illustrate this property by plotting the pressure and adiabatic index for SLy(4) \cite{Chabanat:1997un,Danielewicz:2008cm,Gulminelli:2015csa} in Fig. \eqref{fig:GPPSLY}, QHC19 \cite{Togashi:2017mjp,Baym:2017whm,Baym:2019iky} in Fig. \eqref{fig:GPPQHC}, and MS1b \cite{Mueller:1996pm} in Fig. \eqref{fig:GPPMS1}. It should be noted that the ``exact'' adiabatic index $\gamma$ was computed by numerically differentiating EOS tables. We see that  the adiabatic indices predicted by the GPP formalism improve  upon the constant adiabatic indices predicted by PPs and behave qualitatively very similar to the  numerically computed adiabatic index $\gamma$. This qualitative agreement is striking, given the fact that GPPs still use four independent constant parameters $K_1, \Gamma_1, \Gamma_2,\Gamma_3$.

\begin{figure*}[!htpb]
        \centering
        \begin{subfigure}{0.4\textwidth}
            \centering
            \includegraphics[width=\textwidth]{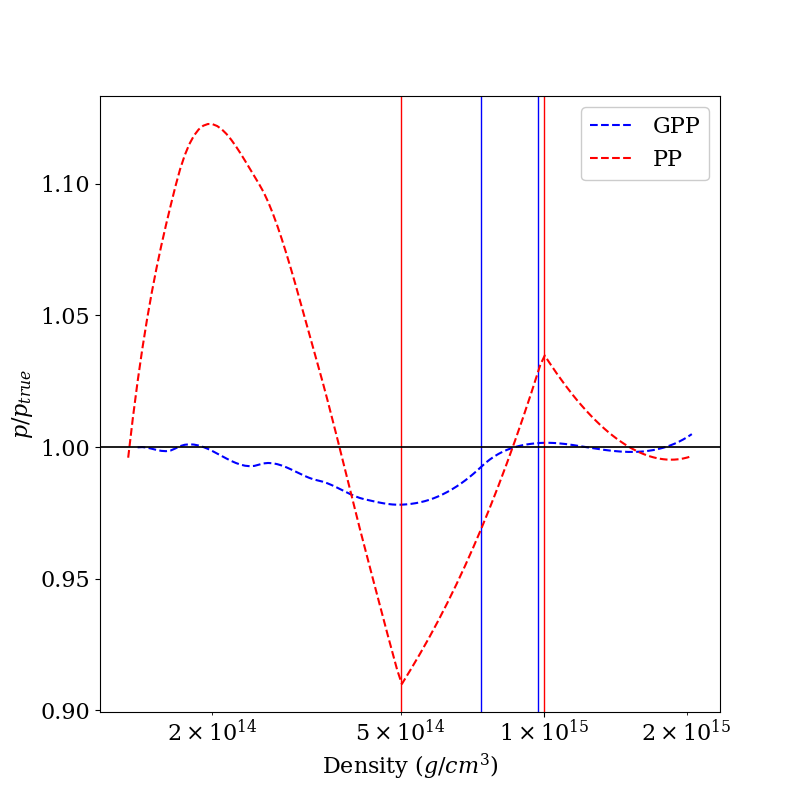}
            \caption{SLy(4)}
        \end{subfigure}
        \hfill
        \begin{subfigure}{0.4\textwidth}
            \centering
            \includegraphics[width=\textwidth]{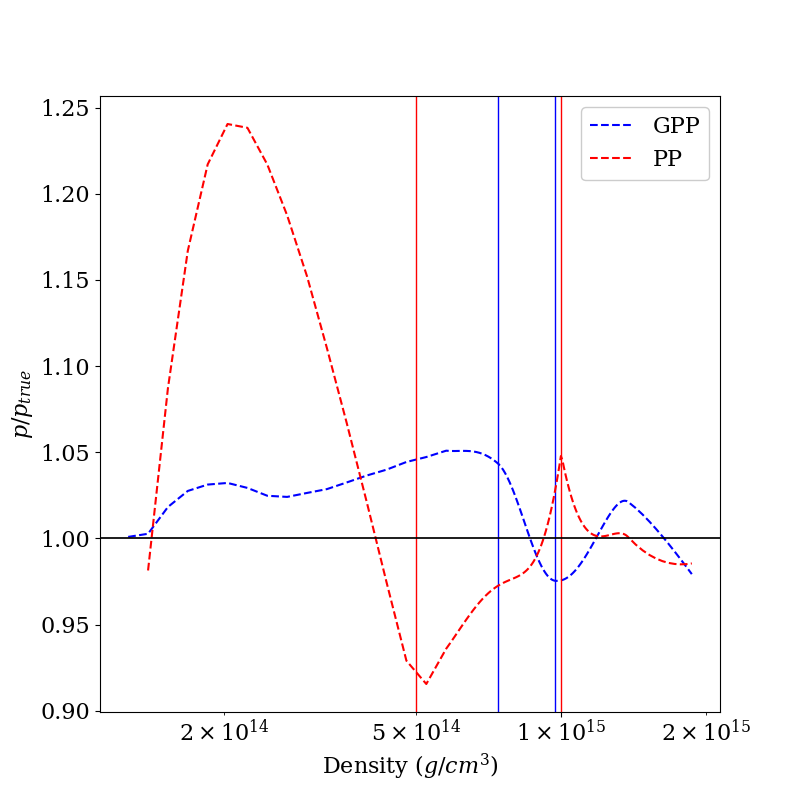}
            \caption{QHC19}
        \end{subfigure}
        \hfill
        \begin{subfigure}{0.4\textwidth}
            \centering
            \includegraphics[width=\textwidth]{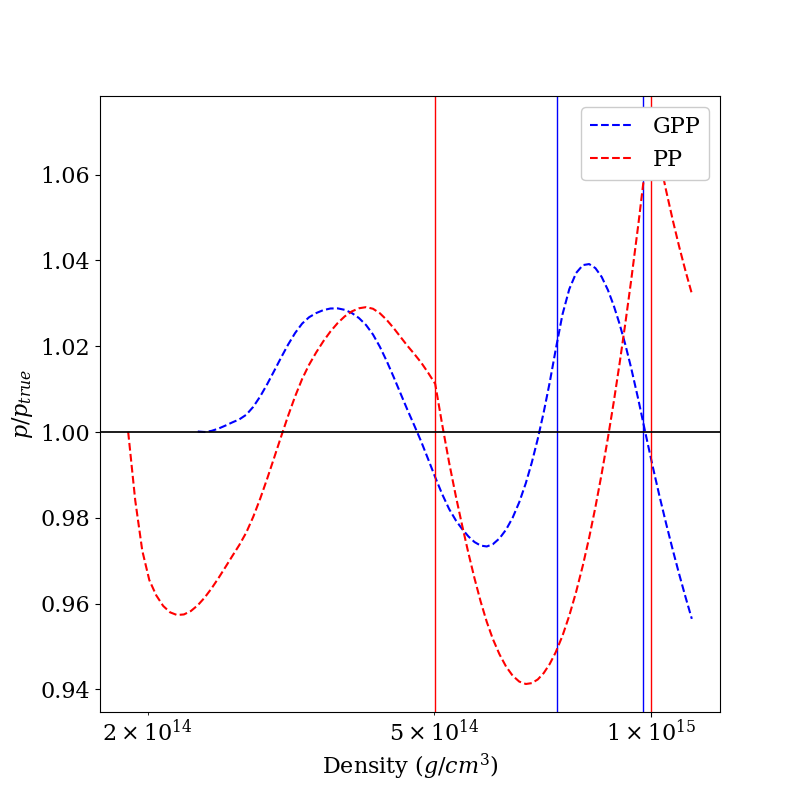}
            \caption{MS1b}
        \end{subfigure}
        
        \caption{\label{fig:WadePlots} The fractional error in pressure as a function of density. The errors in PP fits (red) is larger near the dividing densities. The error of GPPs (blue) is evenly spread out and lower overall.}
\end{figure*}

As we show in Fig. \eqref{fig:WadePlots}, PPs can have higher pointwise errors compared to the original EOS. the errors of GPPs are generally lower and more evenly spread out. We anticipate that this improvement will help with the issues outlined in \cite{Foucart:2019yzo} and \cite{Fasano:2019zwm}, making GPPs more useful for parameter estimation using this formalism.

\subsection{Quasi-normal modes}
Radial mode frequencies directly depend on the sound speed. 
We thus investigated whether a parametrization with continuous sound speed with GPPs might reproduce radial modes more faithfully.
We followed the formalism of \cite{Gondek}
and numerically calculated the fundamental radial $F$-mode frequencies and eigenfunctions for a number of representative cases. 

\begin{figure*}[!htpb]
        \centering
        \begin{subfigure}{0.4\textwidth}
            \centering
            \includegraphics[width=\textwidth]{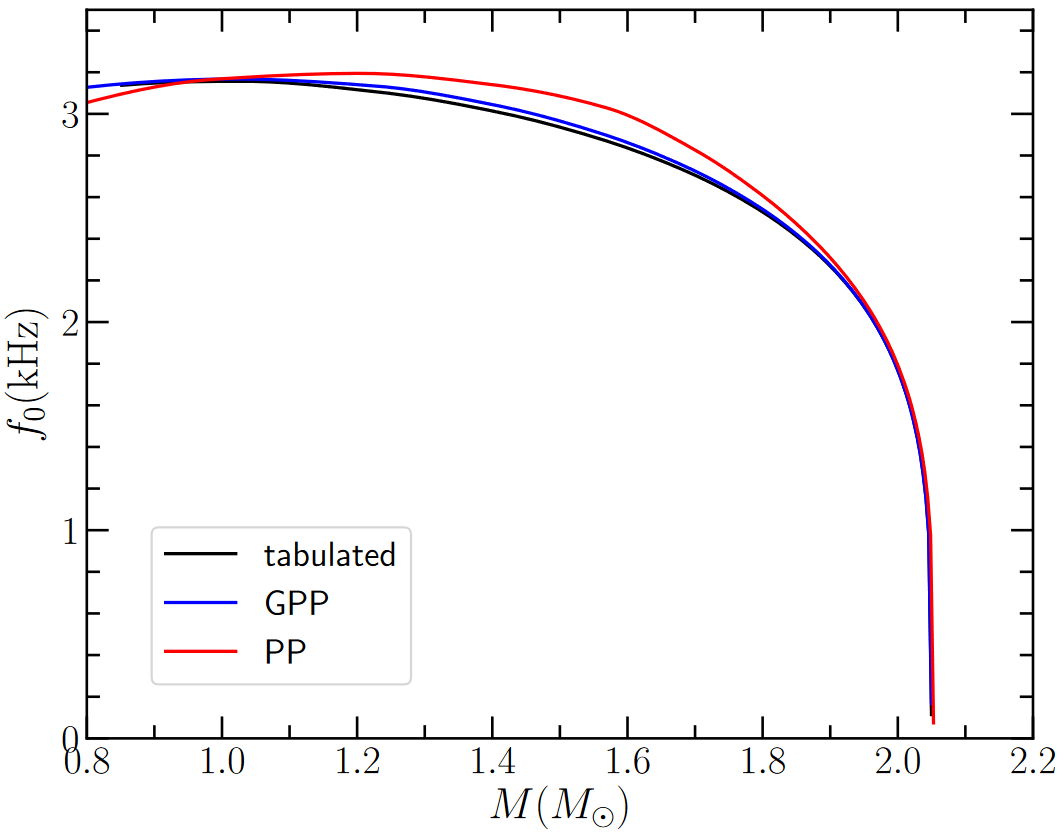} 
            \caption{SLy(4) $F$-mode frequency \label{fig:SLY4-f_vs_M}}
        \end{subfigure}
        \hfill
        \begin{subfigure}{0.4\textwidth}
            \centering
            \includegraphics[width=\textwidth]{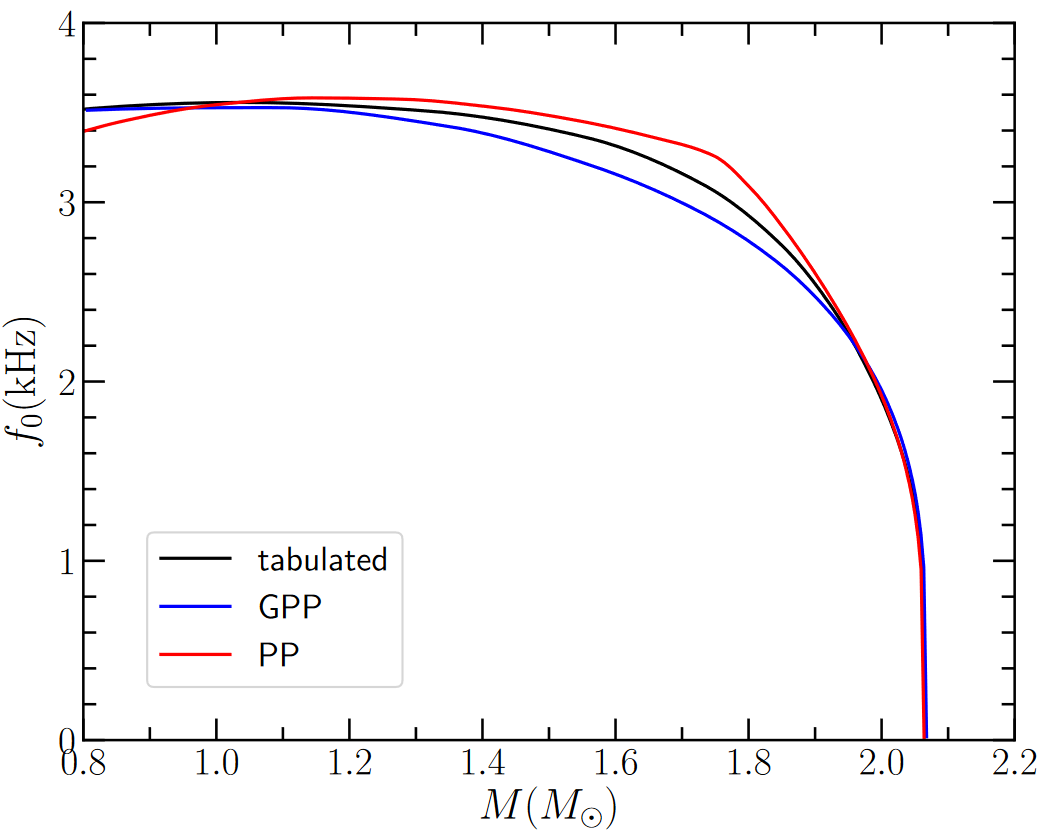} 
            \caption{QHC19 $F$-mode frequency \label{fig:QHC19-f_vs_M}}
        \end{subfigure}
        \hfill
        \begin{subfigure}{0.4\textwidth}
            \centering
            \includegraphics[width=\textwidth]{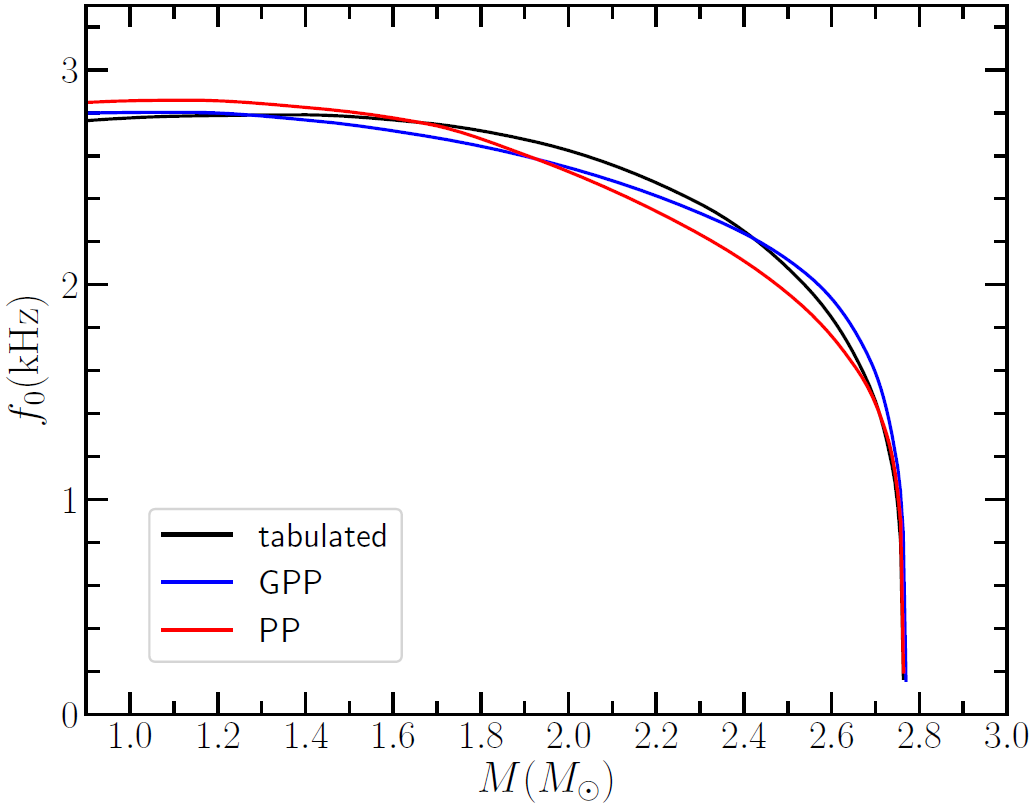} 
            \caption{MS1b $F$-mode frequency \label{fig:MS1B-f_vs_M}}
        \end{subfigure}
              \hfill
        \begin{subfigure}{0.4\textwidth}
            \centering
            \includegraphics[width=\textwidth]{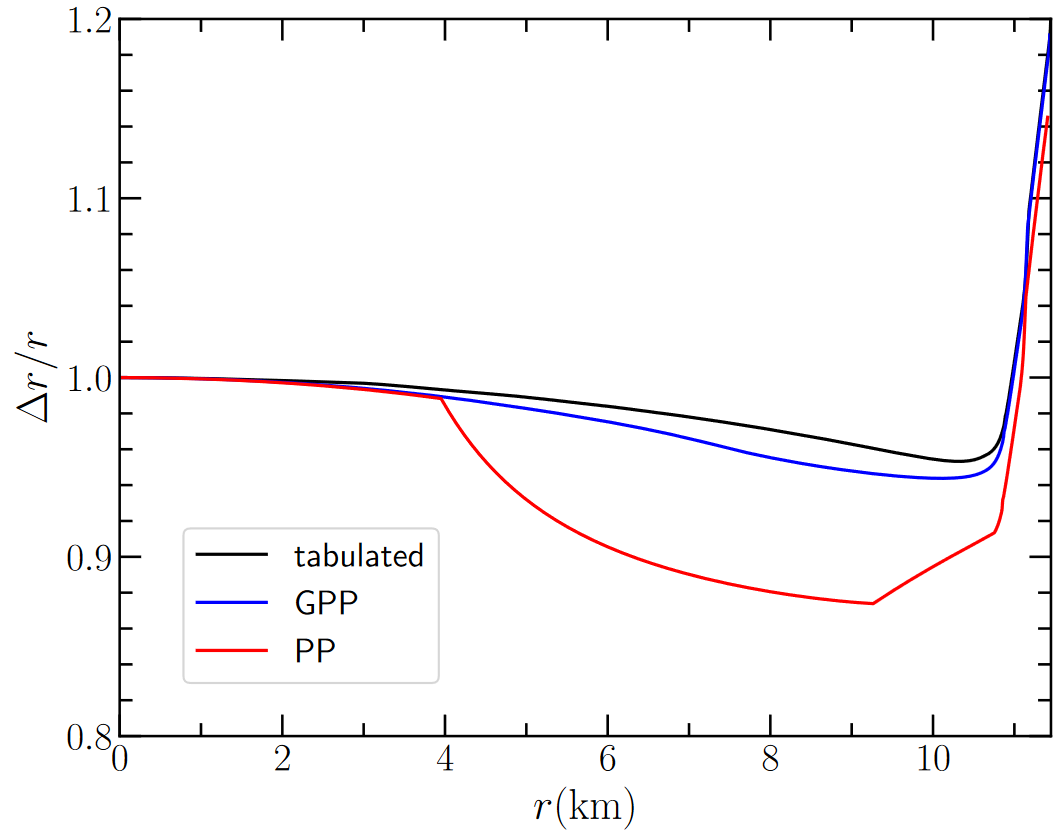} 
            \caption{$F$-mode eigenfunction profile for a particular model with EOS SLy(4)  \label{fig:plot-SLY4-eigen}}
        \end{subfigure}  
        \caption{Radial $F$-mode frequency as a function of stellar mass for EOS SLY4,  QHC19 and MS1b (a-c) and $F$-mode eigenfunction profile for a particular model of EOS SLY4 with central density $\rho_c=1.1\times 10^{15}{\rm g/cm^3}$ (d).}
\end{figure*}

Fig. \ref{fig:SLY4-f_vs_M}  shows the $F$-mode frequency as a function of mass along the sequence of equilibrium models for  EOS SLY(4), calculated with the tabulated version (black line), the PP version (red line) and the new GPP version (blue line). For this EOS,  the $F$-mode frequencies calculated
with the GPP version are much closer to the frequencies calculated with the tabulated version than are the frequencies calculated with the PP version of the EOS.  

The 
comparison of the corresponding eigenfunctions of the relative Lagrangian displacement $\Delta r/r$  for a model with $\rho_c=1.1\times 10^{15}{\rm g/cm^3}$ is shown in Fig. \ref{fig:plot-SLY4-eigen}. It is evident that the GPP eigenfunction
agrees well with the tabulated eigenfunction, whereas the  PP eigenfunction shows larger differences and has a discontinuous first derivative at the locations
where the speed of sound is discontinuous in the PP formalism. 

 For some EOS, such as QHC19 (Fig.~\ref{fig:QHC19-f_vs_M}) or the stiffest
candidate EOS MS1b (Fig.~\ref{fig:MS1B-f_vs_M}), we find that GPPs and PPs
can have a comparable level of agreement with the tabulated version in terms of the eigenfrequency of the $F$-mode. However, the corresponding eigenfunctions
calculated with the PP version still suffer  from noticable kinks at the location where the speed of sound is discontinuous.

%For a soft  EOS such as SLy(4) we find that this is the case, as demonstrated %in  Figs.~\ref{fig:SLY4-f_vs_M} and ~\ref{fig:plot-SLY4-eigen}. For other %EOS, however, such as QHC19 (Fig.~\ref{fig:QHC19-f_vs_M}) or the stiffest %candidate EOS MS1b (Fig.~\ref{fig:MS1B-f_vs_M}), we find that GPPs and PPs %have a comparable level of accuracy. Radial $F$-mode frequency as a function %of stellar mass
%for EOS SLY4,  QHC19 and MS1b (a-c) and $F$-mode eigenfunction profile for
%a particular model of EOS SLY4 with central density $\rho_c=1.1\times 10^{15}{\rm
%g/cm^3}$ (d)

\section{Conclusion}
We have presented a new parameterized EOS for cold, nuclear matter. This formulation was derived by reconsidering the piecewise polytrope formalism in the context of  barotropic thermodynamics. The result was a generalization of classical polytropes that includes an additional integration constant, akin to a cosmological constant. Including this constant allows us to impose differentiability between segments in a piecewise formalism. The result, that we term a generalized piecewise polytrope or GPP, retained the simplicity of standard PPs and had the additional advantage of smooth behavior. 
 
We demonstrated the effectiveness of our formalism by using it to fit microphysically motivated candidate EOSs then comparing the predictions of the formalism to predictions of the original EOS. We found that, overall, it creates an accurate fit to the original EOS and that it tracks the adiabatic index reasonably well. By carefully selecting the dividing densities used in the piecewise formalism, we were able to create fits that accurately reproduced the integral observables predicted by candidate EOSs. 

This new formalism has several potential applications. The first is in numerical relativity simulations. A smooth but algebraically simple representation of a candidate EOS would facilitate calculations that are fast and accurately capture the behavior of the underlying EOS, avoiding artificial reflections and improving convergence near the dividing densities. Secondly, the kinks of PPs near dividing densities and the reduced accuracy for low mass stars sometimes introduce bias in Bayesian parameter estimation from GW observations \cite{Fasano:2019zwm}. Spectral expansions are preferred in parameter estimation for this reason. It would be interesting to compare the performance of our differentiable formalism to a smooth spectral expansion.

In this work, when selecting the dividing densities, we have focused on optimizing the error in faithfully recovering observables from EOS parameters.
The astrophysical goal is to faithfully recover EOS parameters from observables. It may thus be more natural to optimize the error in the direction ``observables $ \rightarrow$ EOS" instead of ``EOS $ \rightarrow$ observables". It is worth exploring whether an ``observables
$ \rightarrow$ EOS" minimization gives a significant decrease in the error with which the EOS is recovered from observation.   However, as argued in \cite{PhysRevD.89.064003,PhysRevD.101.123007}, this is not an exact 1-1 function inversion problem, but a problem statistical in nature, best treated via Bayesian methods.

Finally, we mention, although we do not consider the possibility here, that a piecewise formalism more naturally accommodates first order phase transitions than a spectral expansion. Indeed, our formalism could be modified to include such processes by relaxing or modifying the constraint relations \eqref{eq:ourkequation}-\eqref{eq:ouraequation}. This may be useful in the future if GW measurements become precise enough to explore the possibility of a hadron-quark phase transition in neutron stars \cite{Orsaria_2019,PhysRevLett.122.061101,PhysRevLett.122.061102,doi:10.1063/1.5117803,PhysRevD.97.084038,Chatziioannou:2019yko}.

%https://arxiv.org/pdf/1402.6618.pdf
%https://arxiv.org/pdf/1711.02644.pdf
%https://arxiv.org/pdf/1903.09121.pdf
%https://arxiv.org/pdf/1712.00451.pdf
%https://arxiv.org/pdf/1009.4932.pdf

\section*{Acknowledgements}
We are particularly grateful to John Friedman for helpful discussions and suggestions while performing this work. We also thank Gordon Baym, Charles Gammie, Roland Haas, Vasileios Paschalidis, Cole Miller, J. Ryan Westernacher-Schneider and Leslie Wade for useful suggestions. 
C.M. was supported by the European Union's Horizon 2020 research and innovation programme under the Marie Sk\l odowska-Curie grant agreement No 753115 and by COST  Action  MP1304 NewCompStar. The numerical code for the radial oscillations  was derived from a code  written by P. Kolitsidou for her B.Sc. thesis at AUTh. N.S. was supported by the ARIS facility of GRNET in
Athens (SIMGRAV, SIMDIFF and BNSMERGE allocations) and the Aristotle Cluster at AUTh and
by COST actions CA16214 PHAROS, CA16104
GWVerse, CA17137 G2Net and CA18108 QGMM. 
\appendix
\section{Parameters for Specific EOSs}
To demonstrate that GPPs can accurately reproduce the core region of an EOS, we follow the procedure outlined in \S \ref{sec:OurEOS} to obtain a set of parameters $\{K_1,\Gamma_1,\Gamma_2,\Gamma_3\}$ for each EOS in the LIGO Lab. We take the low density region EOS of each EOS provided with each table as the crust EOS used in Eqs. \eqref{eq:ourcrust1}-\eqref{eq:ourcrust3}. The results are presented in Table \ref{tab:GPP_Fits}.

The matching density $\rho_0$ is more difficult to determine in the GPP formalism than it is in the PP formalism, since it requires numerically computing a thermodynamic derivative (c.f. Eq. \eqref{eq:ourcrust1}). We provide this value in the first column for this reason. Combined with $K_1$ and $\Gamma_1$, this allows $\Lambda_1$ and $a_1$ to be computed from \eqref{eq:ourcrust2} and \eqref{eq:ourcrust3}, respectively. The remaining $K_i$, $\Lambda_i$, and $a_i$ follow from the continuity conditions \eqref{eq:ourkequation}-\eqref{eq:ouraequation}. 

We provide both the residual of the fit in the core region defined by Eq. \eqref{eq:OurLeastSquare} and the observable error defined by \eqref{eq:ObservableError}. We also provided the relative error of three astrophysical quantities predicted by the fit compared to the original EOS: the maximum mass, the radius of a $1.4~M_{\odot}$ model, and the tidal deformability of a $1.4~M_{\odot}$ model. With two exceptions, the predictions of GPPs deviate by less than 1 \% from the predictions of the original candidate.

\begin{table*}
\begin{tabular}{c||c c c c c|c|c||c c|c c|c c}
     EOS & $\rho_0~(\times 10^{14}~\mathrm{g/cm^3})$ & $\log K_1$ & $\Gamma_1$ & $\Gamma_2$ & $\Gamma_3$ & EOS Res. & Obs. Res. & $M_{\rm{max}}$ & \% & $R_{1.4}$ & \% & $\Lambda^{1/5}_{1.4}$ & \%  \\
     \hline
     \hline
     APR & 2.676 & -34.917 & 3.282 & 3.595 & 3.305 & 3.420e-7 & 4.078e-4 & 2.057 & 0.01 & 11.345 & 0.58 & 3.029 & 0.66 \\
     
     BHF & 1.912 & -33.541 & 3.185 & 2.838 & 2.753 & 9.442e-5 & 5.054e-3 & 1.921 & 0.11 & 11.173 & 2.14 & 2.946 & 2.52 \\
     
     FPS & 2.491 & -28.901 & 2.873 & 2.580 & 2.534 & 3.040e-6 & 2.768e-5 & 1.802 & 0.09 & 10.854 & 0.15 & 2.838 & 0.03 \\
     
     H4 & 3.547 & -21.110 & 2.369 & 1.535 & 2.312 & 6.652e-4 & 6.473e-3 & 2.035 & 0.06 & 13.691 & 0.60 & 3.911 & 1.06 \\
     
     KDE0V & 2.730 & -30.351 & 2.974 & 2.788 & 2.808 & 6.820e-8 & 3.089e-5 & 1.961 & 0.01 & 11.431 & 0.18 & 3.020 & 0.22 \\
     
     KDE0V1 & 2.709 & -29.531 & 2.920 & 2.786 & 2.758 & 3.606e-7 & 2.637e-5 & 1.970 & 0.00 & 11.639 & 0.16 & 3.077 & 0.20 \\
     
     MPA1 & 1.781 & -40.301 & 3.661 & 3.044 & 2.580 & 2.731e-6 & 3.080e-5 & 2.465 & 0.03 & 12.467 & 0.20 & 3.469 & 0.15 \\
     
     MS1 & 2.748 & -35.667 & 3.369 & 1.112 & 1.911 & 5.329e-4 & 2.387e-3 & 2.771 & 0.26 & 14.898 & 0.22 & 4.268 & 0.35 \\
     
     MS1b & 2.390 & -34.955 & 3.321 & 1.047 & 1.933 & 5.049e-4 & 2.227e-3 & 2.766 & 0.22 & 14.501 & 0.29 & 4.163 & 0.43 \\
     
     QHC19 & 1.313 & -36.879 & 3.420 & 2.597 & 1.769 & 2.608e-4 & 2.341e-3 & 2.069 & 0.01 & 11.595 & 0.80 & 3.169 & 0.96 \\
     
     RS & 1.771 & -25.150 & 2.636 & 2.749 & 2.638 & 2.592e-6 & 8.330e-5 & 2.117 & 0.00 & 12.945 & 0.17 & 3.604 & 0.27 \\
     
     SK255 & 2.817 & -26.896 & 2.754 & 2.739 & 2.684 & 7.601e-8 & 1.851e-6 & 2.145 & 0.00 & 13.162 & 0.03 & 3.599 & 0.05 \\
     
     SK272 & 1.903 & -27.597 & 2.804 & 2.867 & 2.718 & 4.749e-6 & 2.385e-4 & 2.233 & 0.02 & 13.330 & 0.39 & 3.664 & 0.50 \\
     
     SKI2 & 1.826 & -24.202 & 2.575 & 2.688 & 2.636 & 1.401e-6 & 6.876e-6 & 2.164 & 0.00 & 13.500 & 0.02 & 3.798 & 0.00 \\
     
     SKI3 & 1.865 & -26.457 & 2.729 & 2.552 & 2.757 & 1.486e-5 & 2.492e-4 & 2.241 & 0.02 & 13.571 & 0.25 & 3.813 & 0.37 \\
     
     SKI4 & 1.939 & -31.008 & 3.029 & 2.564 & 2.725 & 1.916e-5 & 6.021e-4 & 2.170 & 0.02 & 12.387 & 0.52 & 3.446 & 0.69 \\
     
     SKI5 & 1.761 & -23.109 & 2.505 & 2.842 & 2.782 & 5.797e-5 & 3.522e-4 & 2.241 & 0.05 & 14.010 & 0.08 & 4.010 & 0.19 \\
     
     SKI6 & 1.943 & -31.089 & 3.036 & 2.556 & 2.736 & 2.368e-5 & 6.122e-4 & 2.191 & 0.02 & 12.503 & 0.49 & 3.474 & 0.66 \\
     
     SKMP & 1.739 & -27.116 & 2.766 & 2.757 & 2.705 & 8.062e-8 & 9.109e-6 & 2.108 & 0.01 & 12.511 & 0.08 & 3.455 & 0.11 \\
     
     SKOP & 1.379 & -26.089 & 2.692 & 2.684 & 2.603 & 2.534e-7 & 1.230e-5 & 1.974 & 0.01 & 12.141 & 0.12 & 3.269 & 0.14 \\
     
     SLY2 & 1.987 & -31.070 & 3.026 & 2.835 & 2.786 & 1.924e-6 & 9.112e-5 & 2.055 & 0.00 & 11.796 & 0.24 & 3.175 & 0.33 \\
     
     SLY230A & 1.739 & -33.385 & 3.184 & 2.807 & 2.678 & 9.863e-6 & 4.067e-4 & 2.100 & 0.02 & 11.845 & 0.53 & 3.214 & 0.63 \\
     
     SLY4 & 1.975 & -31.254 & 3.038 & 2.854 & 2.809 & 2.079e-6 & 9.809e-5 & 2.053 & 0.07 & 11.693 & 0.23 & 3.151 & 0.36 \\
     
     SLY9 & 1.080 & -30.657 & 3.005 & 2.675 & 2.720 & 8.760e-6 & 2.305e-4 & 2.157 & 0.00 & 12.482 & 0.30 & 3.413 & 0.42 \\
     
     WFF1 & 2.817 & -38.158 & 3.489 & 3.850 & 4.073 & 8.157e-5 & 1.361e-3 & 1.926 & 0.43 & 10.419 & 0.82 & 2.748 & 0.74 \\
\end{tabular}
\caption{Fit parameters for a GPP core matched to a tabulated crust. To remove a source of error we use the low density region provided by each EOS table. The residuals defined by Eqs. \eqref{eq:OurLeastSquare} and \eqref{eq:ObservableError} provided as well as percent errors for observable quantities of interest.}
\label{tab:GPP_Fits}
\end{table*}

\section{GPP Fit to the SLy(4) Crust}
As described in the main text, some applications may require an accurate representation of the low-density region of the cold, degenerate EOS. This region has been well-studied both theoretically and experimentally \cite{Togashi:2017mjp,Baym:2017whm}, so accurate models are available. However, the version of SLy(4) in CompOSE only describes densities down to $\sim 10^8~\mathrm{g/cm^3}$. So, for the purpose of creating a low-density crust, we used the low density region of QHC19 from $10^5~\mathrm{g/cm^3}$ to $10^8~\mathrm{g/cm^3}$ then the SLY table from the LIGO lab below $10^5~\mathrm{g/cm^3}$. 

When the adiabatic index was computed from the resulting table, we found that it contained significant jumps near the densities where the original EOSs were joined. These jumps are a numerical artifact that can significantly impact the physics predicted by the EOS. We removed these jumps by deleting points from the joined table until a smooth trend in the adiabatic index was achieved. We worked with this modified table to obtain the GPP fit reported in Table \ref{tab:SLY_GPP_Values}. The dividing densities were selected to create a smooth fit to $p$ vs. $\rho$ and $\gamma$ vs. $\rho$.

It is important to note that GPP parameters are sensitive to the choice of crust. So, the fit parameters obtained by matching to the tabulated crust reported in Table \ref{tab:GPP_Fits} are not valid for our low-density GPP. So, we computed a new set of fit parameters to be used with this crust and report them in Table \ref{tab:GPP_crust_fits}. For convenience, we provide the matching density to be used with each set of parameters. If greater precision is required, the matching densities may be recalculated by
\begin{equation}
    \rho_0 = \bigg(\frac{K_1 \Gamma_1}{K_{\mathrm{crust}} \Gamma_{\mathrm{crust}}}\bigg)^{1/(\Gamma_{\mathrm{crust}} - \Gamma_1)}
\end{equation}
where the subscript ``crust'' denotes the last parameter listed in Table \ref{tab:SLY_GPP_Values}. The remaining core parameters follow from \eqref{eq:ourkequation}-\eqref{eq:ouraequation} as before.

\begin{table*}
    \centering
    \begin{tabular}{c|c|c|c|c}
        $\rho_i~(\mathrm{g/cm^3})$ & $K_i$ (cgs) & $\Gamma_i$ & $\Lambda_i~(\mathrm{g/cm^3})$ & $a_i$ \\
        \hline
        0 & 5.214e-9 & 1.611 & 0 & 0 \\
        6.285e5 & 5.726e-8 & 1.440 & -1.354 & -1.861e-5  \\
        1.826e6 & 1.662e-6 & 1.269 & -6.025e3 & -5.278e-4 \\
        3.350e11 & -7.957e29 & -1.841 & 1.193e9 & 1.035e-2 \\
        5.317e11 & 1.746e-8 & 1.382 & 7.077e8 & 8.208e-3 \\
    \end{tabular}
    \caption{A GPP fit to the SLy(4) Crust found in \cite{Douchin:2001sv}. This fit was designed to accurately reproduce adiabatic index as well as pressure at low densities.}
    \label{tab:SLY_GPP_Values}
\end{table*}

\begin{table*}
    \centering
    \begin{tabular}{c||c c c c c}
    EOS & $\log\rho_0~(\mathrm{g/cm^3})$ & $\log K_1$ & $\Gamma_1$ & $\Gamma_2$ & $\Gamma_3$ \\
     \hline
     \hline
     APR & 14.040 & -33.210 & 3.169 & 3.452 & 3.310 \\
     
     BHF & 14.130 & -35.016 & 3.284 & 2.774 & 2.616 \\
     
     FPS & 14.087 & -32.985 & 3.147 & 2.652 & 2.120 \\
     
     H4 & 13.499 & -23.310 & 2.514 & 2.333 & 1.562 \\
     
     KDE0V & 13.978 & -30.250 & 2.967 & 2.835 & 2.803 \\
     
     KDE0V1 & 13.929 & -29.232 & 2.900 & 2.809 & 2.747 \\
     
     MPA1 & 14.088 & -40.301 & 3.662 & 3.057 & 2.298 \\
     
     MS1 & 13.657 & -30.170 & 2.998 & 2.123 & 1.955 \\
     
     MS1b & 13.795 & -33.774 & 3.241 & 2.136 & 1.963 \\
     
     QHC19 & 14.102 & -36.879 & 3.419 & 2.760 & 2.017 \\
     
     RS & 13.641 & -25.150 & 2.636 & 2.677 & 2.647 \\
     
     SK255 & 13.679 & -25.990 & 2.693 & 2.729 & 2.667 \\
     
     SK272 & 13.732 & -27.597 & 2.804 & 2.793 & 2.733 \\
     
     SKI2 & 13.552 & -24.202 & 2.575 & 2.639 & 2.656 \\
     
     SKI3 & 13.660 & -26.457 & 2.729 & 2.680 & 2.708 \\
     
     SKI4 & 13.907 & -31.008 & 3.029 & 2.759 & 2.651 \\
     
     SKI5 & 13.438 & -23.109 & 2.505 & 2.708 & 2.727 \\
     
     SKI6 & 13.902 & -31.089 & 3.036 & 2.762 & 2.653 \\
     
     SKMP & 13.763 & -27.116 & 2.766 & 2.741 & 2.698 \\
     
     SKOP & 13.761 & -26.089 & 2.693 & 2.660 & 2.579 \\
     
     SLY2 & 13.967 & -31.070 & 3.026 & 2.871 & 2.760 \\
     
     SLY230A & 14.021 & -33.385 & 3.184 & 2.895 & 2.588 \\
     
     SLY4 & 13.980 & -31.350 & 3.045 & 2.884 & 2.773 \\
     
     SLY9 & 13.899 & -30.657 & 3.005 & 2.796 & 2.652 \\
     
     WFF1 & 14.133 & -34.394 & 3.240 & 3.484 & 3.695 \\
    \end{tabular}
    \caption{Fit parameters for the core region of each EOS matched to the SLy(4) crust in Table \ref{tab:SLY_GPP_Values}. These values should provide a self-consistent EOS for use in simulations where accurate crust resolution is desirable.}
    \label{tab:GPP_crust_fits}
\end{table*}

\bibliography{new_references}

\begin{thebibliography}{80}%
\makeatletter
\providecommand \@ifxundefined [1]{%
 \@ifx{#1\undefined}
}%
\providecommand \@ifnum [1]{%
 \ifnum #1\expandafter \@firstoftwo
 \else \expandafter \@secondoftwo
 \fi
}%
\providecommand \@ifx [1]{%
 \ifx #1\expandafter \@firstoftwo
 \else \expandafter \@secondoftwo
 \fi
}%
\providecommand \natexlab [1]{#1}%
\providecommand \enquote  [1]{``#1''}%
\providecommand \bibnamefont  [1]{#1}%
\providecommand \bibfnamefont [1]{#1}%
\providecommand \citenamefont [1]{#1}%
\providecommand \href@noop [0]{\@secondoftwo}%
\providecommand \href [0]{\begingroup \@sanitize@url \@href}%
\providecommand \@href[1]{\@@startlink{#1}\@@href}%
\providecommand \@@href[1]{\endgroup#1\@@endlink}%
\providecommand \@sanitize@url [0]{\catcode `\\12\catcode `\$12\catcode
  `\&12\catcode `\#12\catcode `\^12\catcode `\_12\catcode `\%12\relax}%
\providecommand \@@startlink[1]{}%
\providecommand \@@endlink[0]{}%
\providecommand \url  [0]{\begingroup\@sanitize@url \@url }%
\providecommand \@url [1]{\endgroup\@href {#1}{\urlprefix }}%
\providecommand \urlprefix  [0]{URL }%
\providecommand \Eprint [0]{\href }%
\providecommand \doibase [0]{http://dx.doi.org/}%
\providecommand \selectlanguage [0]{\@gobble}%
\providecommand \bibinfo  [0]{\@secondoftwo}%
\providecommand \bibfield  [0]{\@secondoftwo}%
\providecommand \translation [1]{[#1]}%
\providecommand \BibitemOpen [0]{}%
\providecommand \bibitemStop [0]{}%
\providecommand \bibitemNoStop [0]{.\EOS\space}%
\providecommand \EOS [0]{\spacefactor3000\relax}%
\providecommand \BibitemShut  [1]{\csname bibitem#1\endcsname}%
\let\auto@bib@innerbib\@empty
%</preamble>
\bibitem [{\citenamefont {Oertel}\ \emph {et~al.}(2017)\citenamefont {Oertel},
  \citenamefont {Hempel}, \citenamefont {Kl\"ahn},\ and\ \citenamefont
  {Typel}}]{RevModPhys.89.015007}%
  \BibitemOpen
  \bibfield  {author} {\bibinfo {author} {\bibfnamefont {M.}~\bibnamefont
  {Oertel}}, \bibinfo {author} {\bibfnamefont {M.}~\bibnamefont {Hempel}},
  \bibinfo {author} {\bibfnamefont {T.}~\bibnamefont {Kl\"ahn}}, \ and\
  \bibinfo {author} {\bibfnamefont {S.}~\bibnamefont {Typel}},\ }\href
  {\doibase 10.1103/RevModPhys.89.015007} {\bibfield  {journal} {\bibinfo
  {journal} {Rev. Mod. Phys.}\ }\textbf {\bibinfo {volume} {89}},\ \bibinfo
  {pages} {015007} (\bibinfo {year} {2017})}\BibitemShut {NoStop}%
\bibitem [{\citenamefont {Baym}\ \emph {et~al.}(2018)\citenamefont {Baym},
  \citenamefont {Hatsuda}, \citenamefont {Kojo}, \citenamefont {Powell},
  \citenamefont {Song},\ and\ \citenamefont {Takatsuka}}]{Baym:2017whm}%
  \BibitemOpen
  \bibfield  {author} {\bibinfo {author} {\bibfnamefont {G.}~\bibnamefont
  {Baym}}, \bibinfo {author} {\bibfnamefont {T.}~\bibnamefont {Hatsuda}},
  \bibinfo {author} {\bibfnamefont {T.}~\bibnamefont {Kojo}}, \bibinfo {author}
  {\bibfnamefont {P.~D.}\ \bibnamefont {Powell}}, \bibinfo {author}
  {\bibfnamefont {Y.}~\bibnamefont {Song}}, \ and\ \bibinfo {author}
  {\bibfnamefont {T.}~\bibnamefont {Takatsuka}},\ }\href {\doibase
  10.1088/1361-6633/aaae14} {\bibfield  {journal} {\bibinfo  {journal} {Rept.
  Prog. Phys.}\ }\textbf {\bibinfo {volume} {81}},\ \bibinfo {pages} {056902}
  (\bibinfo {year} {2018})},\ \Eprint {http://arxiv.org/abs/1707.04966}
  {arXiv:1707.04966 [astro-ph.HE]} \BibitemShut {NoStop}%
%%CITATION = ARXIV:1707.04966;%%
\bibitem [{\citenamefont {Antoniadis}\ \emph {et~al.}(2013)\citenamefont
  {Antoniadis} \emph {et~al.}}]{Antoniadis:2013pzd}%
  \BibitemOpen
  \bibfield  {author} {\bibinfo {author} {\bibfnamefont {J.}~\bibnamefont
  {Antoniadis}} \emph {et~al.},\ }\href {\doibase 10.1126/science.1233232}
  {\bibfield  {journal} {\bibinfo  {journal} {Science}\ }\textbf {\bibinfo
  {volume} {340}},\ \bibinfo {pages} {6131} (\bibinfo {year} {2013})},\ \Eprint
  {http://arxiv.org/abs/1304.6875} {arXiv:1304.6875 [astro-ph.HE]} \BibitemShut
  {NoStop}%
%%CITATION = ARXIV:1304.6875;%%
\bibitem [{\citenamefont {Arzoumanian}\ \emph {et~al.}(2018)\citenamefont
  {Arzoumanian} \emph {et~al.}}]{Arzoumanian:2017puf}%
  \BibitemOpen
  \bibfield  {author} {\bibinfo {author} {\bibfnamefont {Z.}~\bibnamefont
  {Arzoumanian}} \emph {et~al.} (\bibinfo {collaboration} {NANOGrav}),\ }\href
  {\doibase 10.3847/1538-4365/aab5b0} {\bibfield  {journal} {\bibinfo
  {journal} {Astrophys. J. Suppl.}\ }\textbf {\bibinfo {volume} {235}},\
  \bibinfo {pages} {37} (\bibinfo {year} {2018})},\ \Eprint
  {http://arxiv.org/abs/1801.01837} {arXiv:1801.01837 [astro-ph.HE]}
  \BibitemShut {NoStop}%
%%CITATION = ARXIV:1801.01837;%%
\bibitem [{\citenamefont {Fortin}\ \emph {et~al.}(2015)\citenamefont {Fortin},
  \citenamefont {Zdunik}, \citenamefont {Haensel},\ and\ \citenamefont
  {Bejger}}]{Fortin:2014mya}%
  \BibitemOpen
  \bibfield  {author} {\bibinfo {author} {\bibfnamefont {M.}~\bibnamefont
  {Fortin}}, \bibinfo {author} {\bibfnamefont {J.~L.}\ \bibnamefont {Zdunik}},
  \bibinfo {author} {\bibfnamefont {P.}~\bibnamefont {Haensel}}, \ and\
  \bibinfo {author} {\bibfnamefont {M.}~\bibnamefont {Bejger}},\ }\href
  {\doibase 10.1051/0004-6361/201424800} {\bibfield  {journal} {\bibinfo
  {journal} {Astron. Astrophys.}\ }\textbf {\bibinfo {volume} {576}},\ \bibinfo
  {pages} {A68} (\bibinfo {year} {2015})},\ \Eprint
  {http://arxiv.org/abs/1408.3052} {arXiv:1408.3052 [astro-ph.SR]} \BibitemShut
  {NoStop}%
%%CITATION = ARXIV:1408.3052;%%
\bibitem [{\citenamefont {Ozel}\ \emph {et~al.}(2016)\citenamefont {Ozel},
  \citenamefont {Psaltis}, \citenamefont {Guver}, \citenamefont {Baym},
  \citenamefont {Heinke},\ and\ \citenamefont {Guillot}}]{Ozel:2015fia}%
  \BibitemOpen
  \bibfield  {author} {\bibinfo {author} {\bibfnamefont {F.}~\bibnamefont
  {Ozel}}, \bibinfo {author} {\bibfnamefont {D.}~\bibnamefont {Psaltis}},
  \bibinfo {author} {\bibfnamefont {T.}~\bibnamefont {Guver}}, \bibinfo
  {author} {\bibfnamefont {G.}~\bibnamefont {Baym}}, \bibinfo {author}
  {\bibfnamefont {C.}~\bibnamefont {Heinke}}, \ and\ \bibinfo {author}
  {\bibfnamefont {S.}~\bibnamefont {Guillot}},\ }\href {\doibase
  10.3847/0004-637X/820/1/28} {\bibfield  {journal} {\bibinfo  {journal}
  {Astrophys. J.}\ }\textbf {\bibinfo {volume} {820}},\ \bibinfo {pages} {28}
  (\bibinfo {year} {2016})},\ \Eprint {http://arxiv.org/abs/1505.05155}
  {arXiv:1505.05155 [astro-ph.HE]} \BibitemShut {NoStop}%
%%CITATION = ARXIV:1505.05155;%%
\bibitem [{\citenamefont {Bogdanov}(2013)}]{Bogdanov:2012md}%
  \BibitemOpen
  \bibfield  {author} {\bibinfo {author} {\bibfnamefont {S.}~\bibnamefont
  {Bogdanov}},\ }\href {\doibase 10.1088/0004-637X/762/2/96} {\bibfield
  {journal} {\bibinfo  {journal} {Astrophys. J.}\ }\textbf {\bibinfo {volume}
  {762}},\ \bibinfo {pages} {96} (\bibinfo {year} {2013})},\ \Eprint
  {http://arxiv.org/abs/1211.6113} {arXiv:1211.6113 [astro-ph.HE]} \BibitemShut
  {NoStop}%
%%CITATION = ARXIV:1211.6113;%%
\bibitem [{\citenamefont {Verbiest}\ \emph {et~al.}(2008)\citenamefont
  {Verbiest}, \citenamefont {Bailes}, \citenamefont {van Straten},
  \citenamefont {Hobbs}, \citenamefont {Edwards}, \citenamefont {Manchester},
  \citenamefont {Bhat}, \citenamefont {Sarkissian}, \citenamefont {Jacoby},\
  and\ \citenamefont {Kulkarni}}]{Verbiest:2008gy}%
  \BibitemOpen
  \bibfield  {author} {\bibinfo {author} {\bibfnamefont {J.~P.~W.}\
  \bibnamefont {Verbiest}}, \bibinfo {author} {\bibfnamefont {M.}~\bibnamefont
  {Bailes}}, \bibinfo {author} {\bibfnamefont {W.}~\bibnamefont {van Straten}},
  \bibinfo {author} {\bibfnamefont {G.~B.}\ \bibnamefont {Hobbs}}, \bibinfo
  {author} {\bibfnamefont {R.~T.}\ \bibnamefont {Edwards}}, \bibinfo {author}
  {\bibfnamefont {R.~N.}\ \bibnamefont {Manchester}}, \bibinfo {author}
  {\bibfnamefont {N.~D.~R.}\ \bibnamefont {Bhat}}, \bibinfo {author}
  {\bibfnamefont {J.~M.}\ \bibnamefont {Sarkissian}}, \bibinfo {author}
  {\bibfnamefont {B.~A.}\ \bibnamefont {Jacoby}}, \ and\ \bibinfo {author}
  {\bibfnamefont {S.~R.}\ \bibnamefont {Kulkarni}},\ }\href {\doibase
  10.1086/529576} {\bibfield  {journal} {\bibinfo  {journal} {Astrophys. J.}\
  }\textbf {\bibinfo {volume} {679}},\ \bibinfo {pages} {675} (\bibinfo {year}
  {2008})},\ \Eprint {http://arxiv.org/abs/0801.2589} {arXiv:0801.2589
  [astro-ph]} \BibitemShut {NoStop}%
%%CITATION = ARXIV:0801.2589;%%
\bibitem [{\citenamefont {Haensel}\ \emph {et~al.}(2009)\citenamefont
  {Haensel}, \citenamefont {Zdunik}, \citenamefont {Bejger},\ and\
  \citenamefont {Lattimer}}]{Haensel_2009}%
  \BibitemOpen
  \bibfield  {author} {\bibinfo {author} {\bibfnamefont {P.}~\bibnamefont
  {Haensel}}, \bibinfo {author} {\bibfnamefont {J.~L.}\ \bibnamefont {Zdunik}},
  \bibinfo {author} {\bibfnamefont {M.}~\bibnamefont {Bejger}}, \ and\ \bibinfo
  {author} {\bibfnamefont {J.~M.}\ \bibnamefont {Lattimer}},\ }\href {\doibase
  10.1051/0004-6361/200811605} {\bibfield  {journal} {\bibinfo  {journal}
  {Astronomy \& Astrophysics}\ }\textbf {\bibinfo {volume} {502}},\ \bibinfo
  {pages} {605} (\bibinfo {year} {2009})}\BibitemShut {NoStop}%
\bibitem [{\citenamefont {Abbott}\ \emph {et~al.}(2017)\citenamefont {Abbott}
  \emph {et~al.}}]{PhysRevLett.119.161101}%
  \BibitemOpen
  \bibfield  {author} {\bibinfo {author} {\bibfnamefont {B.~P.}\ \bibnamefont
  {Abbott}} \emph {et~al.} (\bibinfo {collaboration} {LIGO Scientific
  Collaboration and Virgo Collaboration}),\ }\href {\doibase
  10.1103/PhysRevLett.119.161101} {\bibfield  {journal} {\bibinfo  {journal}
  {Phys. Rev. Lett.}\ }\textbf {\bibinfo {volume} {119}},\ \bibinfo {pages}
  {161101} (\bibinfo {year} {2017})}\BibitemShut {NoStop}%
\bibitem [{\citenamefont {Abbott}\ \emph
  {et~al.}(2018{\natexlab{a}})\citenamefont {Abbott} \emph
  {et~al.}}]{PhysRevLett.121.161101}%
  \BibitemOpen
  \bibfield  {author} {\bibinfo {author} {\bibfnamefont {B.~P.}\ \bibnamefont
  {Abbott}} \emph {et~al.} (\bibinfo {collaboration} {The LIGO Scientific
  Collaboration and the Virgo Collaboration}),\ }\href {\doibase
  10.1103/PhysRevLett.121.161101} {\bibfield  {journal} {\bibinfo  {journal}
  {Phys. Rev. Lett.}\ }\textbf {\bibinfo {volume} {121}},\ \bibinfo {pages}
  {161101} (\bibinfo {year} {2018}{\natexlab{a}})}\BibitemShut {NoStop}%
\bibitem [{\citenamefont {Abbott}\ \emph {et~al.}(2020)\citenamefont {Abbott}
  \emph {et~al.}}]{Abbott_2020}%
  \BibitemOpen
  \bibfield  {author} {\bibinfo {author} {\bibfnamefont {B.~P.}\ \bibnamefont
  {Abbott}} \emph {et~al.},\ }\href {\doibase 10.1088/1361-6382/ab5f7c}
  {\bibfield  {journal} {\bibinfo  {journal} {Classical and Quantum Gravity}\
  }\textbf {\bibinfo {volume} {37}},\ \bibinfo {pages} {045006} (\bibinfo
  {year} {2020})}\BibitemShut {NoStop}%
\bibitem [{\citenamefont {Annala}\ \emph
  {et~al.}(2018{\natexlab{a}})\citenamefont {Annala}, \citenamefont {Gorda},
  \citenamefont {Kurkela},\ and\ \citenamefont
  {Vuorinen}}]{PhysRevLett.120.172703}%
  \BibitemOpen
  \bibfield  {author} {\bibinfo {author} {\bibfnamefont {E.}~\bibnamefont
  {Annala}}, \bibinfo {author} {\bibfnamefont {T.}~\bibnamefont {Gorda}},
  \bibinfo {author} {\bibfnamefont {A.}~\bibnamefont {Kurkela}}, \ and\
  \bibinfo {author} {\bibfnamefont {A.}~\bibnamefont {Vuorinen}},\ }\href
  {\doibase 10.1103/PhysRevLett.120.172703} {\bibfield  {journal} {\bibinfo
  {journal} {Phys. Rev. Lett.}\ }\textbf {\bibinfo {volume} {120}},\ \bibinfo
  {pages} {172703} (\bibinfo {year} {2018}{\natexlab{a}})}\BibitemShut
  {NoStop}%
\bibitem [{\citenamefont {Wysocki}\ \emph {et~al.}(2020)\citenamefont
  {Wysocki}, \citenamefont {O'Shaughnessy}, \citenamefont {Wade},\ and\
  \citenamefont {Lange}}]{wysocki2020inferring}%
  \BibitemOpen
  \bibfield  {author} {\bibinfo {author} {\bibfnamefont {D.}~\bibnamefont
  {Wysocki}}, \bibinfo {author} {\bibfnamefont {R.}~\bibnamefont
  {O'Shaughnessy}}, \bibinfo {author} {\bibfnamefont {L.}~\bibnamefont {Wade}},
  \ and\ \bibinfo {author} {\bibfnamefont {J.}~\bibnamefont {Lange}},\
  }\href@noop {} {\enquote {\bibinfo {title} {Inferring the neutron star
  equation of state simultaneously with the population of merging neutron
  stars},}\ } (\bibinfo {year} {2020}),\ \Eprint
  {http://arxiv.org/abs/2001.01747} {arXiv:2001.01747 [gr-qc]} \BibitemShut
  {NoStop}%
\bibitem [{\citenamefont {Buonanno}\ \emph {et~al.}(2009)\citenamefont
  {Buonanno}, \citenamefont {Iyer}, \citenamefont {Ochsner}, \citenamefont
  {Pan},\ and\ \citenamefont {Sathyaprakash}}]{Buonanno:2009zt}%
  \BibitemOpen
  \bibfield  {author} {\bibinfo {author} {\bibfnamefont {A.}~\bibnamefont
  {Buonanno}}, \bibinfo {author} {\bibfnamefont {B.~R.}\ \bibnamefont {Iyer}},
  \bibinfo {author} {\bibfnamefont {E.}~\bibnamefont {Ochsner}}, \bibinfo
  {author} {\bibfnamefont {Y.}~\bibnamefont {Pan}}, \ and\ \bibinfo {author}
  {\bibfnamefont {B.~S.}\ \bibnamefont {Sathyaprakash}},\ }\href {\doibase
  10.1103/PhysRevD.80.084043} {\bibfield  {journal} {\bibinfo  {journal} {Phys.
  Rev.}\ }\textbf {\bibinfo {volume} {D80}},\ \bibinfo {pages} {084043}
  (\bibinfo {year} {2009})},\ \Eprint {http://arxiv.org/abs/0907.0700}
  {arXiv:0907.0700 [gr-qc]} \BibitemShut {NoStop}%
%%CITATION = ARXIV:0907.0700;%%
\bibitem [{\citenamefont {Flanagan}\ and\ \citenamefont
  {Hinderer}(2008)}]{Flanagan:2007ix}%
  \BibitemOpen
  \bibfield  {author} {\bibinfo {author} {\bibfnamefont {E.~E.}\ \bibnamefont
  {Flanagan}}\ and\ \bibinfo {author} {\bibfnamefont {T.}~\bibnamefont
  {Hinderer}},\ }\href {\doibase 10.1103/PhysRevD.77.021502} {\bibfield
  {journal} {\bibinfo  {journal} {Phys. Rev.}\ }\textbf {\bibinfo {volume}
  {D77}},\ \bibinfo {pages} {021502(R)} (\bibinfo {year} {2008})},\ \Eprint
  {http://arxiv.org/abs/0709.1915} {arXiv:0709.1915 [astro-ph]} \BibitemShut
  {NoStop}%
%%CITATION = ARXIV:0709.1915;%%
\bibitem [{\citenamefont {Yagi}\ and\ \citenamefont
  {Yunes}(2013)}]{Yagi:2013awa}%
  \BibitemOpen
  \bibfield  {author} {\bibinfo {author} {\bibfnamefont {K.}~\bibnamefont
  {Yagi}}\ and\ \bibinfo {author} {\bibfnamefont {N.}~\bibnamefont {Yunes}},\
  }\href {\doibase 10.1103/PhysRevD.88.023009} {\bibfield  {journal} {\bibinfo
  {journal} {Phys. Rev.}\ }\textbf {\bibinfo {volume} {D88}},\ \bibinfo {pages}
  {023009} (\bibinfo {year} {2013})},\ \Eprint {http://arxiv.org/abs/1303.1528}
  {arXiv:1303.1528 [gr-qc]} \BibitemShut {NoStop}%
%%CITATION = ARXIV:1303.1528;%%
\bibitem [{\citenamefont {Van~Oeveren}\ and\ \citenamefont
  {Friedman}(2017)}]{VanOeveren:2017xkv}%
  \BibitemOpen
  \bibfield  {author} {\bibinfo {author} {\bibfnamefont {E.~D.}\ \bibnamefont
  {Van~Oeveren}}\ and\ \bibinfo {author} {\bibfnamefont {J.~L.}\ \bibnamefont
  {Friedman}},\ }\href {\doibase 10.1103/PhysRevD.95.083014} {\bibfield
  {journal} {\bibinfo  {journal} {Phys. Rev.}\ }\textbf {\bibinfo {volume}
  {D95}},\ \bibinfo {pages} {083014} (\bibinfo {year} {2017})},\ \Eprint
  {http://arxiv.org/abs/1701.03797} {arXiv:1701.03797 [gr-qc]} \BibitemShut
  {NoStop}%
%%CITATION = ARXIV:1701.03797;%%
\bibitem [{\citenamefont {Wade}\ \emph {et~al.}(2014)\citenamefont {Wade},
  \citenamefont {Creighton}, \citenamefont {Ochsner}, \citenamefont {Lackey},
  \citenamefont {Farr}, \citenamefont {Littenberg},\ and\ \citenamefont
  {Raymond}}]{Wade:2014vqa}%
  \BibitemOpen
  \bibfield  {author} {\bibinfo {author} {\bibfnamefont {L.}~\bibnamefont
  {Wade}}, \bibinfo {author} {\bibfnamefont {J.~D.~E.}\ \bibnamefont
  {Creighton}}, \bibinfo {author} {\bibfnamefont {E.}~\bibnamefont {Ochsner}},
  \bibinfo {author} {\bibfnamefont {B.~D.}\ \bibnamefont {Lackey}}, \bibinfo
  {author} {\bibfnamefont {B.~F.}\ \bibnamefont {Farr}}, \bibinfo {author}
  {\bibfnamefont {T.~B.}\ \bibnamefont {Littenberg}}, \ and\ \bibinfo {author}
  {\bibfnamefont {V.}~\bibnamefont {Raymond}},\ }\href {\doibase
  10.1103/PhysRevD.89.103012} {\bibfield  {journal} {\bibinfo  {journal} {Phys.
  Rev.}\ }\textbf {\bibinfo {volume} {D89}},\ \bibinfo {pages} {103012}
  (\bibinfo {year} {2014})},\ \Eprint {http://arxiv.org/abs/1402.5156}
  {arXiv:1402.5156 [gr-qc]} \BibitemShut {NoStop}%
%%CITATION = ARXIV:1402.5156;%%
\bibitem [{\citenamefont {Vivanco}\ \emph {et~al.}(2019)\citenamefont
  {Vivanco}, \citenamefont {Smith}, \citenamefont {Thrane}, \citenamefont
  {Lasky}, \citenamefont {Talbot},\ and\ \citenamefont
  {Raymond}}]{Vivanco:2019qnt}%
  \BibitemOpen
  \bibfield  {author} {\bibinfo {author} {\bibfnamefont {F.~H.}\ \bibnamefont
  {Vivanco}}, \bibinfo {author} {\bibfnamefont {R.}~\bibnamefont {Smith}},
  \bibinfo {author} {\bibfnamefont {E.}~\bibnamefont {Thrane}}, \bibinfo
  {author} {\bibfnamefont {P.~D.}\ \bibnamefont {Lasky}}, \bibinfo {author}
  {\bibfnamefont {C.}~\bibnamefont {Talbot}}, \ and\ \bibinfo {author}
  {\bibfnamefont {V.}~\bibnamefont {Raymond}},\ }\href@noop {} {\  (\bibinfo
  {year} {2019})},\ \Eprint {http://arxiv.org/abs/1909.02698} {arXiv:1909.02698
  [gr-qc]} \BibitemShut {NoStop}%
%%CITATION = ARXIV:1909.02698;%%
\bibitem [{\citenamefont {Read}\ \emph
  {et~al.}(2009{\natexlab{a}})\citenamefont {Read}, \citenamefont {Lackey},
  \citenamefont {Owen},\ and\ \citenamefont {Friedman}}]{Read:2008iy}%
  \BibitemOpen
  \bibfield  {author} {\bibinfo {author} {\bibfnamefont {J.~S.}\ \bibnamefont
  {Read}}, \bibinfo {author} {\bibfnamefont {B.~D.}\ \bibnamefont {Lackey}},
  \bibinfo {author} {\bibfnamefont {B.~J.}\ \bibnamefont {Owen}}, \ and\
  \bibinfo {author} {\bibfnamefont {J.~L.}\ \bibnamefont {Friedman}},\ }\href
  {\doibase 10.1103/PhysRevD.79.124032} {\bibfield  {journal} {\bibinfo
  {journal} {Phys. Rev.}\ }\textbf {\bibinfo {volume} {D79}},\ \bibinfo {pages}
  {124032} (\bibinfo {year} {2009}{\natexlab{a}})},\ \Eprint
  {http://arxiv.org/abs/0812.2163} {arXiv:0812.2163 [astro-ph]} \BibitemShut
  {NoStop}%
%%CITATION = ARXIV:0812.2163;%%
\bibitem [{\citenamefont {Read}\ \emph
  {et~al.}(2009{\natexlab{b}})\citenamefont {Read}, \citenamefont {Markakis},
  \citenamefont {Shibata}, \citenamefont {Uryu}, \citenamefont {Creighton},\
  and\ \citenamefont {Friedman}}]{Read:2009yp}%
  \BibitemOpen
  \bibfield  {author} {\bibinfo {author} {\bibfnamefont {J.~S.}\ \bibnamefont
  {Read}}, \bibinfo {author} {\bibfnamefont {C.}~\bibnamefont {Markakis}},
  \bibinfo {author} {\bibfnamefont {M.}~\bibnamefont {Shibata}}, \bibinfo
  {author} {\bibfnamefont {K.}~\bibnamefont {Uryu}}, \bibinfo {author}
  {\bibfnamefont {J.~D.~E.}\ \bibnamefont {Creighton}}, \ and\ \bibinfo
  {author} {\bibfnamefont {J.~L.}\ \bibnamefont {Friedman}},\ }\href {\doibase
  10.1103/PhysRevD.79.124033} {\bibfield  {journal} {\bibinfo  {journal} {Phys.
  Rev.}\ }\textbf {\bibinfo {volume} {D79}},\ \bibinfo {pages} {124033}
  (\bibinfo {year} {2009}{\natexlab{b}})},\ \Eprint
  {http://arxiv.org/abs/0901.3258} {arXiv:0901.3258 [gr-qc]} \BibitemShut
  {NoStop}%
%%CITATION = ARXIV:0901.3258;%%
\bibitem [{\citenamefont {Markakis}\ \emph {et~al.}(2009)\citenamefont
  {Markakis}, \citenamefont {Read}, \citenamefont {Shibata}, \citenamefont
  {Ury{\=u}}, \citenamefont {Creighton}, \citenamefont {Friedman},\ and\
  \citenamefont {Lackey}}]{markakisNeutronStarEquation2009}%
  \BibitemOpen
  \bibfield  {author} {\bibinfo {author} {\bibfnamefont {C.}~\bibnamefont
  {Markakis}}, \bibinfo {author} {\bibfnamefont {J.~S.}\ \bibnamefont {Read}},
  \bibinfo {author} {\bibfnamefont {M.}~\bibnamefont {Shibata}}, \bibinfo
  {author} {\bibfnamefont {K.}~\bibnamefont {Ury{\=u}}}, \bibinfo {author}
  {\bibfnamefont {J.~D.~E.}\ \bibnamefont {Creighton}}, \bibinfo {author}
  {\bibfnamefont {J.~L.}\ \bibnamefont {Friedman}}, \ and\ \bibinfo {author}
  {\bibfnamefont {B.~D.}\ \bibnamefont {Lackey}},\ }\href {\doibase
  10.1088/1742-6596/189/1/012024} {\bibfield  {journal} {\bibinfo  {journal}
  {Journal of Physics: Conference Series}\ }\textbf {\bibinfo {volume} {189}},\
  \bibinfo {pages} {012024} (\bibinfo {year} {2009})}\BibitemShut {NoStop}%
\bibitem [{\citenamefont {Markakis}\ \emph {et~al.}(2012)\citenamefont
  {Markakis}, \citenamefont {Read}, \citenamefont {Shibata}, \citenamefont
  {Uryu}, \citenamefont {Creighton},\ and\ \citenamefont
  {Friedman}}]{markakisInferringNeutronStar2012}%
  \BibitemOpen
  \bibfield  {author} {\bibinfo {author} {\bibfnamefont {C.}~\bibnamefont
  {Markakis}}, \bibinfo {author} {\bibfnamefont {J.~S.}\ \bibnamefont {Read}},
  \bibinfo {author} {\bibfnamefont {M.}~\bibnamefont {Shibata}}, \bibinfo
  {author} {\bibfnamefont {K.}~\bibnamefont {Uryu}}, \bibinfo {author}
  {\bibfnamefont {J.~D.~E.}\ \bibnamefont {Creighton}}, \ and\ \bibinfo
  {author} {\bibfnamefont {J.~L.}\ \bibnamefont {Friedman}},\ }\bibfield
  {booktitle} {\emph {\bibinfo {booktitle} {The {{Twelfth Marcel Grossmann
  Meeting}}}},\ }\href {\doibase 10.1142/9789814374552_0046} {\ ,\ \bibinfo
  {pages} {743} (\bibinfo {year} {2012})}\BibitemShut {NoStop}%
\bibitem [{\citenamefont {Read}\ \emph {et~al.}(2013)\citenamefont {Read},
  \citenamefont {Baiotti}, \citenamefont {Creighton}, \citenamefont {Friedman},
  \citenamefont {Giacomazzo}, \citenamefont {Kyutoku}, \citenamefont
  {Markakis}, \citenamefont {Rezzolla}, \citenamefont {Shibata},\ and\
  \citenamefont {Taniguchi}}]{readMatterEffectsBinary2013}%
  \BibitemOpen
  \bibfield  {author} {\bibinfo {author} {\bibfnamefont {J.~S.}\ \bibnamefont
  {Read}}, \bibinfo {author} {\bibfnamefont {L.}~\bibnamefont {Baiotti}},
  \bibinfo {author} {\bibfnamefont {J.~D.~E.}\ \bibnamefont {Creighton}},
  \bibinfo {author} {\bibfnamefont {J.~L.}\ \bibnamefont {Friedman}}, \bibinfo
  {author} {\bibfnamefont {B.}~\bibnamefont {Giacomazzo}}, \bibinfo {author}
  {\bibfnamefont {K.}~\bibnamefont {Kyutoku}}, \bibinfo {author} {\bibfnamefont
  {C.}~\bibnamefont {Markakis}}, \bibinfo {author} {\bibfnamefont
  {L.}~\bibnamefont {Rezzolla}}, \bibinfo {author} {\bibfnamefont
  {M.}~\bibnamefont {Shibata}}, \ and\ \bibinfo {author} {\bibfnamefont
  {K.}~\bibnamefont {Taniguchi}},\ }\href {\doibase 10.1103/PhysRevD.88.044042}
  {\bibfield  {journal} {\bibinfo  {journal} {Physical Review D}\ }\textbf
  {\bibinfo {volume} {88}},\ \bibinfo {pages} {044042} (\bibinfo {year}
  {2013})}\BibitemShut {NoStop}%
\bibitem [{\citenamefont {{Vuille}}\ and\ \citenamefont
  {{Ipser}}(1999)}]{Vuille:1999}%
  \BibitemOpen
  \bibfield  {author} {\bibinfo {author} {\bibfnamefont {C.}~\bibnamefont
  {{Vuille}}}\ and\ \bibinfo {author} {\bibfnamefont {J.}~\bibnamefont
  {{Ipser}}},\ }in\ \href {\doibase 10.1063/1.1301564} {\emph {\bibinfo
  {booktitle} {General Relativity and Relativistic Astrophysics}}},\ \bibinfo
  {series} {American Institute of Physics Conference Series}, Vol.\ \bibinfo
  {volume} {493},\ \bibinfo {editor} {edited by\ \bibinfo {editor}
  {\bibfnamefont {C.~P.}\ \bibnamefont {{Burgess}}}\ and\ \bibinfo {editor}
  {\bibfnamefont {R.~C.}\ \bibnamefont {{Myers}}}}\ (\bibinfo {year} {1999})\
  pp.\ \bibinfo {pages} {60--62}\BibitemShut {NoStop}%
\bibitem [{\citenamefont {Hinderer}(2008)}]{Hinderer:2007mb}%
  \BibitemOpen
  \bibfield  {author} {\bibinfo {author} {\bibfnamefont {T.}~\bibnamefont
  {Hinderer}},\ }\href {\doibase 10.1086/533487} {\bibfield  {journal}
  {\bibinfo  {journal} {Astrophys. J.}\ }\textbf {\bibinfo {volume} {677}},\
  \bibinfo {pages} {1216} (\bibinfo {year} {2008})},\ \Eprint
  {http://arxiv.org/abs/0711.2420} {arXiv:0711.2420 [astro-ph]} \BibitemShut
  {NoStop}%
%%CITATION = ARXIV:0711.2420;%%
\bibitem [{\citenamefont {Kastaun}(2008)}]{Kastaun:2008jr}%
  \BibitemOpen
  \bibfield  {author} {\bibinfo {author} {\bibfnamefont {W.}~\bibnamefont
  {Kastaun}},\ }\href {\doibase 10.1103/PhysRevD.77.124019} {\bibfield
  {journal} {\bibinfo  {journal} {Phys. Rev.}\ }\textbf {\bibinfo {volume}
  {D77}},\ \bibinfo {pages} {124019} (\bibinfo {year} {2008})},\ \Eprint
  {http://arxiv.org/abs/0804.1151} {arXiv:0804.1151 [astro-ph]} \BibitemShut
  {NoStop}%
%%CITATION = ARXIV:0804.1151;%%
\bibitem [{\citenamefont {{Westernacher-Schneider}}\ \emph
  {et~al.}(2020)\citenamefont {{Westernacher-Schneider}}, \citenamefont
  {Markakis},\ and\ \citenamefont
  {Tsao}}]{westernacher-schneiderHamiltonJacobiHydrodynamics2020}%
  \BibitemOpen
  \bibfield  {author} {\bibinfo {author} {\bibfnamefont {J.~R.}\ \bibnamefont
  {{Westernacher-Schneider}}}, \bibinfo {author} {\bibfnamefont
  {C.}~\bibnamefont {Markakis}}, \ and\ \bibinfo {author} {\bibfnamefont
  {B.~J.}\ \bibnamefont {Tsao}},\ }\href {\doibase 10.1088/1361-6382/ab93e9}
  {\bibfield  {journal} {\bibinfo  {journal} {Classical and Quantum Gravity}\
  }\textbf {\bibinfo {volume} {37}},\ \bibinfo {pages} {155005} (\bibinfo
  {year} {2020})}\BibitemShut {NoStop}%
\bibitem [{\citenamefont {Christodoulou}\ and\ \citenamefont
  {Miao}(2014)}]{demetrios_christodoulou_compressible_2014}%
  \BibitemOpen
  \bibfield  {author} {\bibinfo {author} {\bibfnamefont {D.}~\bibnamefont
  {Christodoulou}}\ and\ \bibinfo {author} {\bibfnamefont {S.}~\bibnamefont
  {Miao}},\ }\href@noop {} {\emph {\bibinfo {title} {Compressible {Flow} and
  {Euler}'s {Equations}}}},\ \bibinfo {series} {Surveys of {Modern}
  {Mathematics}}, Vol.~\bibinfo {volume} {9}\ (\bibinfo  {publisher}
  {International Press of Boston, Inc},\ \bibinfo {address} {Boston},\ \bibinfo
  {year} {2014})\BibitemShut {NoStop}%
\bibitem [{\citenamefont
  {Markakis}(2014)}]{markakisHamiltonianHydrodynamicsIrrotational2014}%
  \BibitemOpen
  \bibfield  {author} {\bibinfo {author} {\bibfnamefont {C.}~\bibnamefont
  {Markakis}},\ }\href@noop {} {\  (\bibinfo {year} {2014})},\ \Eprint
  {http://arxiv.org/abs/1410.7777} {arXiv:1410.7777} \BibitemShut {NoStop}%
\bibitem [{\citenamefont {Lindblom}(2010)}]{Lindblom:2010bb}%
  \BibitemOpen
  \bibfield  {author} {\bibinfo {author} {\bibfnamefont {L.}~\bibnamefont
  {Lindblom}},\ }\href {\doibase 10.1103/PhysRevD.82.103011} {\bibfield
  {journal} {\bibinfo  {journal} {Phys. Rev.}\ }\textbf {\bibinfo {volume}
  {D82}},\ \bibinfo {pages} {103011} (\bibinfo {year} {2010})},\ \Eprint
  {http://arxiv.org/abs/1009.0738} {arXiv:1009.0738 [astro-ph.HE]} \BibitemShut
  {NoStop}%
%%CITATION = ARXIV:1009.0738;%%
\bibitem [{\citenamefont {Abbott}\ \emph
  {et~al.}(2018{\natexlab{b}})\citenamefont {Abbott} \emph
  {et~al.}}]{Abbott:2018exr}%
  \BibitemOpen
  \bibfield  {author} {\bibinfo {author} {\bibfnamefont {B.~P.}\ \bibnamefont
  {Abbott}} \emph {et~al.} (\bibinfo {collaboration} {LIGO Scientific,
  Virgo}),\ }\href {\doibase 10.1103/PhysRevLett.121.161101} {\bibfield
  {journal} {\bibinfo  {journal} {Phys. Rev. Lett.}\ }\textbf {\bibinfo
  {volume} {121}},\ \bibinfo {pages} {161101} (\bibinfo {year}
  {2018}{\natexlab{b}})},\ \Eprint {http://arxiv.org/abs/1805.11581}
  {arXiv:1805.11581 [gr-qc]} \BibitemShut {NoStop}%
%%CITATION = ARXIV:1805.11581;%%
\bibitem [{\citenamefont {Foucart}\ \emph {et~al.}(2019)\citenamefont
  {Foucart}, \citenamefont {Duez}, \citenamefont {Gudinas}, \citenamefont
  {Hebert}, \citenamefont {Kidder}, \citenamefont {Pfeiffer},\ and\
  \citenamefont {Scheel}}]{Foucart:2019yzo}%
  \BibitemOpen
  \bibfield  {author} {\bibinfo {author} {\bibfnamefont {F.}~\bibnamefont
  {Foucart}}, \bibinfo {author} {\bibfnamefont {M.~D.}\ \bibnamefont {Duez}},
  \bibinfo {author} {\bibfnamefont {A.}~\bibnamefont {Gudinas}}, \bibinfo
  {author} {\bibfnamefont {F.}~\bibnamefont {Hebert}}, \bibinfo {author}
  {\bibfnamefont {L.~E.}\ \bibnamefont {Kidder}}, \bibinfo {author}
  {\bibfnamefont {H.~P.}\ \bibnamefont {Pfeiffer}}, \ and\ \bibinfo {author}
  {\bibfnamefont {M.~A.}\ \bibnamefont {Scheel}},\ }\href@noop {} {\  (\bibinfo
  {year} {2019})},\ \Eprint {http://arxiv.org/abs/1908.05277} {arXiv:1908.05277
  [gr-qc]} \BibitemShut {NoStop}%
%%CITATION = ARXIV:1908.05277;%%
\bibitem [{\citenamefont {Shibata}(1998)}]{Shibata:1998um}%
  \BibitemOpen
  \bibfield  {author} {\bibinfo {author} {\bibfnamefont {M.}~\bibnamefont
  {Shibata}},\ }\href {\doibase 10.1103/PhysRevD.58.024012} {\bibfield
  {journal} {\bibinfo  {journal} {Phys. Rev.}\ }\textbf {\bibinfo {volume}
  {D58}},\ \bibinfo {pages} {024012} (\bibinfo {year} {1998})},\ \Eprint
  {http://arxiv.org/abs/gr-qc/9803085} {arXiv:gr-qc/9803085 [gr-qc]}
  \BibitemShut {NoStop}%
%%CITATION = GR-QC/9803085;%%
\bibitem [{\citenamefont {Gourgoulhon}\ \emph {et~al.}(2001)\citenamefont
  {Gourgoulhon}, \citenamefont {Grandclement}, \citenamefont {Taniguchi},
  \citenamefont {Marck},\ and\ \citenamefont {Bonazzola}}]{Gourgoulhon:2000nn}%
  \BibitemOpen
  \bibfield  {author} {\bibinfo {author} {\bibfnamefont {E.}~\bibnamefont
  {Gourgoulhon}}, \bibinfo {author} {\bibfnamefont {P.}~\bibnamefont
  {Grandclement}}, \bibinfo {author} {\bibfnamefont {K.}~\bibnamefont
  {Taniguchi}}, \bibinfo {author} {\bibfnamefont {J.-A.}\ \bibnamefont
  {Marck}}, \ and\ \bibinfo {author} {\bibfnamefont {S.}~\bibnamefont
  {Bonazzola}},\ }\href {\doibase 10.1103/PhysRevD.63.064029} {\bibfield
  {journal} {\bibinfo  {journal} {Phys. Rev.}\ }\textbf {\bibinfo {volume}
  {D63}},\ \bibinfo {pages} {064029} (\bibinfo {year} {2001})},\ \Eprint
  {http://arxiv.org/abs/gr-qc/0007028} {arXiv:gr-qc/0007028 [gr-qc]}
  \BibitemShut {NoStop}%
%%CITATION = GR-QC/0007028;%%
\bibitem [{\citenamefont {Taniguchi}\ \emph {et~al.}(2001)\citenamefont
  {Taniguchi}, \citenamefont {Gourgoulhon},\ and\ \citenamefont
  {Bonazzola}}]{Taniguchi:2001qv}%
  \BibitemOpen
  \bibfield  {author} {\bibinfo {author} {\bibfnamefont {K.}~\bibnamefont
  {Taniguchi}}, \bibinfo {author} {\bibfnamefont {E.}~\bibnamefont
  {Gourgoulhon}}, \ and\ \bibinfo {author} {\bibfnamefont {S.}~\bibnamefont
  {Bonazzola}},\ }\href {\doibase 10.1103/PhysRevD.64.064012} {\bibfield
  {journal} {\bibinfo  {journal} {Phys. Rev.}\ }\textbf {\bibinfo {volume}
  {D64}},\ \bibinfo {pages} {064012} (\bibinfo {year} {2001})},\ \Eprint
  {http://arxiv.org/abs/gr-qc/0103041} {arXiv:gr-qc/0103041 [gr-qc]}
  \BibitemShut {NoStop}%
%%CITATION = GR-QC/0103041;%%
\bibitem [{\citenamefont {Visser}(1998)}]{Visser:1997ux}%
  \BibitemOpen
  \bibfield  {author} {\bibinfo {author} {\bibfnamefont {M.}~\bibnamefont
  {Visser}},\ }\href {\doibase 10.1088/0264-9381/15/6/024} {\bibfield
  {journal} {\bibinfo  {journal} {Class. Quant. Grav.}\ }\textbf {\bibinfo
  {volume} {15}},\ \bibinfo {pages} {1767} (\bibinfo {year} {1998})},\ \Eprint
  {http://arxiv.org/abs/gr-qc/9712010} {arXiv:gr-qc/9712010 [gr-qc]}
  \BibitemShut {NoStop}%
%%CITATION = GR-QC/9712010;%%
\bibitem [{\citenamefont {LeVeque}(2002)}]{levequeFiniteVolumeMethods2002}%
  \BibitemOpen
  \bibfield  {author} {\bibinfo {author} {\bibfnamefont {R.~J.}\ \bibnamefont
  {LeVeque}},\ }\href {\doibase 10.1017/CBO9780511791253} {\emph {\bibinfo
  {title} {Finite {{Volume Methods}} for {{Hyperbolic Problems}}}}},\ Cambridge
  {{Texts}} in {{Applied Mathematics}}\ (\bibinfo  {publisher} {{Cambridge
  University Press}},\ \bibinfo {address} {{Cambridge}},\ \bibinfo {year}
  {2002})\BibitemShut {NoStop}%
\bibitem [{\citenamefont {Voss}(2005)}]{vossExactRiemannSolution2005}%
  \BibitemOpen
  \bibfield  {author} {\bibinfo {author} {\bibfnamefont {A.}~\bibnamefont
  {Voss}},\ }\href@noop {} {\emph {\bibinfo {title} {Exact {{Riemann}} Solution
  for the {{Euler}} Equations with Nonconvex and Nonsmooth Equation of State,
  PhD Dissertation}}}\ (\bibinfo {address} {{Aachen}},\ \bibinfo {year}
  {2005})\BibitemShut {NoStop}%
\bibitem [{\citenamefont {Paschalidis}\ \emph {et~al.}(2011)\citenamefont
  {Paschalidis}, \citenamefont {Etienne}, \citenamefont {Liu},\ and\
  \citenamefont {Shapiro}}]{PhysRevD.83.064002}%
  \BibitemOpen
  \bibfield  {author} {\bibinfo {author} {\bibfnamefont {V.}~\bibnamefont
  {Paschalidis}}, \bibinfo {author} {\bibfnamefont {Z.}~\bibnamefont
  {Etienne}}, \bibinfo {author} {\bibfnamefont {Y.~T.}\ \bibnamefont {Liu}}, \
  and\ \bibinfo {author} {\bibfnamefont {S.~L.}\ \bibnamefont {Shapiro}},\
  }\href {\doibase 10.1103/PhysRevD.83.064002} {\bibfield  {journal} {\bibinfo
  {journal} {Phys. Rev. D}\ }\textbf {\bibinfo {volume} {83}},\ \bibinfo
  {pages} {064002} (\bibinfo {year} {2011})}\BibitemShut {NoStop}%
\bibitem [{\citenamefont {Annala}\ \emph {et~al.}(2020)\citenamefont {Annala},
  \citenamefont {Gorda}, \citenamefont {Kurkela}, \citenamefont
  {N{\"a}ttil{\"a}},\ and\ \citenamefont
  {Vuorinen}}]{annalaEvidenceQuarkmatterCores2020}%
  \BibitemOpen
  \bibfield  {author} {\bibinfo {author} {\bibfnamefont {E.}~\bibnamefont
  {Annala}}, \bibinfo {author} {\bibfnamefont {T.}~\bibnamefont {Gorda}},
  \bibinfo {author} {\bibfnamefont {A.}~\bibnamefont {Kurkela}}, \bibinfo
  {author} {\bibfnamefont {J.}~\bibnamefont {N{\"a}ttil{\"a}}}, \ and\ \bibinfo
  {author} {\bibfnamefont {A.}~\bibnamefont {Vuorinen}},\ }\href {\doibase
  10.1038/s41567-020-0914-9} {\bibfield  {journal} {\bibinfo  {journal} {Nature
  Physics}\ ,\ \bibinfo {pages} {1}} (\bibinfo {year} {2020})}\BibitemShut
  {NoStop}%
\bibitem [{\citenamefont {Fasano}\ \emph {et~al.}(2019)\citenamefont {Fasano},
  \citenamefont {Abdelsalhin}, \citenamefont {Maselli},\ and\ \citenamefont
  {Ferrari}}]{Fasano:2019zwm}%
  \BibitemOpen
  \bibfield  {author} {\bibinfo {author} {\bibfnamefont {M.}~\bibnamefont
  {Fasano}}, \bibinfo {author} {\bibfnamefont {T.}~\bibnamefont {Abdelsalhin}},
  \bibinfo {author} {\bibfnamefont {A.}~\bibnamefont {Maselli}}, \ and\
  \bibinfo {author} {\bibfnamefont {V.}~\bibnamefont {Ferrari}},\ }\href@noop
  {} {\  (\bibinfo {year} {2019})},\ \Eprint {http://arxiv.org/abs/1902.05078}
  {arXiv:1902.05078 [astro-ph.HE]} \BibitemShut {NoStop}%
%%CITATION = ARXIV:1902.05078;%%
\bibitem [{\citenamefont {Gamba}\ \emph
  {et~al.}(2019{\natexlab{a}})\citenamefont {Gamba}, \citenamefont {Read},\
  and\ \citenamefont {Wade}}]{Gamba:2019kwu}%
  \BibitemOpen
  \bibfield  {author} {\bibinfo {author} {\bibfnamefont {R.}~\bibnamefont
  {Gamba}}, \bibinfo {author} {\bibfnamefont {J.~S.}\ \bibnamefont {Read}}, \
  and\ \bibinfo {author} {\bibfnamefont {L.~E.}\ \bibnamefont {Wade}},\
  }\href@noop {} {\  (\bibinfo {year} {2019}{\natexlab{a}})},\ \Eprint
  {http://arxiv.org/abs/1902.04616} {arXiv:1902.04616 [gr-qc]} \BibitemShut
  {NoStop}%
%%CITATION = ARXIV:1902.04616;%%
\bibitem [{\citenamefont {{{\"O}zel}}\ and\ \citenamefont
  {{Psaltis}}(2009)}]{Ozel:2018}%
  \BibitemOpen
  \bibfield  {author} {\bibinfo {author} {\bibfnamefont {F.}~\bibnamefont
  {{{\"O}zel}}}\ and\ \bibinfo {author} {\bibfnamefont {D.}~\bibnamefont
  {{Psaltis}}},\ }\href {\doibase 10.1103/PhysRevD.80.103003} {\bibfield
  {journal} {\bibinfo  {journal} {\prd}\ }\textbf {\bibinfo {volume} {80}},\
  \bibinfo {eid} {103003} (\bibinfo {year} {2009})},\ \Eprint
  {http://arxiv.org/abs/0905.1959} {arXiv:0905.1959 [astro-ph.HE]} \BibitemShut
  {NoStop}%
\bibitem [{\citenamefont {Lattimer}\ and\ \citenamefont
  {Prakash}(2016)}]{Lattimer:2015nhk}%
  \BibitemOpen
  \bibfield  {author} {\bibinfo {author} {\bibfnamefont {J.~M.}\ \bibnamefont
  {Lattimer}}\ and\ \bibinfo {author} {\bibfnamefont {M.}~\bibnamefont
  {Prakash}},\ }\href {\doibase 10.1016/j.physrep.2015.12.005} {\bibfield
  {journal} {\bibinfo  {journal} {Phys. Rept.}\ }\textbf {\bibinfo {volume}
  {621}},\ \bibinfo {pages} {127} (\bibinfo {year} {2016})},\ \Eprint
  {http://arxiv.org/abs/1512.07820} {arXiv:1512.07820 [astro-ph.SR]}
  \BibitemShut {NoStop}%
\bibitem [{\citenamefont {Lattimer}(2017)}]{Lattimer:2017}%
  \BibitemOpen
  \bibfield  {author} {\bibinfo {author} {\bibfnamefont {J.~M.}\ \bibnamefont
  {Lattimer}},\ }\href {\doibase 10.1142/S0218301317400146} {\bibfield
  {journal} {\bibinfo  {journal} {International Journal of Modern Physics E}\
  }\textbf {\bibinfo {volume} {26}},\ \bibinfo {pages} {1740014} (\bibinfo
  {year} {2017})},\ \Eprint
  {http://arxiv.org/abs/https://doi.org/10.1142/S0218301317400146}
  {https://doi.org/10.1142/S0218301317400146} \BibitemShut {NoStop}%
\bibitem [{\citenamefont {De}\ \emph {et~al.}(2018)\citenamefont {De},
  \citenamefont {Finstad}, \citenamefont {Lattimer}, \citenamefont {Brown},
  \citenamefont {Berger},\ and\ \citenamefont {Biwer}}]{De:2018uhw}%
  \BibitemOpen
  \bibfield  {author} {\bibinfo {author} {\bibfnamefont {S.}~\bibnamefont
  {De}}, \bibinfo {author} {\bibfnamefont {D.}~\bibnamefont {Finstad}},
  \bibinfo {author} {\bibfnamefont {J.~M.}\ \bibnamefont {Lattimer}}, \bibinfo
  {author} {\bibfnamefont {D.~A.}\ \bibnamefont {Brown}}, \bibinfo {author}
  {\bibfnamefont {E.}~\bibnamefont {Berger}}, \ and\ \bibinfo {author}
  {\bibfnamefont {C.~M.}\ \bibnamefont {Biwer}},\ }\href {\doibase
  10.1103/PhysRevLett.121.091102} {\bibfield  {journal} {\bibinfo  {journal}
  {Phys. Rev. Lett.}\ }\textbf {\bibinfo {volume} {121}},\ \bibinfo {pages}
  {091102} (\bibinfo {year} {2018})},\ \bibinfo {note} {[Erratum:
  Phys.Rev.Lett. 121, 259902 (2018)]},\ \Eprint
  {http://arxiv.org/abs/1804.08583} {arXiv:1804.08583 [astro-ph.HE]}
  \BibitemShut {NoStop}%
\bibitem [{\citenamefont {Carney}\ \emph {et~al.}(2018)\citenamefont {Carney},
  \citenamefont {Wade},\ and\ \citenamefont {Irwin}}]{Carney:2018sdv}%
  \BibitemOpen
  \bibfield  {author} {\bibinfo {author} {\bibfnamefont {M.~F.}\ \bibnamefont
  {Carney}}, \bibinfo {author} {\bibfnamefont {L.~E.}\ \bibnamefont {Wade}}, \
  and\ \bibinfo {author} {\bibfnamefont {B.~S.}\ \bibnamefont {Irwin}},\ }\href
  {\doibase 10.1103/PhysRevD.98.063004} {\bibfield  {journal} {\bibinfo
  {journal} {Phys. Rev. D}\ }\textbf {\bibinfo {volume} {98}},\ \bibinfo
  {pages} {063004} (\bibinfo {year} {2018})},\ \Eprint
  {http://arxiv.org/abs/1805.11217} {arXiv:1805.11217 [gr-qc]} \BibitemShut
  {NoStop}%
\bibitem [{\citenamefont {Burrows}(2013)}]{Burrows:2012ew}%
  \BibitemOpen
  \bibfield  {author} {\bibinfo {author} {\bibfnamefont {A.}~\bibnamefont
  {Burrows}},\ }\href {\doibase 10.1103/RevModPhys.85.245} {\bibfield
  {journal} {\bibinfo  {journal} {Rev. Mod. Phys.}\ }\textbf {\bibinfo {volume}
  {85}},\ \bibinfo {pages} {245} (\bibinfo {year} {2013})},\ \Eprint
  {http://arxiv.org/abs/1210.4921} {arXiv:1210.4921 [astro-ph.SR]} \BibitemShut
  {NoStop}%
%%CITATION = ARXIV:1210.4921;%%
\bibitem [{\citenamefont {Janka}\ \emph {et~al.}(2016)\citenamefont {Janka},
  \citenamefont {Melson},\ and\ \citenamefont {Summa}}]{Janka:2016fox}%
  \BibitemOpen
  \bibfield  {author} {\bibinfo {author} {\bibfnamefont {H.~T.}\ \bibnamefont
  {Janka}}, \bibinfo {author} {\bibfnamefont {T.}~\bibnamefont {Melson}}, \
  and\ \bibinfo {author} {\bibfnamefont {A.}~\bibnamefont {Summa}},\ }\href
  {\doibase 10.1146/annurev-nucl-102115-044747} {\bibfield  {journal} {\bibinfo
   {journal} {Ann. Rev. Nucl. Part. Sci.}\ }\textbf {\bibinfo {volume} {66}},\
  \bibinfo {pages} {341} (\bibinfo {year} {2016})},\ \Eprint
  {http://arxiv.org/abs/1602.05576} {arXiv:1602.05576 [astro-ph.SR]}
  \BibitemShut {NoStop}%
%%CITATION = ARXIV:1602.05576;%%
\bibitem [{\citenamefont {Suwa}\ \emph {et~al.}(2018)\citenamefont {Suwa},
  \citenamefont {Yoshida}, \citenamefont {Shibata}, \citenamefont {Umeda},\
  and\ \citenamefont {Takahashi}}]{Suwa:2018uni}%
  \BibitemOpen
  \bibfield  {author} {\bibinfo {author} {\bibfnamefont {Y.}~\bibnamefont
  {Suwa}}, \bibinfo {author} {\bibfnamefont {T.}~\bibnamefont {Yoshida}},
  \bibinfo {author} {\bibfnamefont {M.}~\bibnamefont {Shibata}}, \bibinfo
  {author} {\bibfnamefont {H.}~\bibnamefont {Umeda}}, \ and\ \bibinfo {author}
  {\bibfnamefont {K.}~\bibnamefont {Takahashi}},\ }\href {\doibase
  10.1093/mnras/sty2460} {\bibfield  {journal} {\bibinfo  {journal} {Mon. Not.
  Roy. Astron. Soc.}\ }\textbf {\bibinfo {volume} {481}},\ \bibinfo {pages}
  {3305} (\bibinfo {year} {2018})},\ \Eprint {http://arxiv.org/abs/1808.02328}
  {arXiv:1808.02328 [astro-ph.HE]} \BibitemShut {NoStop}%
%%CITATION = ARXIV:1808.02328;%%
\bibitem [{\citenamefont {Botev}\ \emph {et~al.}(2016)\citenamefont {Botev},
  \citenamefont {Lever},\ and\ \citenamefont {Barber}}]{aleks2016nesterovs}%
  \BibitemOpen
  \bibfield  {author} {\bibinfo {author} {\bibfnamefont {A.}~\bibnamefont
  {Botev}}, \bibinfo {author} {\bibfnamefont {G.}~\bibnamefont {Lever}}, \ and\
  \bibinfo {author} {\bibfnamefont {D.}~\bibnamefont {Barber}},\ }\href@noop {}
  {\enquote {\bibinfo {title} {Nesterov's accelerated gradient and momentum as
  approximations to regularised update descent},}\ } (\bibinfo {year} {2016}),\
  \Eprint {http://arxiv.org/abs/1607.01981} {arXiv:1607.01981 [stat.ML]}
  \BibitemShut {NoStop}%
\bibitem [{\citenamefont {Goodfellow}\ \emph {et~al.}(2016)\citenamefont
  {Goodfellow}, \citenamefont {Bengio},\ and\ \citenamefont
  {Courville}}]{goodfellow_deep_2016}%
  \BibitemOpen
  \bibfield  {author} {\bibinfo {author} {\bibfnamefont {I.}~\bibnamefont
  {Goodfellow}}, \bibinfo {author} {\bibfnamefont {Y.}~\bibnamefont {Bengio}},
  \ and\ \bibinfo {author} {\bibfnamefont {A.}~\bibnamefont {Courville}},\
  }\href {http://www.deeplearningbook.org} {\emph {\bibinfo {title} {Deep
  {Learning}}}}\ (\bibinfo  {publisher} {MIT Press},\ \bibinfo {year}
  {2016})\BibitemShut {NoStop}%
\bibitem [{\citenamefont {Typel}\ \emph {et~al.}(2020)\citenamefont {Typel},
  \citenamefont {Oertel},\ and\ \citenamefont {Klaehn}}]{typelEoSCatalog}%
  \BibitemOpen
  \bibfield  {author} {\bibinfo {author} {\bibfnamefont {S.}~\bibnamefont
  {Typel}}, \bibinfo {author} {\bibfnamefont {M.}~\bibnamefont {Oertel}}, \
  and\ \bibinfo {author} {\bibfnamefont {T.}~\bibnamefont {Klaehn}},\
  }\href@noop {} {\emph {\bibinfo {title} {EoS Catalog}}} (\bibinfo {year}
  {accessed August 3, 2020}),\ \bibinfo {note}
  {\url{https://compose.obspm.fr/}}\BibitemShut {NoStop}%
\bibitem [{\citenamefont {Ozel}(2020)}]{ozelEoSCatalog}%
  \BibitemOpen
  \bibfield  {author} {\bibinfo {author} {\bibfnamefont {F.}~\bibnamefont
  {Ozel}},\ }\href@noop {} {\emph {\bibinfo {title} {EoS Catalog}}} (\bibinfo
  {year} {accessed August 3, 2020}),\ \bibinfo {note}
  {\url{http://xtreme.as.arizona.edu/NeutronStars/data/eos\_tables.tar}}\BibitemShut
  {NoStop}%
\bibitem [{\citenamefont {Kyutoku}\ \emph {et~al.}(2010)\citenamefont
  {Kyutoku}, \citenamefont {Shibata},\ and\ \citenamefont
  {Taniguchi}}]{Kyutoku:2010zd}%
  \BibitemOpen
  \bibfield  {author} {\bibinfo {author} {\bibfnamefont {K.}~\bibnamefont
  {Kyutoku}}, \bibinfo {author} {\bibfnamefont {M.}~\bibnamefont {Shibata}}, \
  and\ \bibinfo {author} {\bibfnamefont {K.}~\bibnamefont {Taniguchi}},\ }\href
  {\doibase 10.1103/PhysRevD.82.044049, 10.1103/PhysRevD.84.049902} {\bibfield
  {journal} {\bibinfo  {journal} {Phys. Rev.}\ }\textbf {\bibinfo {volume}
  {D82}},\ \bibinfo {pages} {044049} (\bibinfo {year} {2010})},\ \bibinfo
  {note} {[Erratum: Phys. Rev.D84,049902(2011)]},\ \Eprint
  {http://arxiv.org/abs/1008.1460} {arXiv:1008.1460 [astro-ph.HE]} \BibitemShut
  {NoStop}%
%%CITATION = ARXIV:1008.1460;%%
\bibitem [{\citenamefont {Kyutoku}\ \emph {et~al.}(2011)\citenamefont
  {Kyutoku}, \citenamefont {Okawa}, \citenamefont {Shibata},\ and\
  \citenamefont {Taniguchi}}]{Kyutoku:2011vz}%
  \BibitemOpen
  \bibfield  {author} {\bibinfo {author} {\bibfnamefont {K.}~\bibnamefont
  {Kyutoku}}, \bibinfo {author} {\bibfnamefont {H.}~\bibnamefont {Okawa}},
  \bibinfo {author} {\bibfnamefont {M.}~\bibnamefont {Shibata}}, \ and\
  \bibinfo {author} {\bibfnamefont {K.}~\bibnamefont {Taniguchi}},\ }\href
  {\doibase 10.1103/PhysRevD.84.064018} {\bibfield  {journal} {\bibinfo
  {journal} {Phys. Rev.}\ }\textbf {\bibinfo {volume} {D84}},\ \bibinfo {pages}
  {064018} (\bibinfo {year} {2011})},\ \Eprint {http://arxiv.org/abs/1108.1189}
  {arXiv:1108.1189 [astro-ph.HE]} \BibitemShut {NoStop}%
%%CITATION = ARXIV:1108.1189;%%
\bibitem [{\citenamefont {Hotokezaka}\ \emph {et~al.}(2013)\citenamefont
  {Hotokezaka}, \citenamefont {Kiuchi}, \citenamefont {Kyutoku}, \citenamefont
  {Okawa}, \citenamefont {Sekiguchi}, \citenamefont {Shibata},\ and\
  \citenamefont {Taniguchi}}]{Hotokezaka:2012ze}%
  \BibitemOpen
  \bibfield  {author} {\bibinfo {author} {\bibfnamefont {K.}~\bibnamefont
  {Hotokezaka}}, \bibinfo {author} {\bibfnamefont {K.}~\bibnamefont {Kiuchi}},
  \bibinfo {author} {\bibfnamefont {K.}~\bibnamefont {Kyutoku}}, \bibinfo
  {author} {\bibfnamefont {H.}~\bibnamefont {Okawa}}, \bibinfo {author}
  {\bibfnamefont {Y.-i.}\ \bibnamefont {Sekiguchi}}, \bibinfo {author}
  {\bibfnamefont {M.}~\bibnamefont {Shibata}}, \ and\ \bibinfo {author}
  {\bibfnamefont {K.}~\bibnamefont {Taniguchi}},\ }\href {\doibase
  10.1103/PhysRevD.87.024001} {\bibfield  {journal} {\bibinfo  {journal} {Phys.
  Rev.}\ }\textbf {\bibinfo {volume} {D87}},\ \bibinfo {pages} {024001}
  (\bibinfo {year} {2013})},\ \Eprint {http://arxiv.org/abs/1212.0905}
  {arXiv:1212.0905 [astro-ph.HE]} \BibitemShut {NoStop}%
%%CITATION = ARXIV:1212.0905;%%
\bibitem [{\citenamefont {Siegel}\ and\ \citenamefont
  {Metzger}(2017)}]{Siegel:2017nub}%
  \BibitemOpen
  \bibfield  {author} {\bibinfo {author} {\bibfnamefont {D.~M.}\ \bibnamefont
  {Siegel}}\ and\ \bibinfo {author} {\bibfnamefont {B.~D.}\ \bibnamefont
  {Metzger}},\ }\href {\doibase 10.1103/PhysRevLett.119.231102} {\bibfield
  {journal} {\bibinfo  {journal} {Phys. Rev. Lett.}\ }\textbf {\bibinfo
  {volume} {119}},\ \bibinfo {pages} {231102} (\bibinfo {year} {2017})},\
  \Eprint {http://arxiv.org/abs/1705.05473} {arXiv:1705.05473 [astro-ph.HE]}
  \BibitemShut {NoStop}%
%%CITATION = ARXIV:1705.05473;%%
\bibitem [{\citenamefont {Fern{\'a}ndez}\ \emph {et~al.}(2019)\citenamefont
  {Fern{\'a}ndez}, \citenamefont {Tchekhovskoy}, \citenamefont {Quataert},
  \citenamefont {Foucart},\ and\ \citenamefont {Kasen}}]{Fernandez:2018kax}%
  \BibitemOpen
  \bibfield  {author} {\bibinfo {author} {\bibfnamefont {R.}~\bibnamefont
  {Fern{\'a}ndez}}, \bibinfo {author} {\bibfnamefont {A.}~\bibnamefont
  {Tchekhovskoy}}, \bibinfo {author} {\bibfnamefont {E.}~\bibnamefont
  {Quataert}}, \bibinfo {author} {\bibfnamefont {F.}~\bibnamefont {Foucart}}, \
  and\ \bibinfo {author} {\bibfnamefont {D.}~\bibnamefont {Kasen}},\ }\href
  {\doibase 10.1093/mnras/sty2932} {\bibfield  {journal} {\bibinfo  {journal}
  {Mon. Not. Roy. Astron. Soc.}\ }\textbf {\bibinfo {volume} {482}},\ \bibinfo
  {pages} {3373} (\bibinfo {year} {2019})},\ \Eprint
  {http://arxiv.org/abs/1808.00461} {arXiv:1808.00461 [astro-ph.HE]}
  \BibitemShut {NoStop}%
%%CITATION = ARXIV:1808.00461;%%
\bibitem [{\citenamefont {Shibata}\ and\ \citenamefont
  {Hotokezaka}(2019)}]{Shibata:2019wef}%
  \BibitemOpen
  \bibfield  {author} {\bibinfo {author} {\bibfnamefont {M.}~\bibnamefont
  {Shibata}}\ and\ \bibinfo {author} {\bibfnamefont {K.}~\bibnamefont
  {Hotokezaka}},\ }\href {\doibase 10.1146/annurev-nucl-101918-023625} {\
  (\bibinfo {year} {2019}),\ 10.1146/annurev-nucl-101918-023625},\ \Eprint
  {http://arxiv.org/abs/1908.02350} {arXiv:1908.02350 [astro-ph.HE]}
  \BibitemShut {NoStop}%
%%CITATION = ARXIV:1908.02350;%%
\bibitem [{\citenamefont {Gamba}\ \emph
  {et~al.}(2019{\natexlab{b}})\citenamefont {Gamba}, \citenamefont {Read},\
  and\ \citenamefont {Wade}}]{Gamba_2019}%
  \BibitemOpen
  \bibfield  {author} {\bibinfo {author} {\bibfnamefont {R.}~\bibnamefont
  {Gamba}}, \bibinfo {author} {\bibfnamefont {J.~S.}\ \bibnamefont {Read}}, \
  and\ \bibinfo {author} {\bibfnamefont {L.~E.}\ \bibnamefont {Wade}},\ }\href
  {\doibase 10.1088/1361-6382/ab5ba4} {\bibfield  {journal} {\bibinfo
  {journal} {Classical and Quantum Gravity}\ }\textbf {\bibinfo {volume}
  {37}},\ \bibinfo {pages} {025008} (\bibinfo {year}
  {2019}{\natexlab{b}})}\BibitemShut {NoStop}%
\bibitem [{\citenamefont {Annala}\ \emph
  {et~al.}(2018{\natexlab{b}})\citenamefont {Annala}, \citenamefont {Gorda},
  \citenamefont {Kurkela},\ and\ \citenamefont {Vuorinen}}]{Annala:2017llu}%
  \BibitemOpen
  \bibfield  {author} {\bibinfo {author} {\bibfnamefont {E.}~\bibnamefont
  {Annala}}, \bibinfo {author} {\bibfnamefont {T.}~\bibnamefont {Gorda}},
  \bibinfo {author} {\bibfnamefont {A.}~\bibnamefont {Kurkela}}, \ and\
  \bibinfo {author} {\bibfnamefont {A.}~\bibnamefont {Vuorinen}},\ }\href
  {\doibase 10.1103/PhysRevLett.120.172703} {\bibfield  {journal} {\bibinfo
  {journal} {Phys. Rev. Lett.}\ }\textbf {\bibinfo {volume} {120}},\ \bibinfo
  {pages} {172703} (\bibinfo {year} {2018}{\natexlab{b}})},\ \Eprint
  {http://arxiv.org/abs/1711.02644} {arXiv:1711.02644 [astro-ph.HE]}
  \BibitemShut {NoStop}%
\bibitem [{\citenamefont {Chabanat}\ \emph {et~al.}(1998)\citenamefont
  {Chabanat}, \citenamefont {Bonche}, \citenamefont {Haensel}, \citenamefont
  {Meyer},\ and\ \citenamefont {Schaeffer}}]{Chabanat:1997un}%
  \BibitemOpen
  \bibfield  {author} {\bibinfo {author} {\bibfnamefont {E.}~\bibnamefont
  {Chabanat}}, \bibinfo {author} {\bibfnamefont {P.}~\bibnamefont {Bonche}},
  \bibinfo {author} {\bibfnamefont {P.}~\bibnamefont {Haensel}}, \bibinfo
  {author} {\bibfnamefont {J.}~\bibnamefont {Meyer}}, \ and\ \bibinfo {author}
  {\bibfnamefont {R.}~\bibnamefont {Schaeffer}},\ }\href {\doibase
  10.1016/S0375-9474(98)00570-3, 10.1016/S0375-9474(98)00180-8} {\bibfield
  {journal} {\bibinfo  {journal} {Nucl. Phys.}\ }\textbf {\bibinfo {volume}
  {A635}},\ \bibinfo {pages} {231} (\bibinfo {year} {1998})},\ \bibinfo {note}
  {[Erratum: Nucl. Phys.A643,441(1998)]}\BibitemShut {NoStop}%
%%CITATION = NUPHA,A635,231;%%
\bibitem [{\citenamefont {Danielewicz}\ and\ \citenamefont
  {Lee}(2009)}]{Danielewicz:2008cm}%
  \BibitemOpen
  \bibfield  {author} {\bibinfo {author} {\bibfnamefont {P.}~\bibnamefont
  {Danielewicz}}\ and\ \bibinfo {author} {\bibfnamefont {J.}~\bibnamefont
  {Lee}},\ }\href {\doibase 10.1016/j.nuclphysa.2008.11.007} {\bibfield
  {journal} {\bibinfo  {journal} {Nucl. Phys.}\ }\textbf {\bibinfo {volume}
  {A818}},\ \bibinfo {pages} {36} (\bibinfo {year} {2009})},\ \Eprint
  {http://arxiv.org/abs/0807.3743} {arXiv:0807.3743 [nucl-th]} \BibitemShut
  {NoStop}%
%%CITATION = ARXIV:0807.3743;%%
\bibitem [{\citenamefont {Gulminelli}\ and\ \citenamefont
  {Raduta}(2015)}]{Gulminelli:2015csa}%
  \BibitemOpen
  \bibfield  {author} {\bibinfo {author} {\bibfnamefont {F.}~\bibnamefont
  {Gulminelli}}\ and\ \bibinfo {author} {\bibfnamefont {A.~R.}\ \bibnamefont
  {Raduta}},\ }\href {\doibase 10.1103/PhysRevC.92.055803} {\bibfield
  {journal} {\bibinfo  {journal} {Phys. Rev.}\ }\textbf {\bibinfo {volume}
  {C92}},\ \bibinfo {pages} {055803} (\bibinfo {year} {2015})},\ \Eprint
  {http://arxiv.org/abs/1504.04493} {arXiv:1504.04493 [nucl-th]} \BibitemShut
  {NoStop}%
%%CITATION = ARXIV:1504.04493;%%
\bibitem [{\citenamefont {Togashi}\ \emph {et~al.}(2017)\citenamefont
  {Togashi}, \citenamefont {Nakazato}, \citenamefont {Takehara}, \citenamefont
  {Yamamuro}, \citenamefont {Suzuki},\ and\ \citenamefont
  {Takano}}]{Togashi:2017mjp}%
  \BibitemOpen
  \bibfield  {author} {\bibinfo {author} {\bibfnamefont {H.}~\bibnamefont
  {Togashi}}, \bibinfo {author} {\bibfnamefont {K.}~\bibnamefont {Nakazato}},
  \bibinfo {author} {\bibfnamefont {Y.}~\bibnamefont {Takehara}}, \bibinfo
  {author} {\bibfnamefont {S.}~\bibnamefont {Yamamuro}}, \bibinfo {author}
  {\bibfnamefont {H.}~\bibnamefont {Suzuki}}, \ and\ \bibinfo {author}
  {\bibfnamefont {M.}~\bibnamefont {Takano}},\ }\href {\doibase
  10.1016/j.nuclphysa.2017.02.010} {\bibfield  {journal} {\bibinfo  {journal}
  {Nucl. Phys.}\ }\textbf {\bibinfo {volume} {A961}},\ \bibinfo {pages} {78}
  (\bibinfo {year} {2017})},\ \Eprint {http://arxiv.org/abs/1702.05324}
  {arXiv:1702.05324 [nucl-th]} \BibitemShut {NoStop}%
%%CITATION = ARXIV:1702.05324;%%
\bibitem [{\citenamefont {Baym}\ \emph {et~al.}(2019)\citenamefont {Baym},
  \citenamefont {Furusawa}, \citenamefont {Hatsuda}, \citenamefont {Kojo},\
  and\ \citenamefont {Togashi}}]{Baym:2019iky}%
  \BibitemOpen
  \bibfield  {author} {\bibinfo {author} {\bibfnamefont {G.}~\bibnamefont
  {Baym}}, \bibinfo {author} {\bibfnamefont {S.}~\bibnamefont {Furusawa}},
  \bibinfo {author} {\bibfnamefont {T.}~\bibnamefont {Hatsuda}}, \bibinfo
  {author} {\bibfnamefont {T.}~\bibnamefont {Kojo}}, \ and\ \bibinfo {author}
  {\bibfnamefont {H.}~\bibnamefont {Togashi}},\ }\href@noop {} {\  (\bibinfo
  {year} {2019})},\ \Eprint {http://arxiv.org/abs/1903.08963} {arXiv:1903.08963
  [astro-ph.HE]} \BibitemShut {NoStop}%
%%CITATION = ARXIV:1903.08963;%%
\bibitem [{\citenamefont {Mueller}\ and\ \citenamefont
  {Serot}(1996)}]{Mueller:1996pm}%
  \BibitemOpen
  \bibfield  {author} {\bibinfo {author} {\bibfnamefont {H.}~\bibnamefont
  {Mueller}}\ and\ \bibinfo {author} {\bibfnamefont {B.~D.}\ \bibnamefont
  {Serot}},\ }\href {\doibase 10.1016/0375-9474(96)00187-X} {\bibfield
  {journal} {\bibinfo  {journal} {Nucl. Phys.}\ }\textbf {\bibinfo {volume}
  {A606}},\ \bibinfo {pages} {508} (\bibinfo {year} {1996})},\ \Eprint
  {http://arxiv.org/abs/nucl-th/9603037} {arXiv:nucl-th/9603037 [nucl-th]}
  \BibitemShut {NoStop}%
%%CITATION = NUCL-TH/9603037;%%
\bibitem [{\citenamefont {Gondek}\ \emph {et~al.}(1997)\citenamefont {Gondek},
  \citenamefont {Haensel},\ and\ \citenamefont {Zdunik}}]{Gondek}%
  \BibitemOpen
  \bibfield  {author} {\bibinfo {author} {\bibfnamefont {D.}~\bibnamefont
  {Gondek}}, \bibinfo {author} {\bibfnamefont {P.}~\bibnamefont {Haensel}}, \
  and\ \bibinfo {author} {\bibfnamefont {J.~L.}\ \bibnamefont {Zdunik}},\
  }\href@noop {} {\bibfield  {journal} {\bibinfo  {journal} {Astron.
  Astrophys.}\ }\textbf {\bibinfo {volume} {325}},\ \bibinfo {pages} {217}
  (\bibinfo {year} {1997})}\BibitemShut {NoStop}%
\bibitem [{\citenamefont {Lindblom}\ and\ \citenamefont
  {Indik}(2014)}]{PhysRevD.89.064003}%
  \BibitemOpen
  \bibfield  {author} {\bibinfo {author} {\bibfnamefont {L.}~\bibnamefont
  {Lindblom}}\ and\ \bibinfo {author} {\bibfnamefont {N.~M.}\ \bibnamefont
  {Indik}},\ }\href {\doibase 10.1103/PhysRevD.89.064003} {\bibfield  {journal}
  {\bibinfo  {journal} {Phys. Rev. D}\ }\textbf {\bibinfo {volume} {89}},\
  \bibinfo {pages} {064003} (\bibinfo {year} {2014})}\BibitemShut {NoStop}%
\bibitem [{\citenamefont {Landry}\ \emph {et~al.}(2020)\citenamefont {Landry},
  \citenamefont {Essick},\ and\ \citenamefont
  {Chatziioannou}}]{PhysRevD.101.123007}%
  \BibitemOpen
  \bibfield  {author} {\bibinfo {author} {\bibfnamefont {P.}~\bibnamefont
  {Landry}}, \bibinfo {author} {\bibfnamefont {R.}~\bibnamefont {Essick}}, \
  and\ \bibinfo {author} {\bibfnamefont {K.}~\bibnamefont {Chatziioannou}},\
  }\href {\doibase 10.1103/PhysRevD.101.123007} {\bibfield  {journal} {\bibinfo
   {journal} {Phys. Rev. D}\ }\textbf {\bibinfo {volume} {101}},\ \bibinfo
  {pages} {123007} (\bibinfo {year} {2020})}\BibitemShut {NoStop}%
\bibitem [{\citenamefont {Orsaria}\ \emph {et~al.}(2019)\citenamefont
  {Orsaria}, \citenamefont {Malfatti}, \citenamefont {Mariani}, \citenamefont
  {Ranea-Sandoval}, \citenamefont {Garc{\'{\i}}a}, \citenamefont {Spinella},
  \citenamefont {Contrera}, \citenamefont {Lugones},\ and\ \citenamefont
  {Weber}}]{Orsaria_2019}%
  \BibitemOpen
  \bibfield  {author} {\bibinfo {author} {\bibfnamefont {M.~G.}\ \bibnamefont
  {Orsaria}}, \bibinfo {author} {\bibfnamefont {G.}~\bibnamefont {Malfatti}},
  \bibinfo {author} {\bibfnamefont {M.}~\bibnamefont {Mariani}}, \bibinfo
  {author} {\bibfnamefont {I.~F.}\ \bibnamefont {Ranea-Sandoval}}, \bibinfo
  {author} {\bibfnamefont {F.}~\bibnamefont {Garc{\'{\i}}a}}, \bibinfo {author}
  {\bibfnamefont {W.~M.}\ \bibnamefont {Spinella}}, \bibinfo {author}
  {\bibfnamefont {G.~A.}\ \bibnamefont {Contrera}}, \bibinfo {author}
  {\bibfnamefont {G.}~\bibnamefont {Lugones}}, \ and\ \bibinfo {author}
  {\bibfnamefont {F.}~\bibnamefont {Weber}},\ }\href {\doibase
  10.1088/1361-6471/ab1d81} {\bibfield  {journal} {\bibinfo  {journal} {Journal
  of Physics G: Nuclear and Particle Physics}\ }\textbf {\bibinfo {volume}
  {46}},\ \bibinfo {pages} {073002} (\bibinfo {year} {2019})}\BibitemShut
  {NoStop}%
\bibitem [{\citenamefont {Most}\ \emph {et~al.}(2019)\citenamefont {Most},
  \citenamefont {Papenfort}, \citenamefont {Dexheimer}, \citenamefont
  {Hanauske}, \citenamefont {Schramm}, \citenamefont {St\"ocker},\ and\
  \citenamefont {Rezzolla}}]{PhysRevLett.122.061101}%
  \BibitemOpen
  \bibfield  {author} {\bibinfo {author} {\bibfnamefont {E.~R.}\ \bibnamefont
  {Most}}, \bibinfo {author} {\bibfnamefont {L.~J.}\ \bibnamefont {Papenfort}},
  \bibinfo {author} {\bibfnamefont {V.}~\bibnamefont {Dexheimer}}, \bibinfo
  {author} {\bibfnamefont {M.}~\bibnamefont {Hanauske}}, \bibinfo {author}
  {\bibfnamefont {S.}~\bibnamefont {Schramm}}, \bibinfo {author} {\bibfnamefont
  {H.}~\bibnamefont {St\"ocker}}, \ and\ \bibinfo {author} {\bibfnamefont
  {L.}~\bibnamefont {Rezzolla}},\ }\href {\doibase
  10.1103/PhysRevLett.122.061101} {\bibfield  {journal} {\bibinfo  {journal}
  {Phys. Rev. Lett.}\ }\textbf {\bibinfo {volume} {122}},\ \bibinfo {pages}
  {061101} (\bibinfo {year} {2019})}\BibitemShut {NoStop}%
\bibitem [{\citenamefont {Bauswein}\ \emph
  {et~al.}(2019{\natexlab{a}})\citenamefont {Bauswein}, \citenamefont
  {Bastian}, \citenamefont {Blaschke}, \citenamefont {Chatziioannou},
  \citenamefont {Clark}, \citenamefont {Fischer},\ and\ \citenamefont
  {Oertel}}]{PhysRevLett.122.061102}%
  \BibitemOpen
  \bibfield  {author} {\bibinfo {author} {\bibfnamefont {A.}~\bibnamefont
  {Bauswein}}, \bibinfo {author} {\bibfnamefont {N.-U.~F.}\ \bibnamefont
  {Bastian}}, \bibinfo {author} {\bibfnamefont {D.~B.}\ \bibnamefont
  {Blaschke}}, \bibinfo {author} {\bibfnamefont {K.}~\bibnamefont
  {Chatziioannou}}, \bibinfo {author} {\bibfnamefont {J.~A.}\ \bibnamefont
  {Clark}}, \bibinfo {author} {\bibfnamefont {T.}~\bibnamefont {Fischer}}, \
  and\ \bibinfo {author} {\bibfnamefont {M.}~\bibnamefont {Oertel}},\ }\href
  {\doibase 10.1103/PhysRevLett.122.061102} {\bibfield  {journal} {\bibinfo
  {journal} {Phys. Rev. Lett.}\ }\textbf {\bibinfo {volume} {122}},\ \bibinfo
  {pages} {061102} (\bibinfo {year} {2019}{\natexlab{a}})}\BibitemShut
  {NoStop}%
\bibitem [{\citenamefont {Bauswein}\ \emph
  {et~al.}(2019{\natexlab{b}})\citenamefont {Bauswein}, \citenamefont
  {Bastian}, \citenamefont {Blaschke}, \citenamefont {Chatziioannou},
  \citenamefont {Clark}, \citenamefont {Fischer}, \citenamefont {Janka},
  \citenamefont {Just}, \citenamefont {Oertel},\ and\ \citenamefont
  {Stergioulas}}]{doi:10.1063/1.5117803}%
  \BibitemOpen
  \bibfield  {author} {\bibinfo {author} {\bibfnamefont {A.}~\bibnamefont
  {Bauswein}}, \bibinfo {author} {\bibfnamefont {N.-U.~F.}\ \bibnamefont
  {Bastian}}, \bibinfo {author} {\bibfnamefont {D.}~\bibnamefont {Blaschke}},
  \bibinfo {author} {\bibfnamefont {K.}~\bibnamefont {Chatziioannou}}, \bibinfo
  {author} {\bibfnamefont {J.~A.}\ \bibnamefont {Clark}}, \bibinfo {author}
  {\bibfnamefont {T.}~\bibnamefont {Fischer}}, \bibinfo {author} {\bibfnamefont
  {H.-T.}\ \bibnamefont {Janka}}, \bibinfo {author} {\bibfnamefont
  {O.}~\bibnamefont {Just}}, \bibinfo {author} {\bibfnamefont {M.}~\bibnamefont
  {Oertel}}, \ and\ \bibinfo {author} {\bibfnamefont {N.}~\bibnamefont
  {Stergioulas}},\ }\href {\doibase 10.1063/1.5117803} {\bibfield  {journal}
  {\bibinfo  {journal} {AIP Conference Proceedings}\ }\textbf {\bibinfo
  {volume} {2127}},\ \bibinfo {pages} {020013} (\bibinfo {year}
  {2019}{\natexlab{b}})},\ \Eprint
  {http://arxiv.org/abs/https://aip.scitation.org/doi/pdf/10.1063/1.5117803}
  {https://aip.scitation.org/doi/pdf/10.1063/1.5117803} \BibitemShut {NoStop}%
\bibitem [{\citenamefont {Paschalidis}\ \emph {et~al.}(2018)\citenamefont
  {Paschalidis}, \citenamefont {Yagi}, \citenamefont {Alvarez-Castillo},
  \citenamefont {Blaschke},\ and\ \citenamefont
  {Sedrakian}}]{PhysRevD.97.084038}%
  \BibitemOpen
  \bibfield  {author} {\bibinfo {author} {\bibfnamefont {V.}~\bibnamefont
  {Paschalidis}}, \bibinfo {author} {\bibfnamefont {K.}~\bibnamefont {Yagi}},
  \bibinfo {author} {\bibfnamefont {D.}~\bibnamefont {Alvarez-Castillo}},
  \bibinfo {author} {\bibfnamefont {D.~B.}\ \bibnamefont {Blaschke}}, \ and\
  \bibinfo {author} {\bibfnamefont {A.}~\bibnamefont {Sedrakian}},\ }\href
  {\doibase 10.1103/PhysRevD.97.084038} {\bibfield  {journal} {\bibinfo
  {journal} {Phys. Rev. D}\ }\textbf {\bibinfo {volume} {97}},\ \bibinfo
  {pages} {084038} (\bibinfo {year} {2018})}\BibitemShut {NoStop}%
\bibitem [{\citenamefont {Chatziioannou}\ and\ \citenamefont
  {Han}(2019)}]{Chatziioannou:2019yko}%
  \BibitemOpen
  \bibfield  {author} {\bibinfo {author} {\bibfnamefont {K.}~\bibnamefont
  {Chatziioannou}}\ and\ \bibinfo {author} {\bibfnamefont {S.}~\bibnamefont
  {Han}},\ }\href@noop {} {\  (\bibinfo {year} {2019})},\ \Eprint
  {http://arxiv.org/abs/1911.07091} {arXiv:1911.07091 [gr-qc]} \BibitemShut
  {NoStop}%
%%CITATION = ARXIV:1911.07091;%%
\bibitem [{\citenamefont {Douchin}\ and\ \citenamefont
  {Haensel}(2001)}]{Douchin:2001sv}%
  \BibitemOpen
  \bibfield  {author} {\bibinfo {author} {\bibfnamefont {F.}~\bibnamefont
  {Douchin}}\ and\ \bibinfo {author} {\bibfnamefont {P.}~\bibnamefont
  {Haensel}},\ }\href {\doibase 10.1051/0004-6361:20011402} {\bibfield
  {journal} {\bibinfo  {journal} {Astron. Astrophys.}\ }\textbf {\bibinfo
  {volume} {380}},\ \bibinfo {pages} {151} (\bibinfo {year} {2001})},\ \Eprint
  {http://arxiv.org/abs/astro-ph/0111092} {arXiv:astro-ph/0111092 [astro-ph]}
  \BibitemShut {NoStop}%
%%CITATION = ASTRO-PH/0111092;%%
\end{thebibliography}%

\end{document}